\documentclass[sigconf,10pt]{acmart}
\pdfoutput=1
 \usepackage[export]{adjustbox}
 \usepackage{url}  
 \usepackage{graphicx}  
 \usepackage{subcaption}
 \usepackage{balance}
 \usepackage[linesnumbered,ruled]{algorithm2e}
 \usepackage{amsmath}

 \usepackage{wrapfig}
 \usepackage{xcolor}
 \usepackage{comment}
 \usepackage{tabularx}
 \usepackage{array, makecell}

 \definecolor{brightgreen}{rgb}{0.0, 0.5, 0.0}
 \definecolor{ao(english)}{rgb}{0.0, 0.5, 0.0}
 \definecolor{amber(sae/ece)}{rgb}{1.0, 0.49, 0.0}
 \definecolor{darkorange}{rgb}{1.0, 0.55, 0.0}
 \definecolor{darkmagenta}{rgb}{0.55, 0.0, 0.55}
 
 \newcommand{\RONE}[1]{\textcolor{black}{#1}}
 \newcommand{\RTHREE}[1]{\textcolor{black}{#1}}
 \newcommand{\RFIVE}[1]{\textcolor{black}{#1}}
 \newcommand{\vamsi}[1]{\textcolor{magenta}{Vamsi: #1}}

 \newcommand{\squishlist}{ 
 	\begin{list}{$\bullet$}
 		{ \setlength{\itemsep}{0pt}      \setlength{\parsep}{3pt} 
 			\setlength{\topsep}{3pt}       \setlength{\partopsep}{0pt}
 			\setlength{\leftmargin}{1.5em} \setlength{\labelwidth}{1em}
 			\setlength{\labelsep}{0.5em} } }
 	
 	\newcommand{\squishend}{
 \end{list}  } 

\usepackage{balance}

\def\BibTeX{{\rm B\kern-.05em{\sc i\kern-.025em b}\kern-.08emT\kern-.1667em\lower.7ex\hbox{E}\kern-.125emX}}
    
\copyrightyear{2020}
\acmYear{2020}
\setcopyright{acmcopyright}
\acmConference[SIGMOD'20]{Proceedings of the 2020 ACM SIGMOD International Conference on Management of Data}{June 14--19, 2020}{Portland, OR, USA}
\acmBooktitle{Proceedings of the 2020 ACM SIGMOD International Conference on Management of Data (SIGMOD'20), June 14--19, 2020, Portland, OR, USA}
\acmPrice{15.00}
\acmDOI{10.1145/3318464.3380597}
\acmISBN{978-1-4503-6735-6/20/06}

\hypersetup{draft}
\begin{document}


\title[A Comprehensive Active Learning Benchmark for Entity Matching]{A Comprehensive Benchmark Framework for Active Learning Methods in Entity Matching}

\author{Vamsi Meduri}
\affiliation{%
  \institution{Arizona State University}
}\email{vmeduri@asu.edu}

\author{Lucian Popa}
\affiliation{%
	\institution{IBM Research, Almaden}
}\email{lpopa@us.ibm.com}

\author{Prithviraj Sen}
\affiliation{%
	\institution{IBM Research, Almaden}
}\email{senp@us.ibm.com}

\author{Mohamed Sarwat}
\affiliation{%
	\institution{Arizona State University}
}\email{msarwat@asu.edu}
%
\renewcommand{\shortauthors}{Meduri, Popa, Sen and Sarwat}

%
\begin{abstract}
	Entity Matching (EM) is a core data cleaning task, aiming to identify different mentions of the same real-world entity. 
	Active learning is one way to address the challenge of scarce labeled data in practice, 
	by dynamically collecting 
	the necessary examples to be labeled by an Oracle and refining the learned model (classifier) upon them. 
	In this paper, we build a unified active learning benchmark framework for EM that allows users to easily combine different learning algorithms with applicable example selection algorithms. 
	The goal of the framework is to enable concrete guidelines for 
	practitioners as to what active learning combinations will work well for EM. 
	Towards this, we perform comprehensive experiments on publicly available EM datasets 
	from product and publication domains to evaluate active learning methods, using a variety of metrics including 
	EM quality, $\#$labels and example selection latencies. 
	Our most surprising result finds that active learning with fewer labels can learn a classifier of comparable quality as supervised learning.
	In fact, for several of the datasets, we show that there is an active learning combination that beats the 
	state-of-the-art supervised learning result. 
	Our framework also includes novel optimizations that improve the quality of the learned model by roughly $9\%$ in terms of F1-score and reduce example selection latencies by up to 10$\times$ without affecting the quality of the model.

\end{abstract}

%
\maketitle
\section{Introduction}
\label{sec:intro}
Entity matching (EM) is an important step in data cleaning where the goal is to link different mentions of the same real-world entity. 
Since many real-world downstream applications can benefit from clean data, improving EM continues to be a topic of fervent research. In particular, a popular approach to EM has been to formulate it as an instance of binary classification: Given relations $D_1$, $D_2$ assign one of \emph{match} or \emph{non-match} to each pair of tuples $r \in D_1, s \in D_2$ where $r$ and $s$ represent entity mentions.

Learning a binary classifier usually entails labeled training data upfront (supervised learning), which is a significant investment in terms of human labeling effort. Active learning \citep{settles:tr09} is a popular alternative that can avoid such prohibitive costs and has a history of application in EM going back almost two decades (early attempts include \citet{SarawagiAL,Tejada:2001}). In contrast to supervised learning, active learning employs an \emph{example selector} that chooses the pair of mentions whose labels refine the quality of the classifier learned thus far. By restricting itself to informative pairs of mentions only, active learning hopes to achieve high quality EM while incurring less human labeling effort. 

While previous work has evaluated supervised learning with classifiers of different flavors on the EM task (e.g., \citep{ERKopcke}) and built frameworks such as \textsf{Magellan}~\cite{Magellan} that enable supervised learning-based EM workflows, the same cannot be said for active learning.  
Lacking comprehensive comparative evaluations, it is difficult to say which combinations of classifiers and example selectors work well on the EM task given that several such combinations have been tried in the past. 
Query-by-committee (QBC)~\citep{seung:colt92,freund:ml97} is a specific example selector which has been tried in conjunction with decision trees~\cite{Tejada:2001}, support vector machines and naive Bayes classifiers~\cite{SarawagiAL}. \citet{MozafariBootstrap} propose to implement QBC in a learner-agnostic manner such that the example selector is completely decoupled from the classifier being used. While this makes implementation easier, the question remains whether or not we can gain improved EM quality if the example selector were learner-aware. 
While QBC has seen sustained use \citep{SarawagiAL,Tejada:2001,MozafariBootstrap}, the active learning literature offers other learner-aware example selectors based on margin \citep{SVMALDaphneKoller} which has not seen much use in EM. \citeauthor{MozafariBootstrap} is the only previous work we are aware of that compares against margin example selector while \citeauthor{SarawagiAL} mention it but do not evaluate it. 

\begin{figure*}[htb]
    \centering
	\begin{subfigure}[t]{0.55\textwidth}
		\centering
		\includegraphics[width=\linewidth]{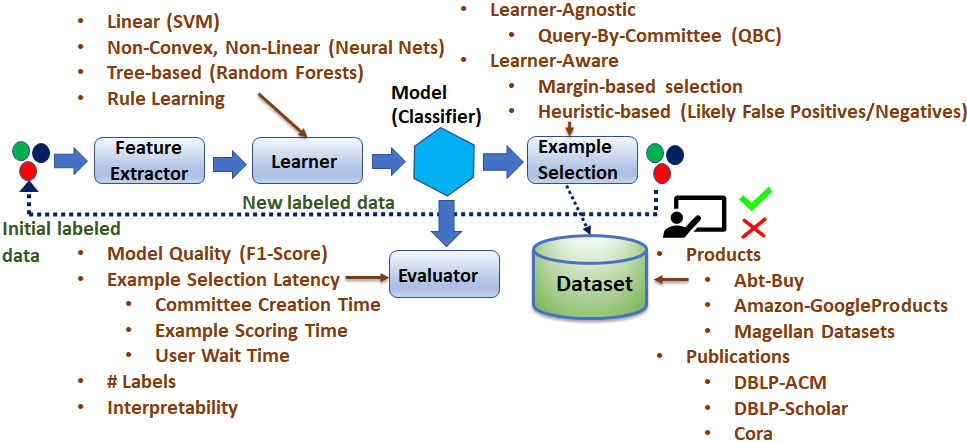}
		\caption{Our Unified Active Learning Benchmark Framework}
		\label{fig:unifiedActive}
	\end{subfigure}
	\begin{subfigure}[t]{0.3\textwidth}
		\centering
		\includegraphics[width=\linewidth]{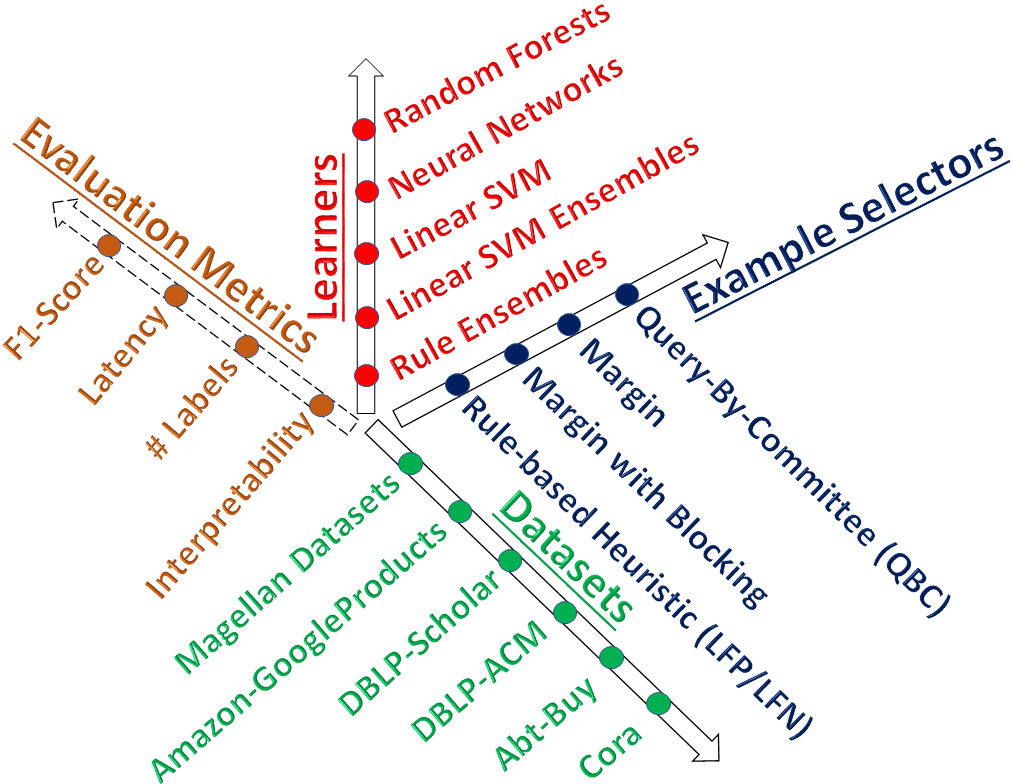}
		\caption{4D view of Unified Active Learning}
		\label{fig:unified3D}
	\end{subfigure}
	\caption{}
	\label{fig:activeLearning}
\end{figure*}

In this paper, we build a comprehensive framework for benchmarking active learning-based EM. Currently, our framework includes representative classifiers of four major types including linear classifiers (e.g., support vector machines), tree-based classifiers (e.g., random forests), non-linear classifiers (e.g., feed-forward neural networks) and rule-based classifiers, and three different types of example selectors including QBC, margin-based and heuristic example selectors (as proposed in \citet{LFPLFN} to learn rule-based classifiers at scale). While QBC is implemented in a learner-agnostic manner with bootstrap \citep{efron:book93, MozafariBootstrap} which allows combining QBC with \emph{any} classifier, margin-based and heuristic example selectors are learner-aware. While not all combinations of classifier and example selector make sense (e.g., some of the heuristic example selectors are explicitly designed for rule-based classifiers), our framework allows maximum plug-and-play ability using which we evaluate active learning approaches on several publicly available EM datasets spread across $product$ and $publication$ domains. 

Our main empirical results show that there is little to choose between margin-based example selection and learner-agnostic QBC in terms of quality of EM achieved, but the former usually results in lower example selection latencies. This situation might change however (depending on the dataset), if we learn a learner-aware ensemble of classifiers with margin-based example selection in which case, EM quality in terms of F1-score might outperform QBC. Our best results appear with learner-aware ensembles and QBC (sans the learner-agnostic committee creation). In particular, learner-aware QBC and random forests, which are natural ensembles of  decision trees, invariably produce the best quality EM approaching F1-scores of $100\%$.  
This highlights the benefits of systematically implementing active learning EM algorithms within the same framework and evaluating them on a level playing field. 

\RONE{Thus, the primary contribution of this work is to build a benchmark framework for active learning-based EM, using which we provide guidelines to practitioners on the combination of learner and example selector that performs the best on various evaluation metrics such as EM quality, latency, $\#$labels and interpretability. Without this framework, we would end up re-implementing the entire, end-to-end active learning pipeline separately for each combination of learner and example selector. Instead, this framework allows for the necessary components to be plugged-in as shown in Fig.~\ref{fig:classHierarchy}, thereby requiring minimum or no changes to the remaining components in the pipeline.}

\RONE{Since both high matching quality and low latency are crucial for active learning scenarios, we propose two general enhancements - learning active ensembles of highly precise classifiers incrementally over several learning iterations for enhanced F1-scores, and blocking to speed up example selection in the context of black-box mathematical models such as Linear SVMs. Given that such enhancements exist to optimize example selection for rule-based classifiers~\cite{Arasu:2010,LFPLFN}, it is essential that these are extended to other learners while evaluating the learners against each other.}

\noindent Following is a summary of our contributions.
\squishlist
	\item We develop a unified active learning benchmark framework that allows users to easily combine multiple learning models with several example selectors for EM. 
    \item We conduct an exhaustive experimental study to compare various active learning methods for EM on multiple, publicly available datasets across two domains using our benchmark framework. Our experiments evaluate different approaches on EM quality, example selection latencies and $\#$labels.
    
    \item Our framework includes various novel optimizations such as being able to learn ensembles of classifiers with active learning. Other classifier-specific optimizations include the usage of blocking with linear classifiers to reduce example selection latencies .
    \item We find that random forests with learner-aware QBC can routinely achieve near-perfect EM quality (progressive F1-score close to $100\%$) on all the datasets we experiment with, while being 10-100x faster w.r.t. example selection latencies than learner-agnostic techniques. 
    
    This significantly improves upon previous EM works, both of the active learning and supervised learning variety, especially on datasets from the product domain.
    \item While previous work \citep{MozafariBootstrap} has reported QBC to outperform margin-based example selection, we find that there is little to choose between the two in terms of EM quality, and that the latter can outperform the former if ensembles are learned.
    \item \RONE{To estimate the effect of labeling errors in crowd-sourcing situations, we evaluate all our active learning approaches with noisy Oracles.} 
    \item We compare rules and tree-based models on an interpretability metric defined by~\citet{conciseERVLDB}. We infer that although tree-based ensembles perform the best on EM quality, they sacrifice interpretability. 
\squishend
\section{Related Work}
\label{sec:related}
In this section, we review other example selectors from the active learning literature, how these relate to EM and prior art from areas related to active learning in EM such as crowdsourcing. Under strong assumptions about data distribution, the earliest active learning algorithms such as \emph{selective sampling} \citep{CohnPassiveAL},  \emph{query-by-committee} (QBC) \citep{seung:colt92,freund:ml97} and margin-based example selection\citep{SVMALDaphneKoller,AggressiveAL} have been shown to either learn the optimal classifier, or reduce the number of candidate classifiers by a fixed fraction with each labeled example.
It is unclear whether such theoretical results hold in practice as QBC and margin-based example selection are reduced to heuristics in this significantly more challenging setting of EM \citep{twofacesAL}. There exist other active learning algorithms such as IWAL (importance weighted active learning) \citep{IWAL} and ConvexHull \citep{activesampling} which either choose a poor objective of label prediction accuracy (instead of F1-score) for EM which is pervasive of class skew or incur excessive labels in practice.

Several prior works on EM have explored the use of crowdsourcing however, the  focus is usually not on learning an EM model but to reduce the number of labels asked from the crowd~\cite{wang:sigmod13,whang:vldb13,vesdapunt:vldb14,wang:sigmod15,CostEffectiveCrowdSourcing,CrowdER,verroios:sigmod17, khan:cikm16}. 
Due to the lack of a reusable EM model, one drawback of such approaches is having to incur costs associated with crowd-sourcing labels every time an instance of EM needs to be solved. 
\RONE{Our framework is meant for learning a non-trivial EM model with active learning and 
we emulate crowdsourcing by modeling imperfect Oracles without label correction methods such as majority voting or label inference.} 
Corleone \citep{Corleone} (and its more scalable version Falcon \citep{Falcon}) 
use random forests due to their interpretable properties to mine the blocking functions automatically, and to perform EM while incurring the least monetary cost for labeling. 
In our experiments, we too pit random forests against rules to compare them in terms of interpretability. But more importantly, our goal underlying the inclusion of random forests into our framework is to find out how well they can perform EM and and how many labeled examples they incur via active learning.

\RONE{Currently, our EM framework includes feed-forward neural networks admittedly simpler than recently proposed deep learning architectures that perform EM with representation learning \citep{MudgalDeepER,kasai:acl19}. We evaluate the performance of non-convex non-linear classifiers against other kinds of (shallow) classifiers when learned with active learning. But currently, as we shall see in our experiments, the EM results acquired with complex architectures are clearly behind our best approaches (see Fig.~\ref{fig:ActivevsSupervisedMagellan} where we compare \citet{MudgalDeepER} with learner-aware QBC on random forests).}
\section{Benchmark Overview}
\label{sec:benchmark}
Figure~\ref{fig:unifiedActive} presents the system architecture of our unified active learning benchmark framework. In contrast to supervised learning which requires a significant amount of upfront training data, active learning requires a limited amount of initial labeled data ($\sim$ 30 examples in our framework) from which the learner produces an initial model. The example selector chooses ambiguous, unlabeled examples that the model finds hard to predict the label for and queries an $Oracle$ (human or ground truth) for those labels. The newly labeled data is added to the cumulative set of training data obtained thus far upon which a refined model is learned. In each active learning iteration, the learned model is evaluated by an evaluator w.r.t. a variety of metrics pertaining to label prediction quality, informative example selection latency, model interpretability and $\#$labels which will be explained in detail. 
We have four basic components in our framework - \emph{feature extractor, learner, example selector} and $evaluator$. We use the Object-Oriented paradigm of inheritance to model each component as a base class and extend it into a child class to support specialized functionalities.

\textbf{Feature Extractor:}  
We apply a blocking function as a pre-processing step to eliminate obvious non-matches among the Cartesian product of record pairs created from the tables to be matched. \RTHREE{We obtain the feature vectors by applying 21 similarity functions from Java \textsf{Simmetrics} library~\cite{Simmetrics} on all the matching schema attributes across the two tables. If one or both of the pre-aligned attributes of a record pair are null or missing, the similarity evaluates to 0. We use the same set of feature vectors across all the classifiers in the framework barring rule-based models from~\citet{LFPLFN} which only support 3 (equality, Jaro-Winkler and Jaccard) out of the 21 similarity functions. While linear, non-convex non-linear and tree-based classifiers use floating point feature vectors (an example dimension can be JaccardSim(left-table.attr,right-table.attr)), rule-based models evaluate each similarity function on a discrete set of thresholds in (0,1] and create Boolean feature dimensions  (e.g., JaccardSim(left-table.attr,right-table.attr)$\ge\tau$ with $\tau$ from 0.1 to 1.0).} 

\begin{figure}[htb]
	\centering
	\includegraphics[width=\linewidth]{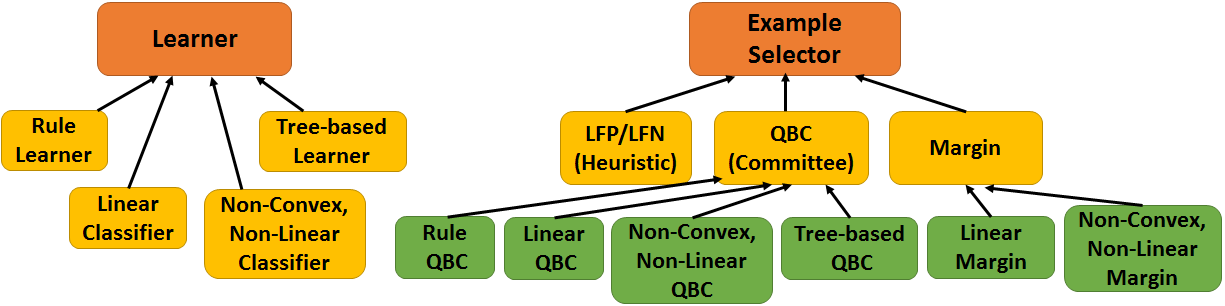}
	\caption{Class Hierarchy of Learners $\&$ Selectors}
	\vspace*{-.3cm}
	\label{fig:classHierarchy}
\end{figure}

\textbf{Learner and Example Selector: } As mentioned in section~\ref{sec:intro}, we support a learner from each of the following diverse categories - linear, non-convex non-linear, tree-based and rule-based classifiers. Figure~\ref{fig:classHierarchy} shows how we derive a sub-class for each classifier from the learner base class. Since the base class hosts the common functionalities across all the learners, each sub-class only needs to contain methods specific to a learner. On similar lines, we support a learner-agnostic example selector and two learner-aware selection strategies. While the learner-agnostic selection strategy of query-by-committee (QBC) can be applied to any classifier, random forests inherently learn a committee of trees in a learner-aware manner. 
 Therefore, a relaxed variant of QBC is applied to such tree-based learners. In contrast, learner-aware selection strategies can work only with specific learners. For instance, margin-based selection is compatible with linear and non-convex non-linear classifiers (and is extended accordingly in Figure~\ref{fig:classHierarchy}) but not with random forests or rules. Heuristic-based technique of LFP/LFN is devised only for the rule-based classifier in~\citet{LFPLFN} and does not have any child classes. Our framework records the compatibilities between specific example selectors and classifiers through the class hierarchy shown in the figure.
 
 \textbf{Evaluator: } We evaluate the active learning methods on quality, latency, $\#$labels and interpretability.

\emph{Quality}: The quality of the model is determined by the usefulness of the examples retrieved by the example selector. In each active learning iteration, once we obtain a refined model, we test it on the entire set of data (both labeled and unlabeled pairs obtained post-blocking). Matching pairs get a label of 1 and non-matching pairs are assigned 0 as the label. Precision, recall and F1-score are computed based on the number of matching pairs predicted accurately. 
\newline
\emph{Latency}: The time taken by an example selector to retrieve the ambiguous examples in each iteration together with the training time of the model on the cumulative set of labeled examples determine the overall user wait time. The example selection time for QBC can be broken down into committee creation time, which is the time taken to create a committee of classifiers and example scoring time which is the time taken to compute the disagreement (entropy) metric for all the unlabeled examples and pick the most ambiguous ones out of them. For learner-aware approaches such as margin and LFP/LFN the latency only comprises the example scoring time as there is no classifier committee to be created. For tree-based approaches, the committee of random trees is created during the training phase. Hence, the example selection time for random forests is the time required to compute the entropy among the committee of trees. This will be further described in the subsequent sections.

\RFIVE{\emph{$\#$Labels}: This is the minimum number of labeled examples required by each active learning method to learn a model that converges to its best achievable quality. If adding more labels no longer changes the quality of the model learned in terms of its Test F1-scores, the model can be deemed to have reached its convergent state. The lower the $\#$labels, the more effective is the active learning strategy used. 
If all the unlabeled examples are required to achieve the best possible classifier, it means that the active learning policy used is ineffective and it is better to resort to supervised learning instead, in such scenarios.}

\emph{Interpretability}: This is a metric that determines how readable and interpretable the model is to the end user. Concise rules are preferred over mathematical models by humans especially in scenarios where explainability takes precedence over model quality or effectiveness. \RFIVE{Interpretability is defined as being inversely proportional to the number of $atoms$ in a rule~\cite{conciseERVLDB}, where an atom is defined as a Boolean predicate that consists of a similarity function applied over an attribute pair accompanied by a threshold.} Since random forests are ensembles of decision trees which consist of similar logical predicates, we compare tree-based approaches to the rule-based models~\cite{LFPLFN} w.r.t. interpretability.
\section{Compared Approaches}
\label{sec:approaches}

In this section, we describe the various active learning methods, i.e., example selection policies and how they are applied to each learner we implement in our benchmark. 
As mentioned in sections~\ref{sec:intro} and~\ref{sec:benchmark}, we categorize the example selectors as being learner-agnostic or learner-aware. While query-by-committee (QBC) is a learner-agnostic approach and can be applied to all classifiers, margin and Likely False Positives/Negatives (LFP/LFN) are learner-aware strategies. 

\subsection{Query-by-committee (QBC)}
\label{sec:QBC}
~\citet{MozafariBootstrap} propose query-by-committee (QBC) as a generic strategy that can be applied to any learner. Variants of it have been proposed earlier in~\citet{SarawagiAL}. QBC formulates the ambiguous example space based on the disagreement among a committee of classifiers regarding the labels of examples.
\begin{figure}[htb]
	\centering
	\includegraphics[width=0.75\linewidth]{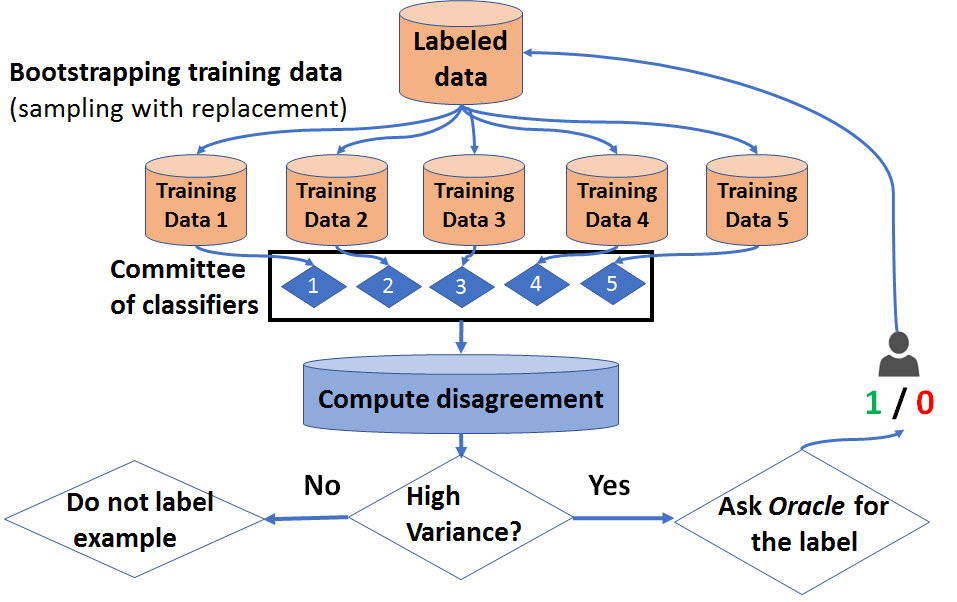}
	\caption{query-by-committee}
	\vspace*{-.2cm}
	\label{fig:QBC}
\end{figure}
As illustrated in Figure~\ref{fig:QBC}, QBC draws \textsf{B} (=5 in the figure) bootstrap training samples with replacement out of the aggregate labeled data from which a committee of \textsf{B} classifiers is learned. Each classifier in the committee predicts the labels of all the unlabeled examples and disagreement is computed based on the entropy among the classifiers upon the assigned label to each example. In lieu of entropy, we use variance defined by~\citet{MozafariBootstrap} 
over an an unlabeled example $Ex_i$ 
as $Variance(Ex_i) = {\frac{P_i}{C}(1-\frac{P_i}{C})}$ where $C$ is \#classifiers in the committee and $P_i$ is \#classifiers which assign a positive class label (matching pair in the context of EM) to $Ex_i$.

\looseness=-1
Examples with the highest variance are passed for labeling after which they are included into the aggregate training set. When several examples have the same measure of high disagreement, a random subset of those examples is selected.  
A way to reduce randomness is to 
increase the \# classifiers in the committee to get fewer examples with the same variance. However there are two hindrances: 1) larger bootstrap committees take longer to train,  2) not every classifier in the committee can be unique as the samples are drawn from the same training set and may contain overlapping examples. In general, larger committees are expected to select more informative examples than smaller committees.

\subsubsection{Tree-based Classifiers}
\label{sec:treeQBC}
As mentioned before, QBC is learner-agnostic and can be applied to all learners such as linear, non-convex non-linear and rule-based classifiers. However, tree-based classifiers such as random forests naturally learn an ensemble of trees in a learner-aware manner during their training phase. Hence, the overhead of creating a committee of classifiers from re-sampled labeled data is unnecessary. We directly use the decision trees in a random forest as the classifier committee to compute the variance on the set of unlabeled examples in each active learning iteration. 
\RTHREE{We use the same settings as the~\textsf{Corleone}~\cite{Corleone} system to implement the learner for random forests in our benchmark framework. Each random forest contains random decision trees of unlimited depth and uses a random subset of log$_2$($Dim$+1) features for node splitting from a total of $Dim$ features. Although Corleone uses 10 decision trees per forest, we allow $\#$trees to be parameterized.}

\subsection{Margin}
\label{sec:margin}
Margin measures the confidence of a classifier based on how far its predicted labels are from the decision boundary. Although the notion of margin has been originally proposed for linear classifiers, non-convex variants of margin~\cite{NonConvexLoss} have also been proposed. We apply margin as an active learning strategy to both linear and non-convex non-linear classifiers. 
\subsubsection{Linear Classifiers}
\label{sec:linearMargin}
In the ML literature, version space of linear learners can be defined as the candidate set of classifiers that can separate the positive from the negative training examples in the aggregate set of labeled data. Margin-based selection sorts unlabeled examples based on their informativeness and selects those examples whose inclusion into the labeled data leads to a drastic reduction of the version space in each active learning iteration. This results in an earlier convergence to the ideal classifier than committee-based strategies. Margin-based selection for linear classifiers has been theoretically proved to aggressively halve the version space in each active learning iteration in the binary classification scenario~\cite{SVMALDaphneKoller} thus terming it as an aggressive strategy while naming committee-based techniques like QBC as passive strategies in the ML literature~\cite{AggressiveAL}. 

Margin for a binary linear classifier is defined as the distance of a feature vector to the separating hyperplane and the strategy picks the unlabeled examples which are closest to the hyperplane. Margin can be approximated by the magnitude of the dot product of a feature vector $X$ with the separating hyperplane unit weight vector $W$ added to normalized bias $b$ as $W.X + b$. The sign of the dot product is ignored because ambiguous examples are chosen from both the classes. It is less likely although possible, that two distinct feature vectors fetch the same dot product, thus making margin-based selection more deterministic than QBC. 
\subsubsection{Non-Convex Non-Linear Classifiers}
\label{sec:nonConvexMargin}
We use a neural network with a single hidden layer as a non-convex non-linear classifier implemented in our framework. During the forward pass of the training phase, we feed the aggregate labeled data at the input layer of the neural network. Given $N$ labeled record pairs each with a feature vector of $Dim$ dimensions and a label of 1 to indicate matching or 0 for non-matching, they are passed to a hidden layer which converts each of these $Dim$-dimensional vectors into $h$-dimensional vectors using an affine combination of hidden-weights with the input features followed by a ReLU activation function. $h$ is the number of neurons in the hidden layer. The intermediate feature vectors from the hidden layer are normalized using a batch-normalization layer~\cite{ioffe:icml15} before passing them to the output layer. At the output layer, the intermediate feature vectors are converted from $h$ dimensions into a single dimension using an affine layer. 
\begin{figure}[htb]
	\centering
	\includegraphics[width=0.6\linewidth]{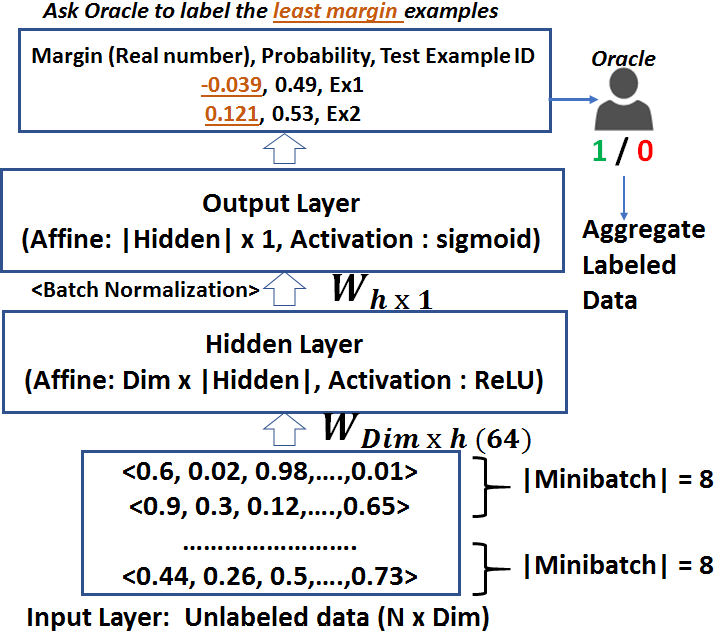}
	\caption{Margin-based selection for Neural Networks}
	\vspace*{-.2cm}
	\label{fig:MarginNN}
\end{figure}
The affine output is termed as the margin (see margin definition for non-convex classifiers~\cite{NonConvexLoss}) which is passed to the sigmoid activation function that emits an output probability. If the output probability > 0.5, the record pair is labeled to be matching else, non-matching. We use L2-loss function and Stochastic Gradient Descent (SGD) with momentum as the optimization function to update the weights during the backpropagation phase. We use 50 epochs and a mini-batch size of 8 during training. For SGD, we use a learning rate of 0.001, a decay constant of 0.99 and a momentum of 0.95. We also use drop-out regularization by turning off half of the hidden nodes randomly during training to prevent overfitting. We could see more stability in the neural network predictions because of batch-normalization and drop-out regularization.

Once a trained neural network is obtained in each active learning iteration, we pass the unlabeled examples to the input layer as shown in Figure~\ref{fig:MarginNN}. At the output layer once we obtain the margin and the output probability, we pass the top-\textsf{K} examples with the least margin to the Oracle for labeling and include them in the labeled data. The ambiguity of an example can be inferred directly from the output probability itself. If it is close to 0.5, the classifier is most ambiguous about its label. This intuitive logic can be used to cross-verify the theoretical margin definition from~\citet{NonConvexLoss}. Since margin obtained from the affine output layer is fed as an input to the sigmoid function, the lower the margin, the closer to 0.5 its sigmoid evaluation would be.

\subsection{Likely False Positives / Negatives (LFP/LFN)}
\label{sec:LFP/LFN}
\begin{figure}[htb]
    \centering
	\includegraphics[width=\linewidth]{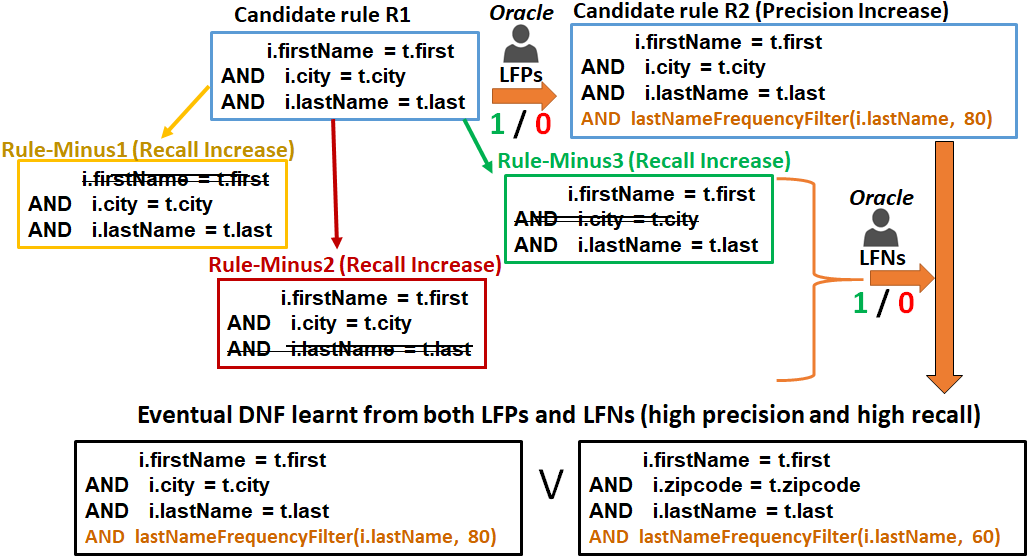}
	\vspace*{-.6cm}
	\caption{LFP/LFN heuristic for Rule-based Learners}
	\vspace*{-.2cm}
	\label{fig:LFPLFN}
\end{figure}
\looseness=-1
LFP/LFN is an example selection heuristic devised for rule-based learning~\cite{LFPLFN}. Active learning is used to learn entity matching rules 
expressed as monotone DNF formulas, that is, disjunctions of conjunctive rules constructed from individual atomic predicates. An example of a conjunctive candidate rule matching user profiles across two distinct social media platforms \textsf{\footnotesize P1} and \textsf{\footnotesize P2} may be \textsf{\footnotesize P1.firstName = P2.FName AND P1.lastName = P2.LName AND P1.city = P2.city}, based on equality of first and last names and cities.
In order to improve precision of the candidate rule, LFP/LFN picks matches predicted by the rule on the unlabeled data that are likely to be non-matches (by using a feature similarity heuristic), and passes these Likely False Positives (LFPs) to the Oracle for labeling.
 
As a result of such labeling, in the next iteration, the system will learn a higher precision rule. For example, a new, more selective predicate may be added to the earlier conjunctive candidate rule: \textsf{\footnotesize lastNameFrequencyFilter(P1.lastName,80)}, filtering out the 
most frequently occurring last names (e.g., in the top 80 percentile). Similarly, LFP/LFN also identifies pairs of records that are {\em not} predicted to be matching by an existing rule but are likely to be matches.
These are the Likely False Negatives (or LFNs) which are again labeled by the Oracle. The LFNs are obtained by executing relaxed variants of the candidate rule $R$ called $Rule$-$Minus$ rules (see Figure~\ref{fig:LFPLFN}); by dropping predicates from $R$, these relaxed rules 
find missed positive examples, and ultimately enhance recall.
New conjunctive rules are thus learned from labeled LFPs and LFNs leading to enhanced precision and recall.
\section{Time and Quality Enhancements}
\label{sec:enhancements}
\looseness=-1
We propose two enhancements, blocking and active ensembles, to improve the runtime and quality of example selection strategies for active learning. 
While generally 
applicable to any underlying active learning model and any example selector, 
we describe them in the context of margin-based selection for linear classifiers.

\subsection{Blocking}\label{sec:blocking}

Blocking has been used for EM~\cite{ERKopcke} to prune, 
out of the Cartesian product of all possible pairs of records, 
those pairs 
that are unlikely to be matches. In contrast to the work so far, we propose blocking for the specific purpose of discovering ambiguous examples for selection strategies without having to compute the ambiguity metric for the entire space of unlabeled data. Hence, unlabeled examples that are unlikely to be ambiguous will be preemptively ignored using blocking.

While most blocking techniques were devised for rule-based learners, \citet{LSHLinearSVM} propose two variants of Locality Sensitive Hashing (LSH) to speed up margin-based selection for linear classifiers. 
Contrary to their approach, the blocking technique we propose forgoes even a full feature vector construction on each unlabeled example and avoids dot product computations aggressively. We apply our learner-aware blocking on top of margin strategy and not for QBC because a majority of the time in QBC is spent in the construction of a classifier committee which is dependent on the already labeled data. So pruning unlabeled data gives meager benefits for QBC.
\begin{figure}[htb]
	\centering
    \includegraphics[width=\linewidth]{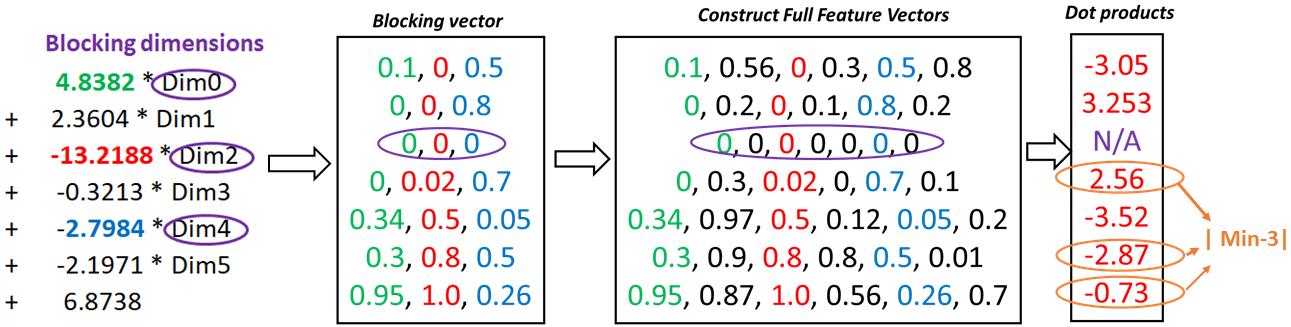}
    \vspace*{-.6cm}
	\caption{Multiple blocking dimensions for SVMs}
	\vspace*{-.2cm}
	\label{fig:blockingDims}
\end{figure}

Margin-based example selection computes ambiguity based on the distance of each unlabeled example from the separating hyperplane.
Our blocking technique skips the margin (dot product) computation for examples whose feature dimensions evaluate to 0 because if all the dimensions of a feature vector $X$ are 0s, $W.X+b = b$ 
i.e., margin equals bias $b$ whose sign decides the class label of $X$ without ambiguity. However, instead of constructing all the feature dimensions for every unlabeled example, we only evaluate the blocking dimension and check if it is equal to 0. We assume that the blocking dimension has the highest predictive power among all the feature dimensions and if it is 0, then all other feature dimensions evaluate to 0. 
The weights of all the feature dimensions are readily available in the weight vector $W$ of the linear classifier; 
a possible blocking dimension is the one with the highest absolute weight.

Since a single blocking dimension may not be predictive enough in determining the values of all remaining feature dimensions, we pick multiple feature dimensions with top-\textsf{K} absolute weights as the blocking dimensions (see Figure~\ref{fig:blockingDims}). 
As we want to prune away high-confidence examples from both the matching and non-matching classes, the top-\textsf{K} (=3 in the figure) blocking dimensions have the largest magnitude in the weight vector disregarding the sign. 
As per the figure, all the blocking dimensions evaluate to 0 for the third example. Hence, we skip it and compute full feature vectors and dot products for all other examples and select those with the least absolute dot products (margin) for labeling.
\subsection{Active Ensemble of Linear Classifiers}\label{sec:ensemble}
In contrast to learning an ensemble using supervised algorithms, the active ensemble is learned incrementally over several active learning iterations. 
In the context of entity matching, active learning for an ensemble of several high precision rules, rather than for a single rule, has been shown to significantly enhance recall~\cite{Arasu:2010, LFPLFN}. Along the same lines, we learn an active ensemble of linear classifiers.

\begin{figure}[htb]
	\centering
	\includegraphics[width=\linewidth]{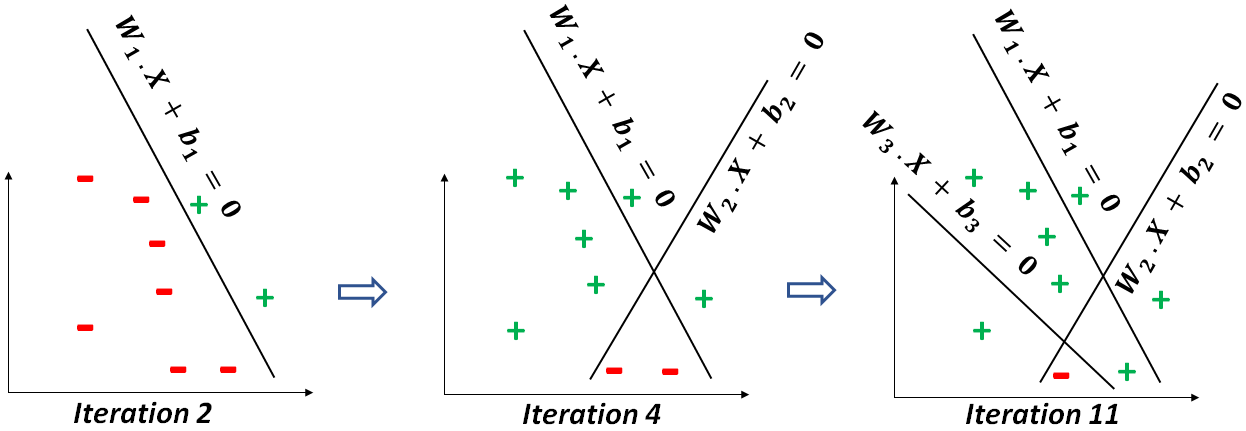}
	\vspace*{-.6cm}
	\caption{Active ensemble of linear classifiers}
	\vspace*{-.2cm}
	\label{fig:ensemble}
\end{figure}
Figure~\ref{fig:ensemble} illustrates the active ensemble into which three linear classifiers are accepted by the time active learning terminates. 
We ensure that distinct classifiers are learned into the ensemble by eliminating the positive label predictions (denoted by + in the figure) or pairs which are predicted to be matching by the accepted classifiers in the ensemble from unlabeled and labeled data before attempting to learn a new classifier. The next model is thus learned from the remaining pool of uncovered examples in the subsequent iterations. Eventually, the union of the positive predictions made by all the accepted linear classifiers in the ensemble are labeled as belonging to the positive class. This can lead to high recall at the expense of losing precision if the accepted classifiers do not exceed a preset precision threshold $\tau$. The precision is computed on the selected examples in each active learning iteration whose labels are provided by the Oracle. If the precision computed on the matches predicted by a candidate linear classifier is $\ge$ $\tau$, it is accepted into the ensemble and its covered examples are removed from the labeled and unlabeled example sets. We set $\tau$ to 0.85 uniformly on all the EM datasets. Ensemble is a general enhancement and can also be applied to QBC but the prohibitively high committee creation times of QBC confined our implementation of ensemble to margin-based strategies.

Similar time and quality enhancement techniques can also be tried for non-convex non-linear models though we have not explored those in this paper. A possible blocking solution for non-linear classifiers would be to include the largest weights for each exponent - $X^n$, $X^{n-1}$,..., $X^2$, $X^1$ as the blocking dimensions. 
Blocking during example selection for rule-based or tree-based models is trivial as the blocking predicate (similarity function and threshold evaluation) can be executed on all the unlabeled examples to prune away non-qualifying examples. 
Active ensemble for neural networks can be applied as discussed in the current section without much of a modification. 
Learning active ensembles for rules already exists in~\citet{LFPLFN}.


\begin{table*}[htb]
	\vspace{-2ex}
	\centering
	\caption{Details of the Public EM Datasets.}
	\vspace{-1ex}
	\label{tab:em_datasets}
	\begin{small}
		\begin{tabular}{|c|l|c|c|c|c|}
			\hline

			{\textbf{\scriptsize Dataset}}&\makecell{{\textbf{\scriptsize Matched Columns}}}&{\textbf{\scriptsize \#Total Pairs}} &{\textbf{\scriptsize \#Post-Blocking Pairs}} &{\textbf{\scriptsize Class skew}}\tabularnewline
			\hline
			\scriptsize Abt-Buy &\makecell{\scriptsize  \{name, description, price\}} &\scriptsize  1.18 M  &\scriptsize  8682 &\scriptsize 0.12 \tabularnewline
			\hline
			\scriptsize Amazon-GoogleProducts &\makecell{\scriptsize  \{name, description, manufacturer, price\}} &\scriptsize  4.39 M  &\scriptsize  14294 &\scriptsize  0.09 \tabularnewline
			\hline
			\scriptsize DBLP-ACM &\makecell{\scriptsize  \{title, authors, venue, year\}} &\scriptsize  6 M  &\scriptsize  11194 &\scriptsize  0.198 \tabularnewline
			\hline
			\scriptsize DBLP-Scholar &\makecell{\scriptsize  \{title, authors, venue, year\}} &\scriptsize  168 M  &\scriptsize  49042 &\scriptsize  0.109 \tabularnewline
			\hline
			\scriptsize Cora & \makecell{\scriptsize \{author, title, venue, address, publisher, editor, date, vol, pgs\}} &\scriptsize  0.97 M  &\scriptsize  114525 &\scriptsize  0.124 \tabularnewline
			\hline
			\scriptsize \RFIVE{Walmart-Amazon} & \makecell{\scriptsize \RFIVE{\{brand, modelno, title, price, dimensions, shipweight,}\\ \scriptsize \RFIVE{orig\_longdescr, shortdescr, longdescr, groupname\}}} &\scriptsize  \RFIVE{56.37 M}  &\scriptsize  \RFIVE{13843} &\scriptsize \RFIVE{0.083} \tabularnewline
			\hline
			\scriptsize \RFIVE{Amazon-BestBuy} & \makecell{\scriptsize \RFIVE{\{brand, title, price, features\}}} &\scriptsize \RFIVE{21.29 M} &\scriptsize \RFIVE{395} &\scriptsize \RFIVE{0.147}
			\tabularnewline
			\hline
			\scriptsize \RFIVE{BeerAdvocate-RateBeer} & \makecell{\scriptsize \RFIVE{\{beer\_name, brew\_factory\_name, style, ABV\}}} &\scriptsize \RFIVE{13.03 M}  &\scriptsize \RFIVE{450} &\scriptsize \RFIVE{0.151} \tabularnewline
			\hline
			\scriptsize \RFIVE{BuyBuyBaby-BabiesRUs} & \makecell{\scriptsize \RFIVE{\{title, price, is\_discounted, category, company\_struct, company\_free,}\\ \RFIVE{\scriptsize brand, weight, length, width, height, fabrics, colors, materials\}}} &\scriptsize \RFIVE{54.5 M}  &\scriptsize \RFIVE{400} &\scriptsize \RFIVE{0.27} \tabularnewline
			\hline
		\end{tabular}
	\end{small}
\end{table*}
\section{Experiments}
\label{sec:exp}
We use a cluster with 24 Intel Xeon 2.4GHz CPUs each containing 6 CPU cores and 99 GB main memory, but a limited Java heap space of 4 GB.
We use Weka~\cite{Weka} for the implementation of SVM and random forests while we use Apache SystemML~\cite{SystemML} for neural networks. As we have mentioned in Section~\ref{sec:intro}, we answer the following questions:
\squishlist
	\item Among the example selection strategies applicable to each classifier, which is the best performing approach w.r.t. both EM prediction quality and latency?
	\item Can active learning methods achieve comparable quality metrics as supervised learning? If so which is the best combination of learner and example selector?
	\item How many labels are required by each active learning method on a dataset to reach a convergent F1-score?
	\item How does rule-based learning~\cite{LFPLFN} compare to tree-based learners on quality and interpretability?
\squishend

\RFIVE{Our experiments can be classified into two broad categories where we assume the presence of either ``perfect'' or ``imperfect (noisy)'' Oracle. We use perfect Oracles without labeling error for our experiments on \textsf{Abt-Buy}, \textsf{Amazon-Google Products} from the \textsf{Products} category and \textsf{DBLP-ACM}, \textsf{DBLP-Scholar} and \textsf{Cora} from the \textsf{Publication} domain. For experiments using noisy Oracles, we choose Abt-Buy, \textsf{Walmart-Amazon} (Products), \textsf{Amazon-BestBuy} (Electronics), \textsf{BeerAdvocate-RateBeer} (Beer) and \textsf{BuyBuyBaby-BabiesRUs} (Baby Products) upon which earlier works like \textsf{Magellan}~\cite{Magellan} and \textsf{DeepMatcher}~\cite{MudgalDeepER} achieve an F1-score of 0.6 - 0.7.}

Each dataset contains left and right tables that produce a Cartesian product of record pairs, whose size is denoted by ``$\#$Total Pairs'' in Table~\ref{tab:em_datasets}. 
To reduce the size of candidate pairs to be matched, during the feature extraction phase (see Section~\ref{sec:benchmark}), we prune away the obvious non-matches using Jaccard similarity function with a numerical threshold in an offline blocking step on the tokenized attributes from each pair.
We set the threshold to 0.1875 to roughly retain the same number of post-blocking pairs as~\citet{MozafariBootstrap,CrowdER} on Abt-Buy, DBLP-ACM and DBLP-Scholar. \RFIVE{We use conservative similarity thresholds of 0.12 on Amazon-GoogleProducts and 0.16 on Cora and Walmart-Amazon to avoid pruning too many non-matching pairs. Due to the unavailability of the entire ground truth for the Amazon-BestBuy, Beer and Baby Products datasets, we use \textsf{Labeled Data L} from~\citet{magellandata} as the set of post-blocking pairs.}
\newline\newline
\textit{Train-Test Splits and Termination Criteria}: \RFIVE{We start active learning with a seed of 30 labeled examples. In each active learning iteration, we query the Oracle (which happens to be the available ground truth on these datasets)  for the labels of a batch of 10 examples chosen from the unlabeled set, upon which the learned model is refined and evaluated on the test set. We use the following settings for train-test splits.} 
\squishlist
\item \RFIVE{We evaluate active learning methods on the test set created from all the post-blocking pairs, while progressively querying the Oracle for a sample of them to be added to the training set in each labeling iteration. While an earlier crowd-sourcing work~\citet{vesdapunt:vldb14} defines progressive recall, we analogously define $progressive$ $F1$ as the test F1-score obtained on post-blocking pairs.}

\item \RFIVE{For active vs. supervised learning experiments, we create the conventional train-test splits (with the same class skew as post-blocking pairs) used in supervised learning scenarios. 80$\%$ of the post-blocking pairs form an unlabeled set, out of which examples are selected in each learning iteration, while the remaining 20$\%$ form a held-out test set upon which the learned models are evaluated. We use this only for the experiments in Fig.~\ref{fig:ActivevsSupervisedMagellan} and~\ref{fig:ActiveVsSupervised}.}
\squishend

\begin{figure*}[htb]
	\centering
	\begin{subfigure}[t]{0.2\textwidth}
		\centering
		\includegraphics[width=\linewidth]{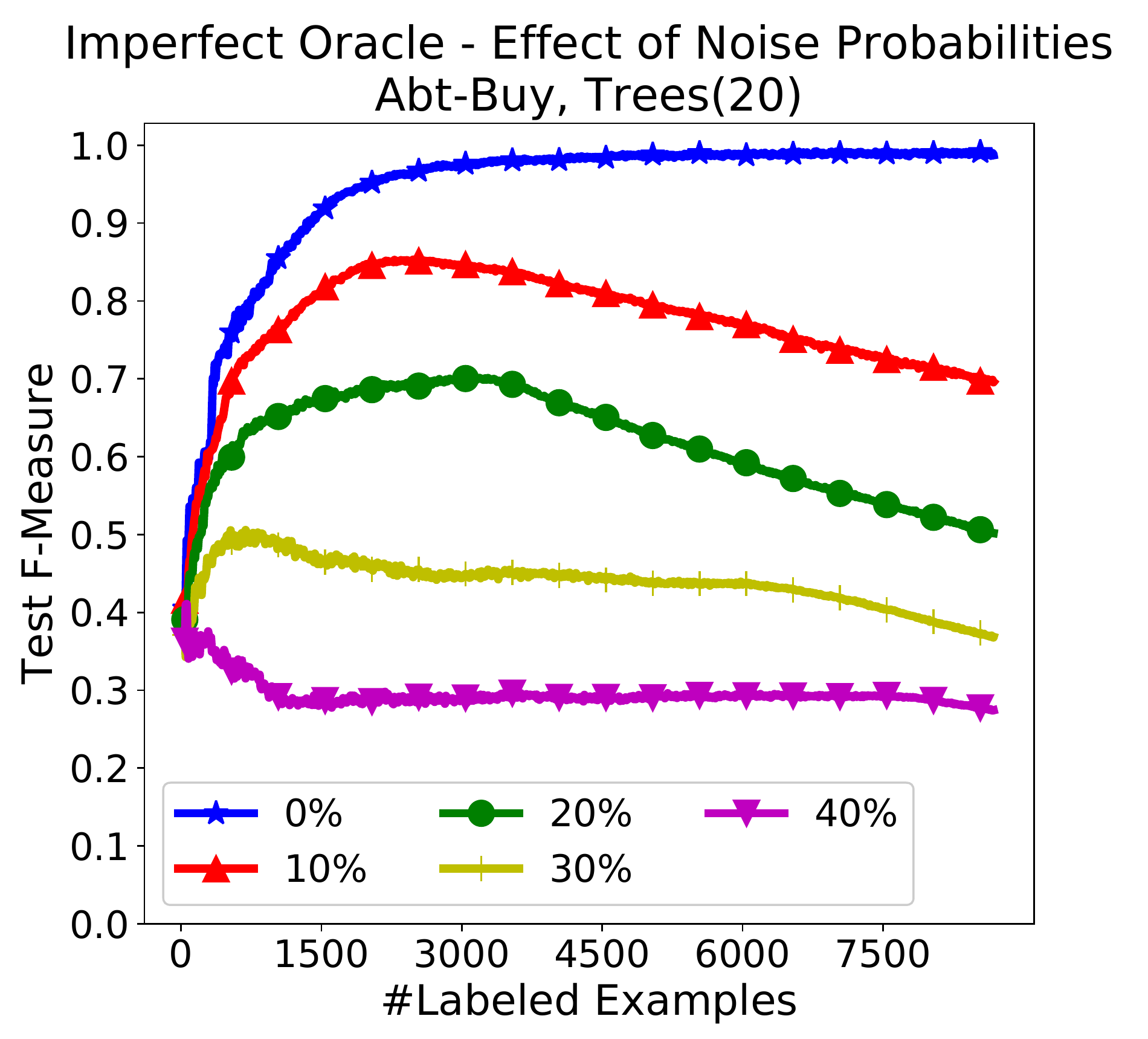}
		\caption{Tree-based}
		\label{fig:Abt-Buy-Noisy-RF-F1}
	\end{subfigure}
	\begin{subfigure}[t]{0.2\textwidth}
		\captionsetup{singlelinecheck = false, format= hang, justification=raggedright, font=footnotesize, labelsep=space}
		\centering
		\includegraphics[width=\linewidth]{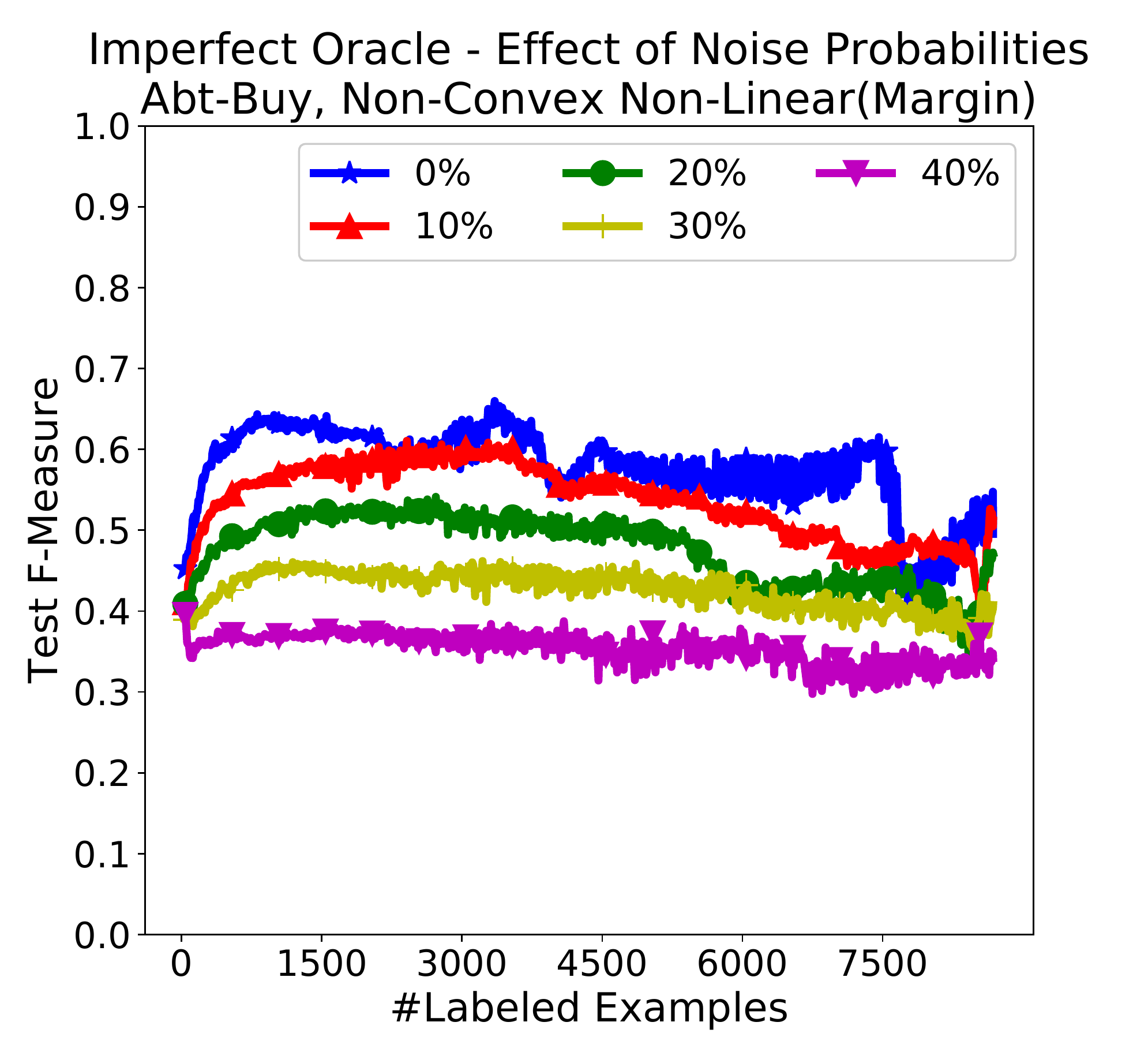}
		\caption{Non-Convex Non-Linear}
		\label{fig:Abt-Buy-Noisy-NN-F1}
	\end{subfigure}
	\begin{subfigure}[t]{0.2\textwidth}
		\centering
		\includegraphics[width=\linewidth]{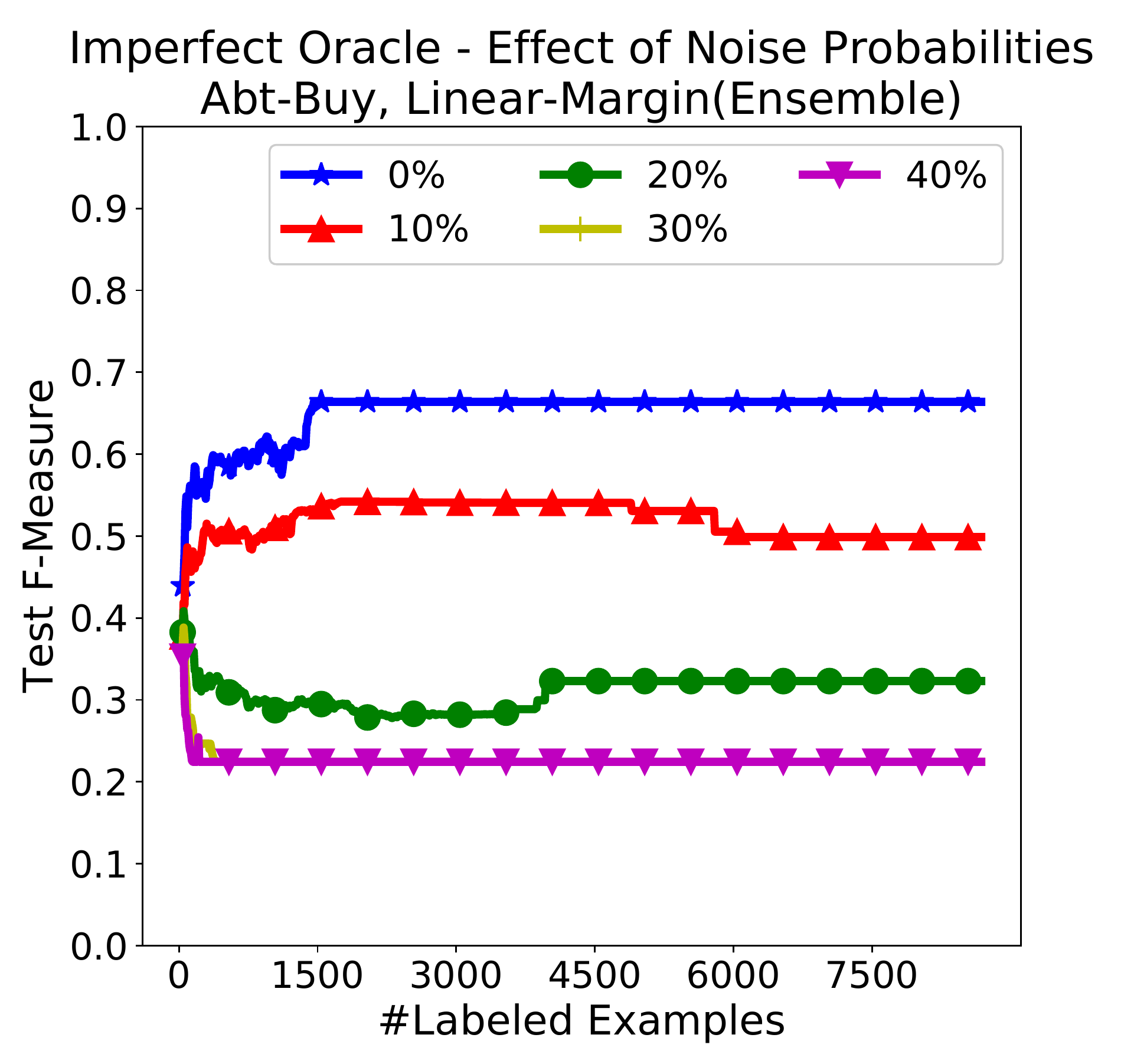}
		\caption{Linear (Ensembles)}
		\label{fig:Abt-Buy-Noisy-SVMEnsemble-F1}
	\end{subfigure}
	\begin{subfigure}[t]{0.2\textwidth}
		\centering
		\includegraphics[width=\linewidth]{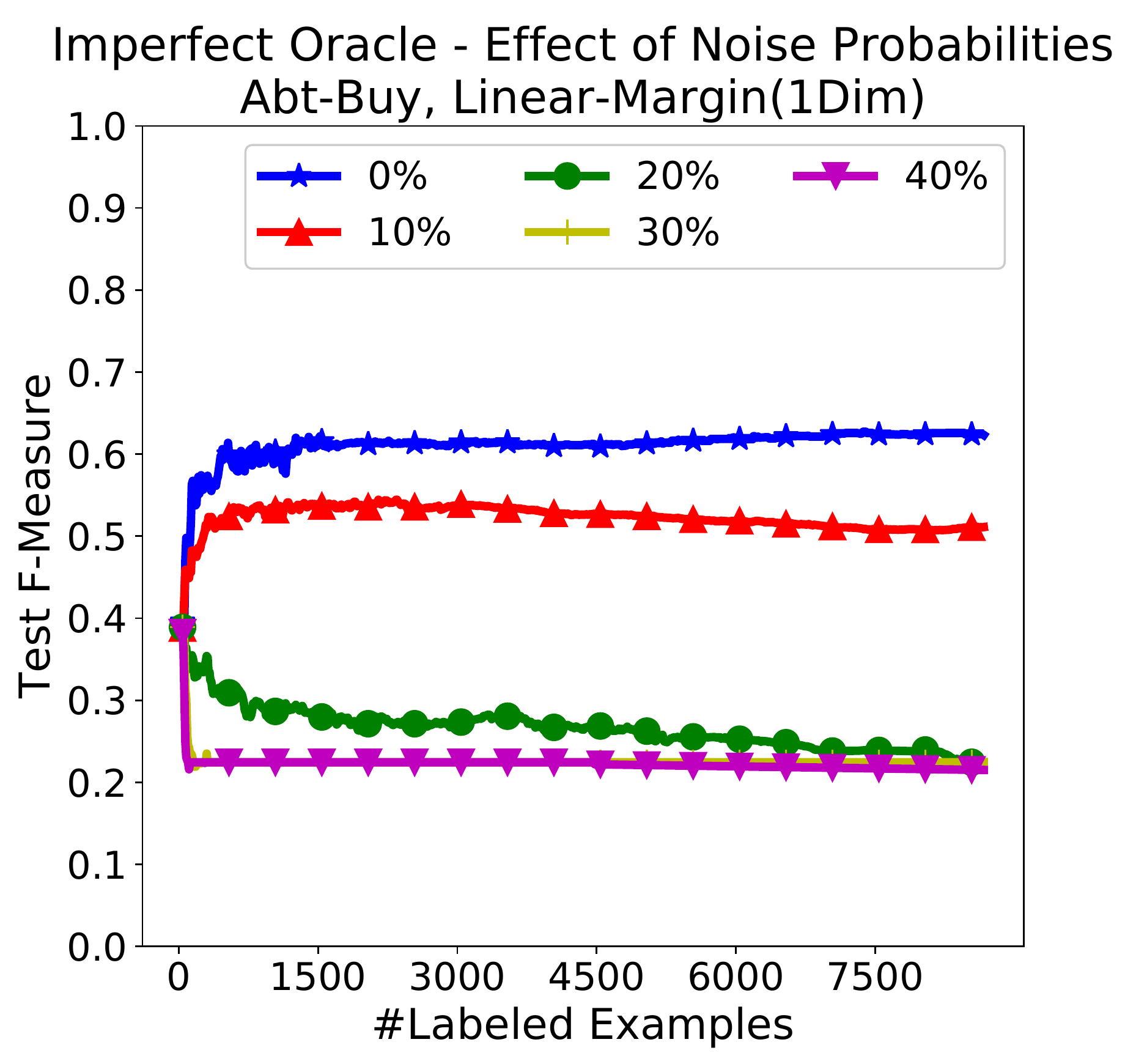}
		\caption{Linear (Blocking)}
		\label{fig:Abt-Buy-Noisy-SVMBlocking-F1}
	\end{subfigure}
	\vspace*{-.4cm}
	\caption{Active Learning using a Probabilistically Noisy Oracle \textit{(Abt-Buy, Progressive F1-Scores)}}
	\vspace*{-.4cm}
	\label{fig:noisyOracle}
\end{figure*}

\begin{figure*}[htb]
	\centering
	\begin{subfigure}[t]{0.2\textwidth}
		\centering
		\includegraphics[width=\linewidth]{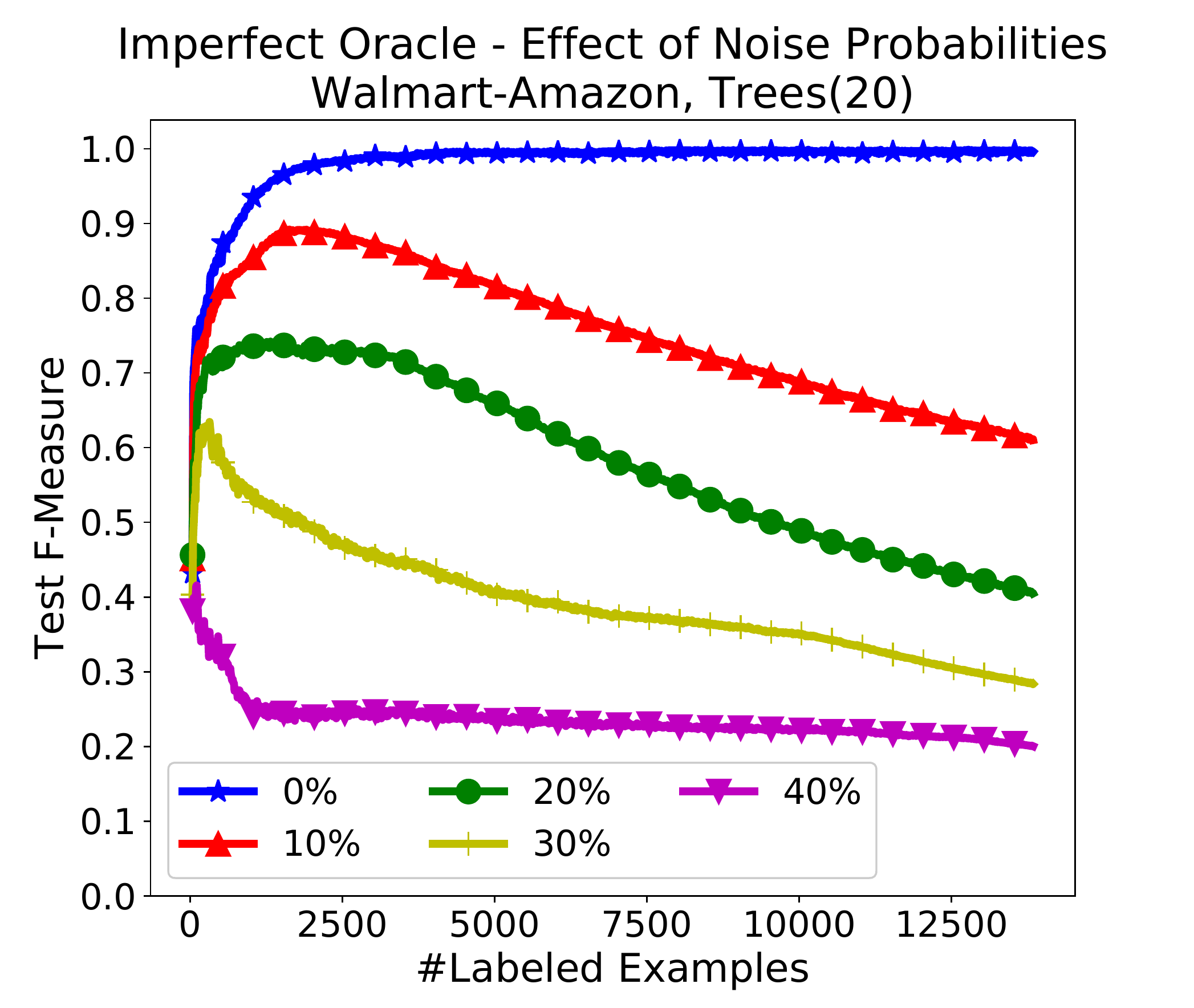}
		\caption{Walmart-Amazon}
		\label{fig:Walmart-Amazon-Noisy-RF-F1}
	\end{subfigure}
	\begin{subfigure}[t]{0.2\textwidth}
		\centering
		\includegraphics[width=\linewidth]{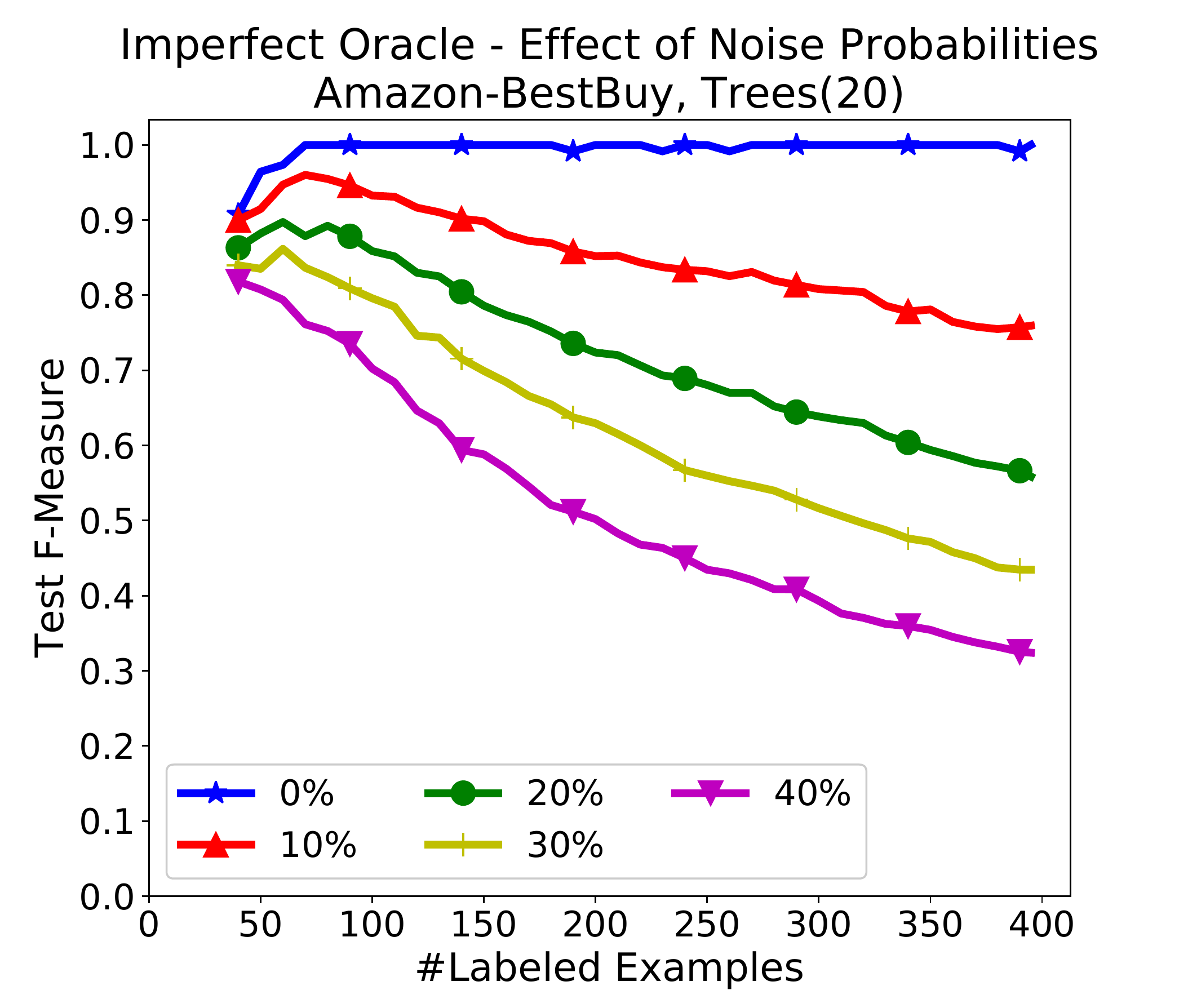}
		\caption{Amazon-BestBuy}
		\label{fig:Amazon-BestBuy-Noisy-RF-F1}
	\end{subfigure}
	\begin{subfigure}[t]{0.2\textwidth}
		\captionsetup{singlelinecheck = false, format= hang, justification=raggedright, font=footnotesize, labelsep=space}
		\centering
		\includegraphics[width=\linewidth]{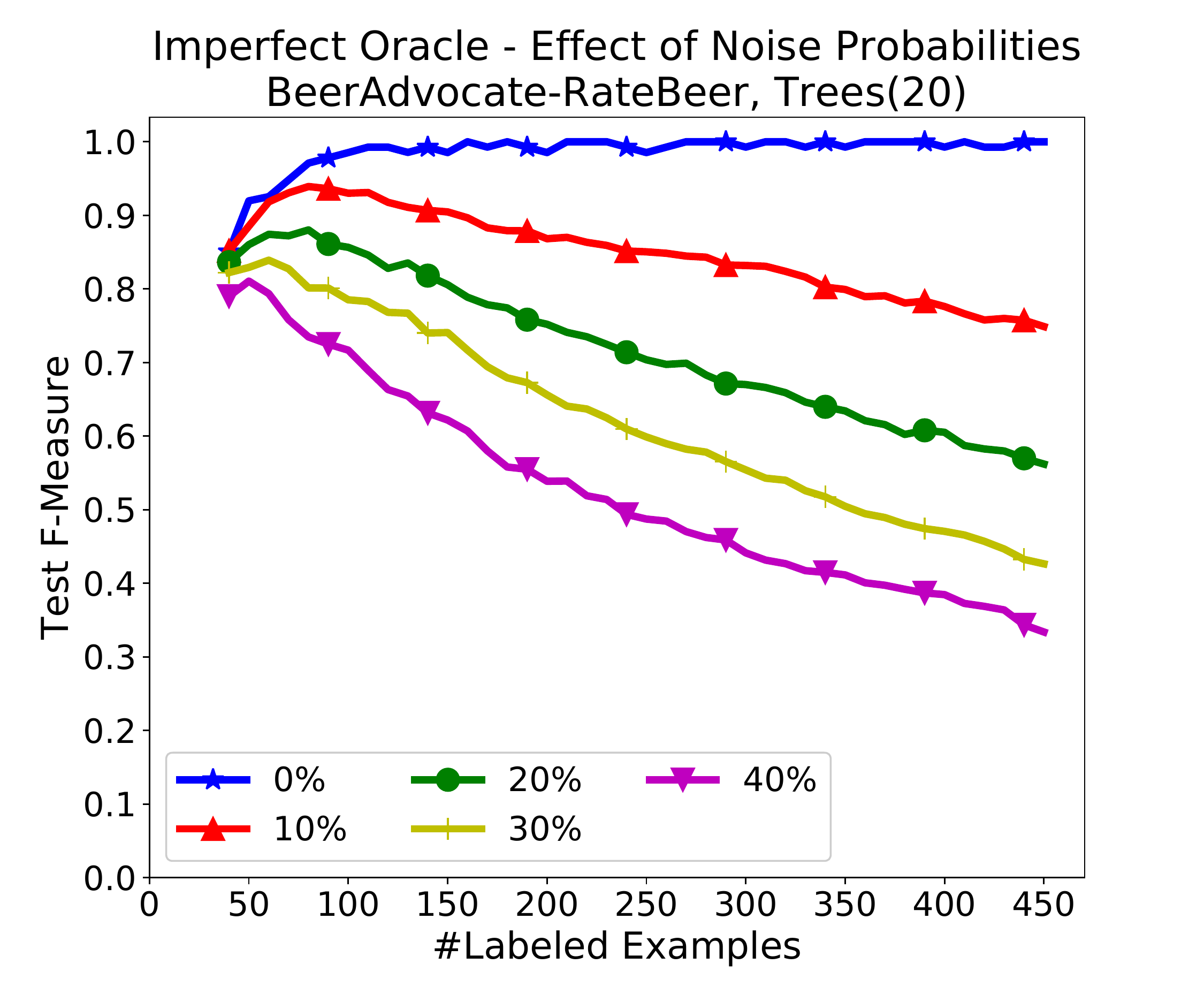}
		\caption{BeerAdvocate-RateBeer}
		\label{fig:BeerAdvocate-RateBeer-Noisy-RF-F1}
	\end{subfigure}
	\begin{subfigure}[t]{0.2\textwidth}
		\captionsetup{singlelinecheck = false, format= hang, justification=raggedright, font=footnotesize, labelsep=space}
		\centering
		\includegraphics[width=\linewidth]{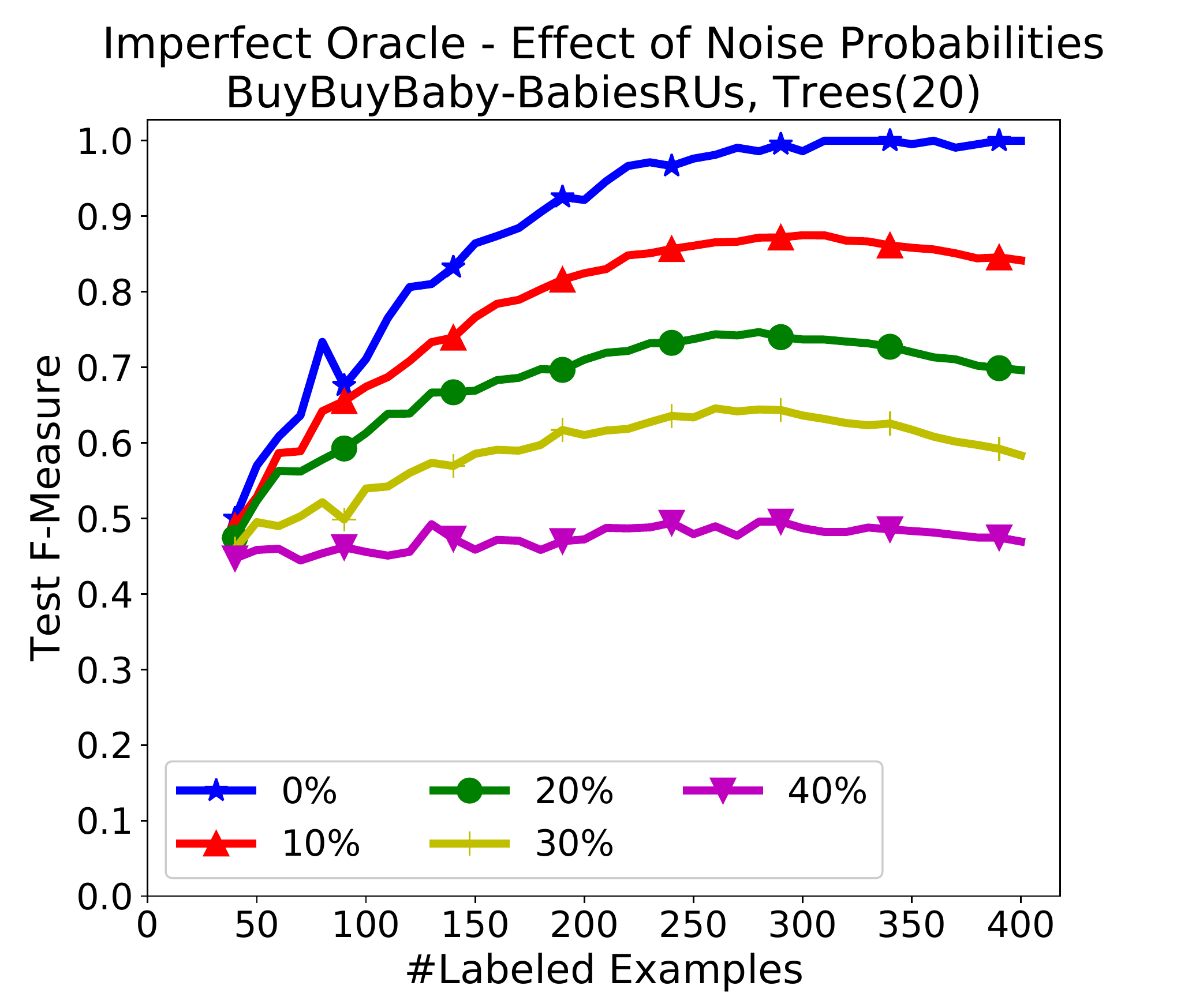}
		\caption{BuyBuyBaby-BabiesRUs}
		\label{fig:BuyBuyBaby-BabiesRUs-Noisy-RF-F1}
	\end{subfigure}
	\vspace*{-.4cm}
	\caption{Tree Ensembles on Magellan/DeepMatcher Datasets \textit{(Noisy Oracles, Progressive F1)}}
	\vspace*{-.4cm}
	\label{fig:noisyOracleMagellan}
\end{figure*}

\RFIVE{The termination criteria differ between perfect and imperfect Oracles. In the case of perfect Oracles, once an active learning method achieves a convergent F1-score, there is little change to it with the addition of more examples. In contrast, in the case of imperfect Oracles, the addition of more examples leads to deteriorating F1-scores because of an added amount of noisy labels. Therefore, we terminate active learning with perfect Oracles in Fig.~\ref{fig:Abt-Buy-QBCvsMargin} to~\ref{fig:Classifiers-userWaitTime} as soon as either one of the approaches achieves a near-perfect (close to 1.0) F1-score or all the examples are labeled. In the experiments on noisy Oracles from Fig.~\ref{fig:noisyOracle} to~\ref{fig:ActiveVsSupervised}, the termination criterion is the exhaustion of all unlabeled examples. Rule-based learners terminate as soon as no likely false positives (LFPs) or likely false negatives (LFNs) are found among the selected examples. This results in no new rules being discovered, that leads to early termination.} We present the results on perfect Oracles in Section~\ref{sec:expSec1} while Section~\ref{sec:expNoisy} contains those on noisy Oracles. Section~\ref{sec:expsec3} covers interpretability.

\subsection{Comparison of Selectors $\&$ Classifiers}
\label{sec:expSec1}
In this section, we assume the presence of a perfect Oracle with no labeling error. We first compare various example selectors applicable to each classifier. Subsequently we compare the best strategies from each family of classifiers in order to understand the combination of classifier and example selector that works best on a majority of the datasets.
\begin{figure}[htb]
	\centering
	\begin{subfigure}[t]{0.155\textwidth}
		\captionsetup{singlelinecheck = false, format= hang, justification=raggedright, font=footnotesize, labelsep=space}
		\centering
		\includegraphics[width=\linewidth]{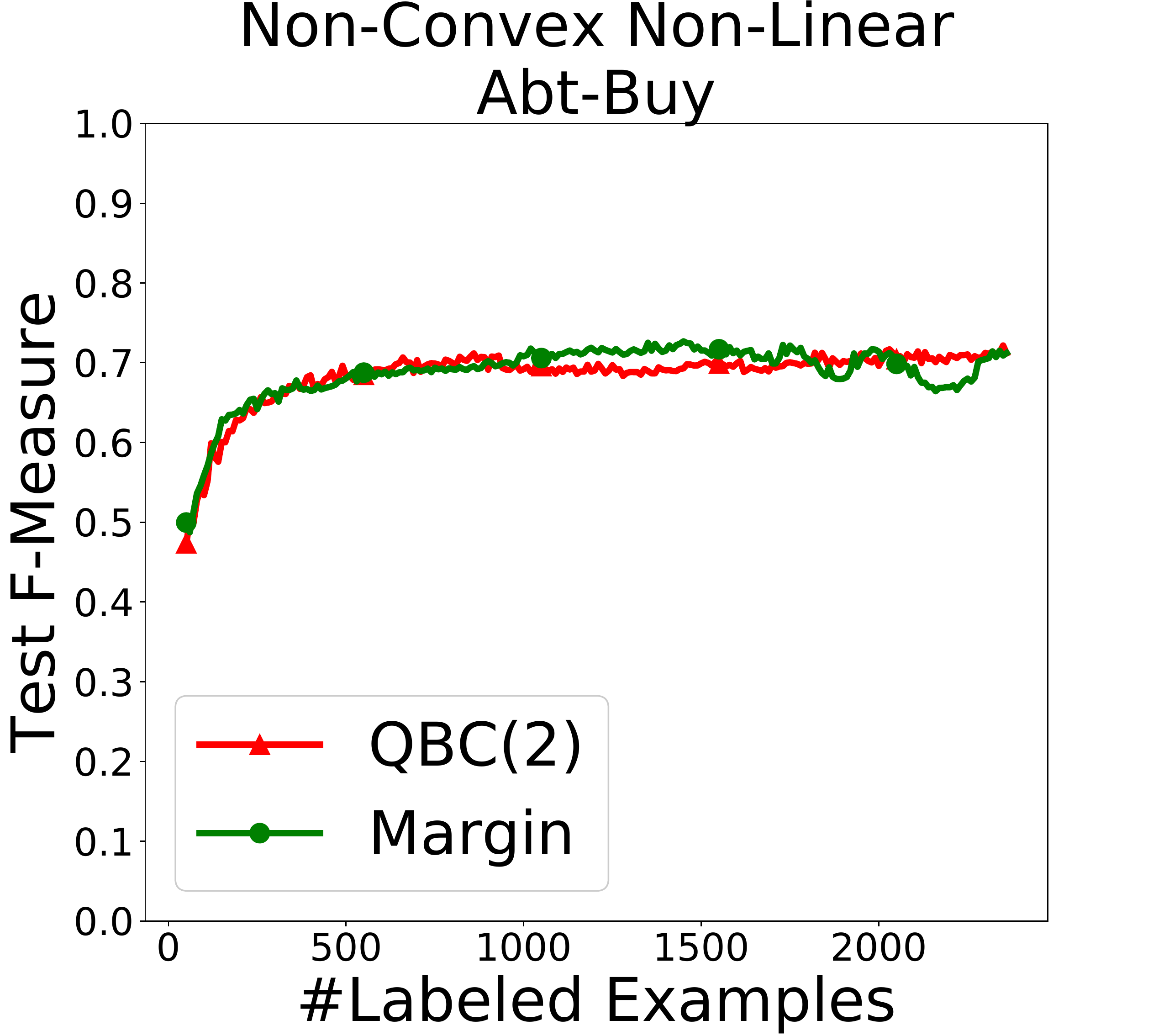}
		\caption{Non-Convex Non-Linear}
		\label{fig:NN-Abt-Buy-F1}
	\end{subfigure}
	\begin{subfigure}[t]{0.155\textwidth}
		\centering
		\includegraphics[width=\linewidth]{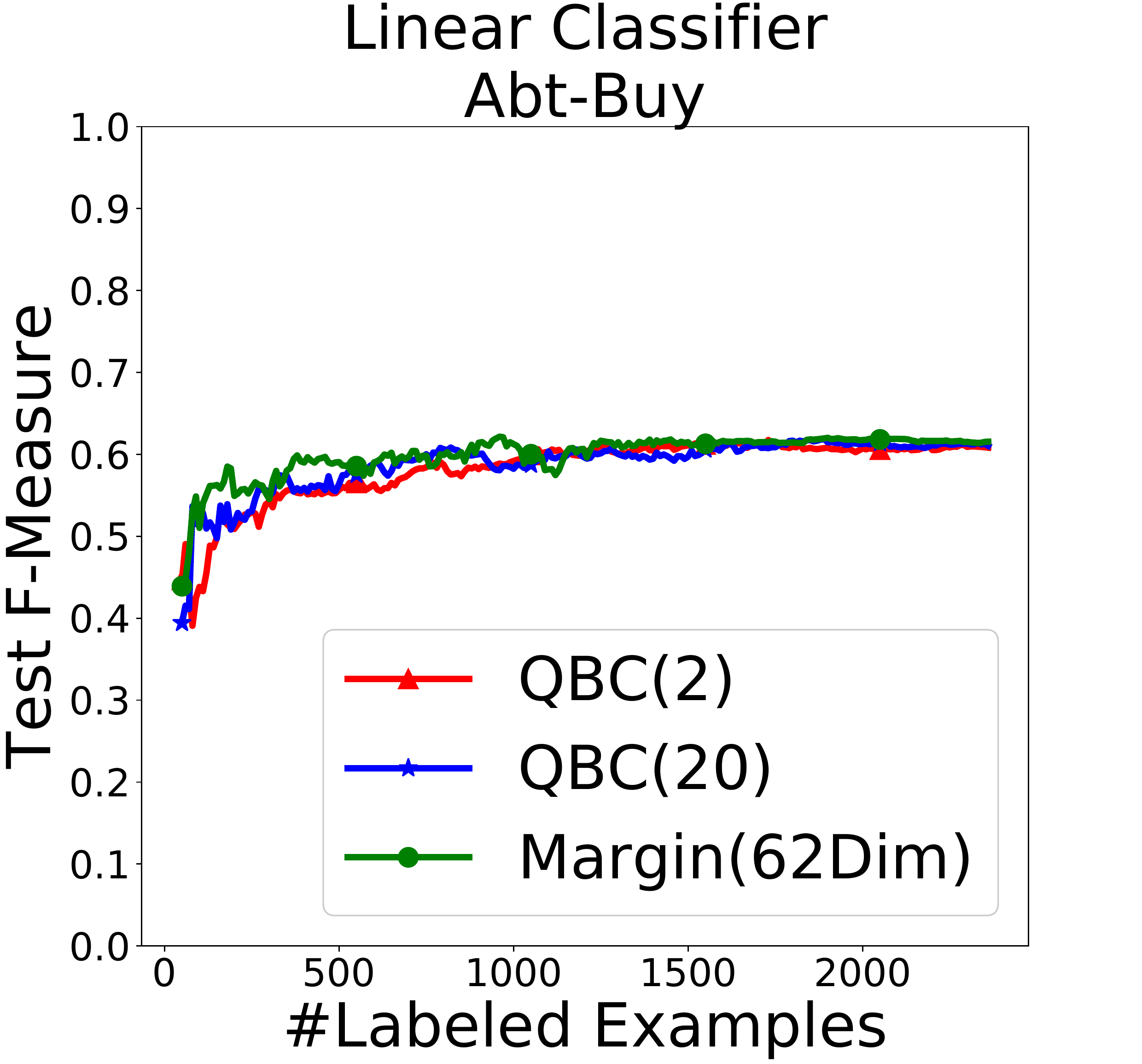}
		\caption{Linear}
		\label{fig:SVM-Abt-Buy-F1}
	\end{subfigure}
	\begin{subfigure}[t]{0.155\textwidth}
		\centering
		\includegraphics[width=\linewidth]{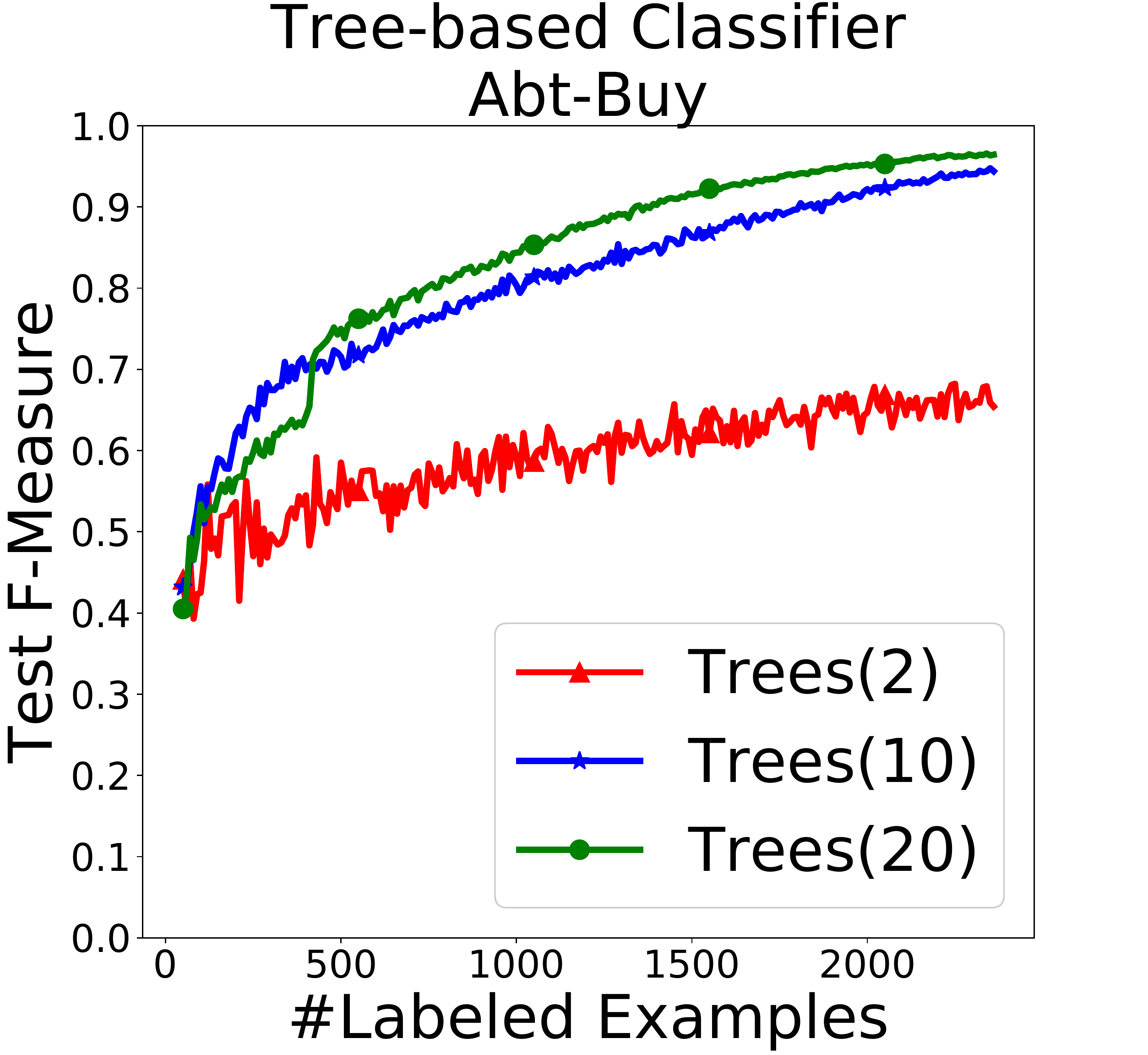}
		\caption{Tree-based}
		\label{fig:RF-Abt-Buy-F1}
	\end{subfigure}
	\vspace*{-.4cm}
	\caption{QBC vs. Margin \textit{(Progressive F1, Abt-Buy)}}
	\vspace*{-.4cm}
	\label{fig:Abt-Buy-QBCvsMargin}
\end{figure}
\begin{figure}[htb]
	\centering
	\begin{subfigure}[t]{0.155\textwidth}
		\captionsetup{singlelinecheck = false, format= hang, justification=raggedright, font=footnotesize, labelsep=space}
		\centering
		\includegraphics[width=\linewidth]{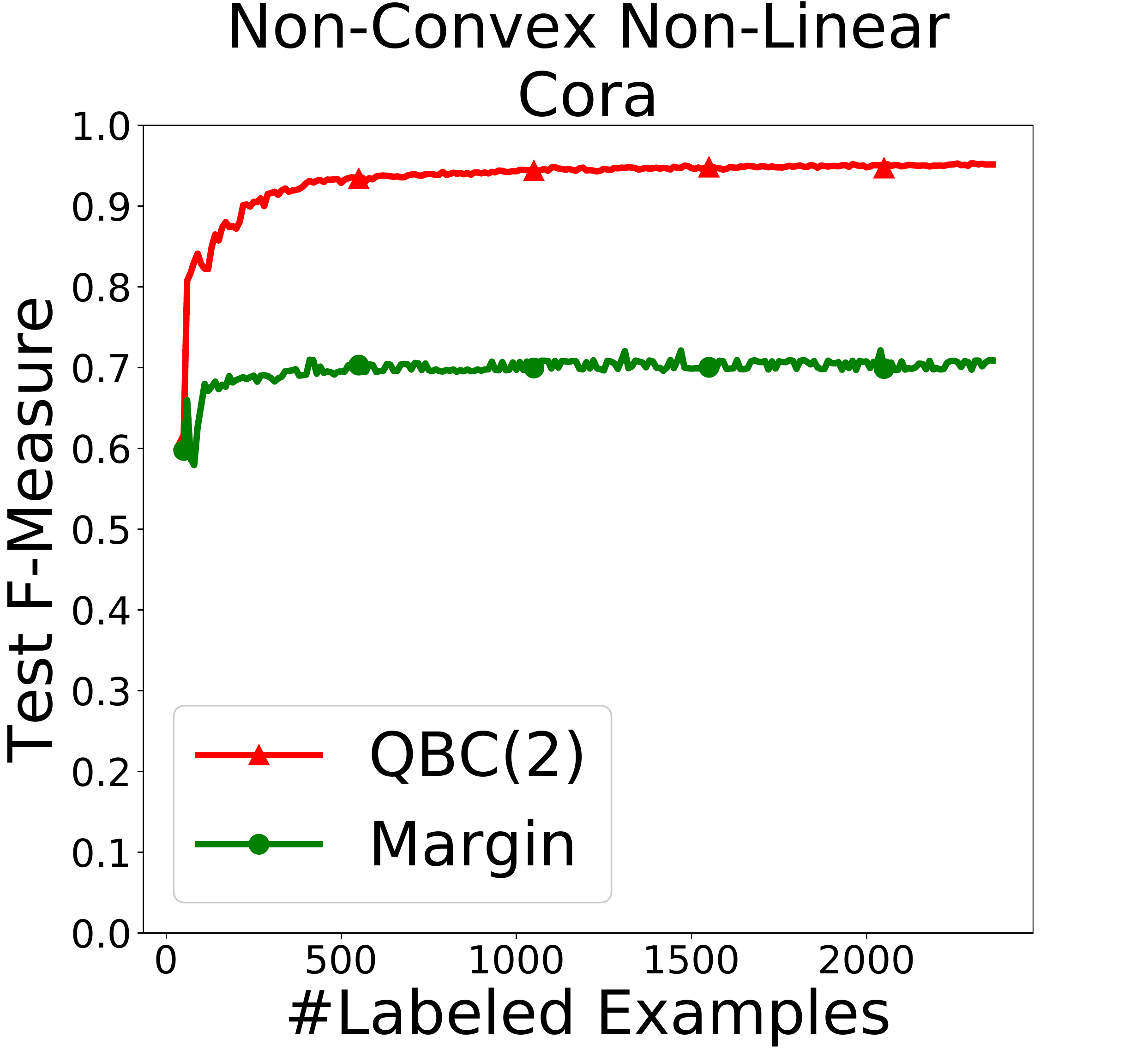}
		\caption{Non-Convex Non-Linear}
		\label{fig:NN-Cora-F1}
	\end{subfigure}
	\begin{subfigure}[t]{0.155\textwidth}
		\centering
		\includegraphics[width=\linewidth]{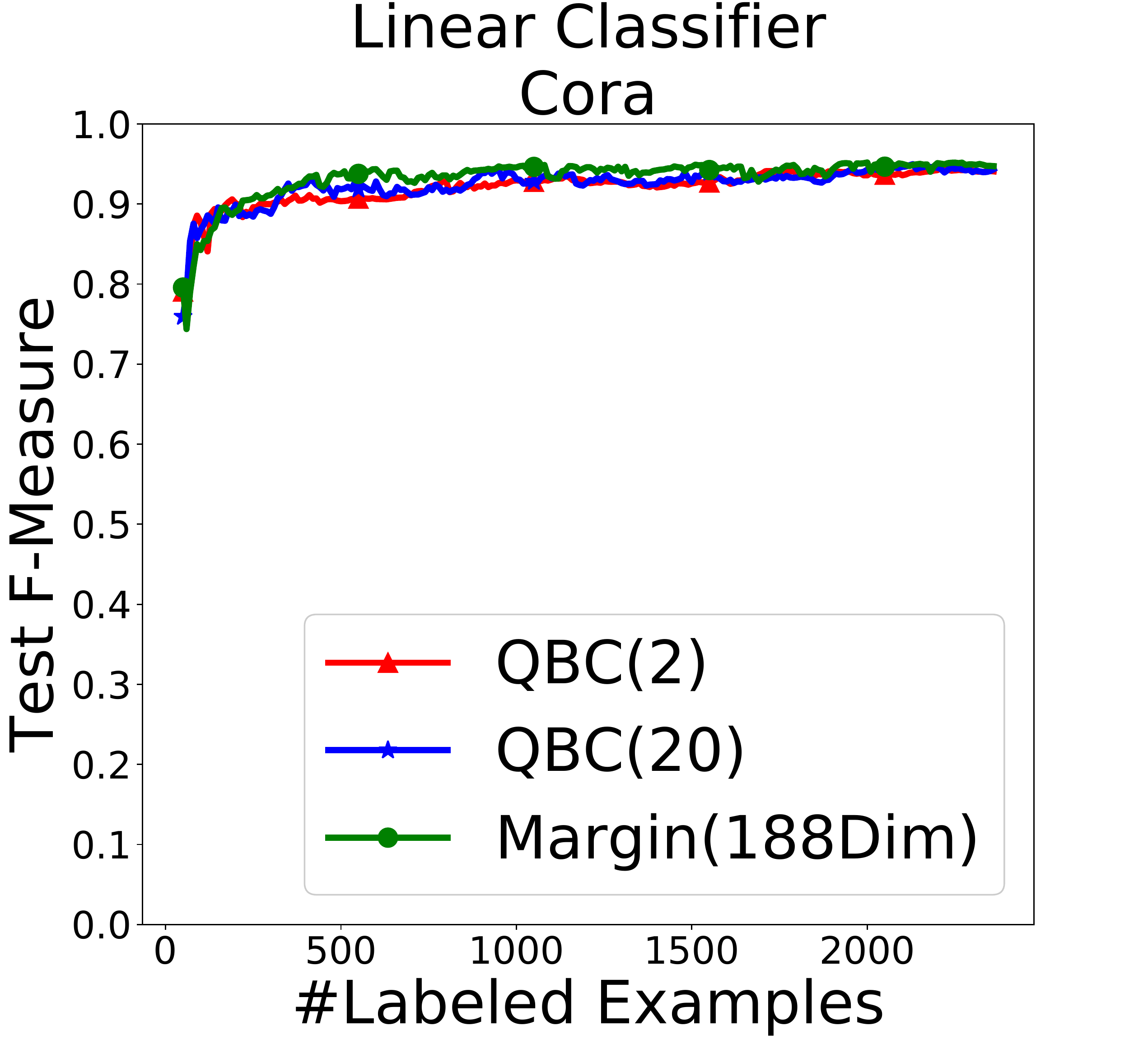}
		\caption{Linear}
		\label{fig:SVM-Cora-F1}
	\end{subfigure}
	\begin{subfigure}[t]{0.155\textwidth}
		\centering
		\includegraphics[width=\linewidth]{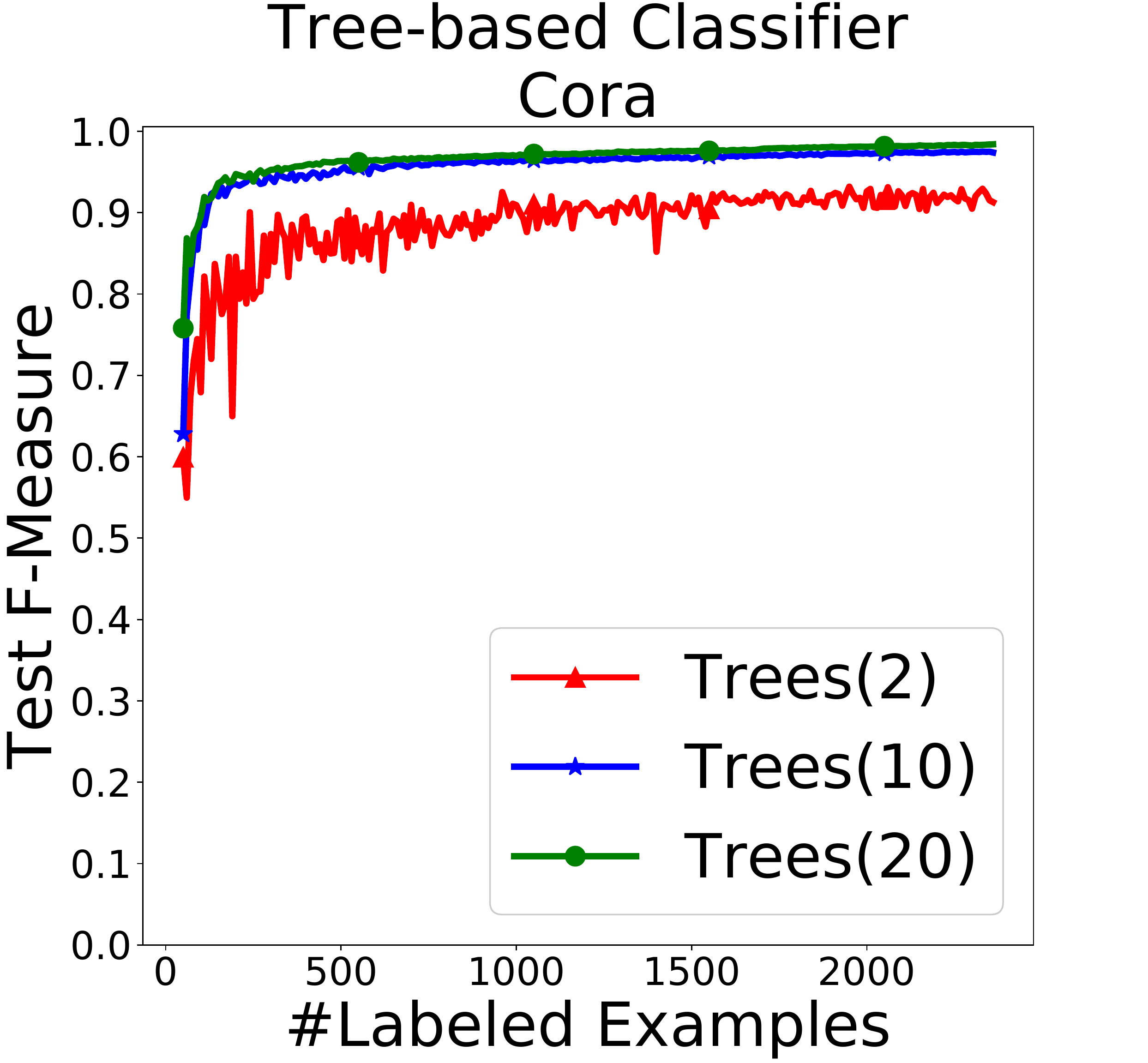}
		\caption{Tree-based}
		\label{fig:RF-Cora-F1}
	\end{subfigure}
	\vspace*{-.4cm}
	\caption{QBC vs. Margin \textit{(Progressive F1, Cora)}}
	\vspace*{-.4cm}
	\label{fig:Cora-QBCvsMargin}
\end{figure}

\begin{figure*}[htb]
	\centering
	\begin{subfigure}[t]{0.22\textwidth}
		\captionsetup{singlelinecheck = false, format= hang, justification=raggedright, font=footnotesize, labelsep=space}
		\centering
		\includegraphics[width=\linewidth]{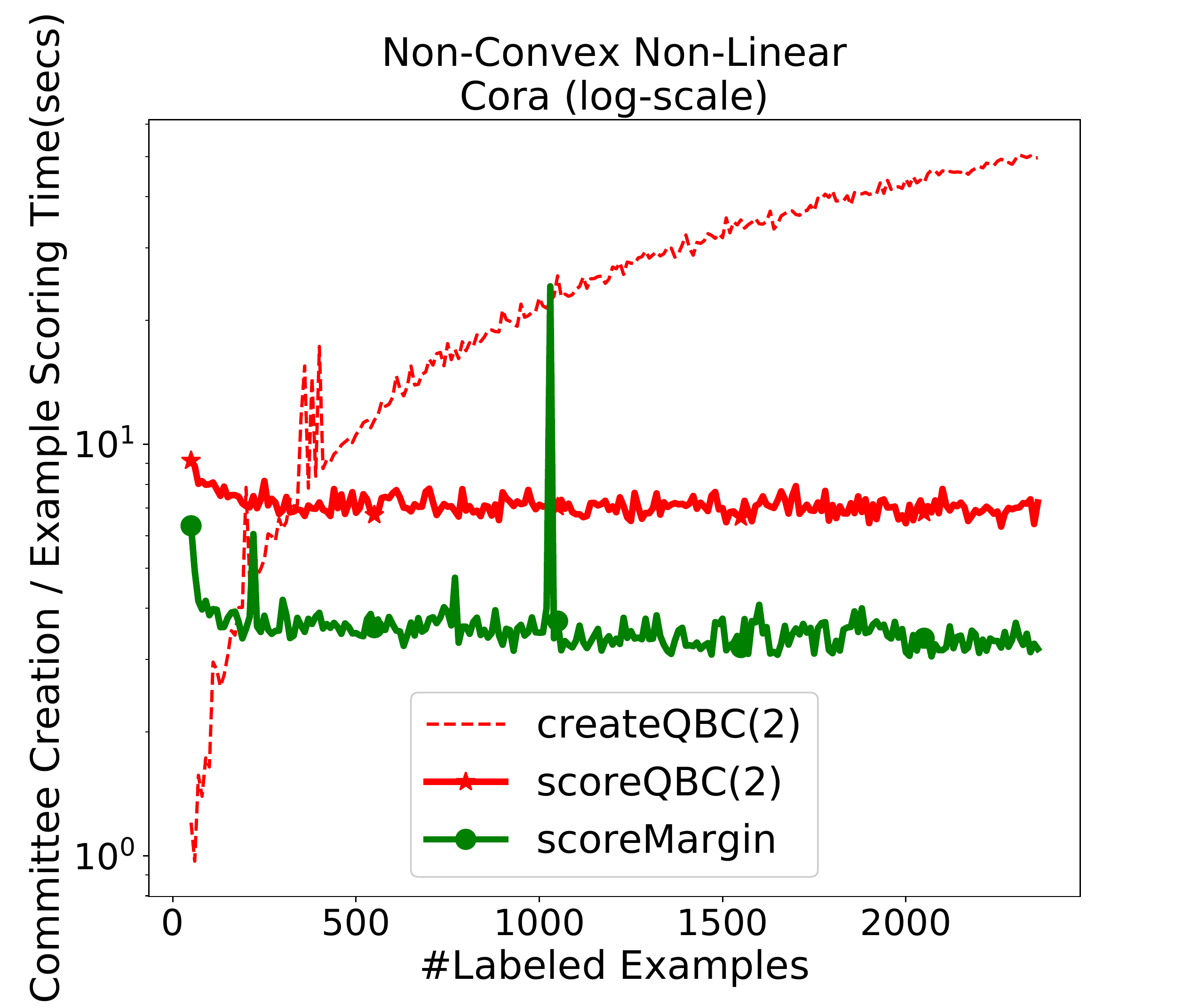}
		\caption{Non-Convex Non-Linear}
		\label{fig:NN-Cora-exSelTime}
	\end{subfigure}
	\begin{subfigure}[t]{0.22\textwidth}
		\centering
		\includegraphics[width=\linewidth]{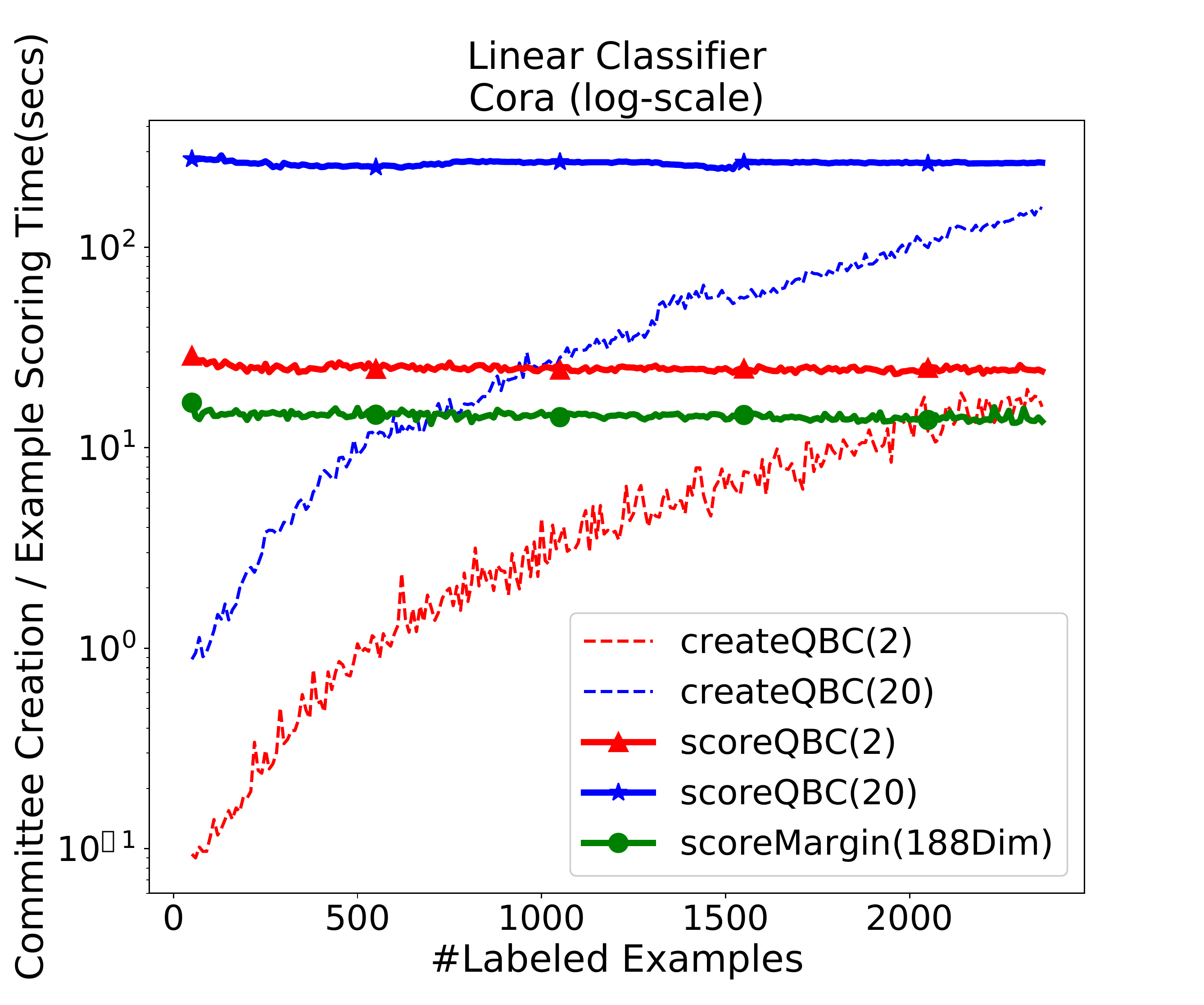}
		\caption{Linear}
		\label{fig:SVM-Cora-exSelTime}
	\end{subfigure}
	\begin{subfigure}[t]{0.22\textwidth}
		\centering
		\includegraphics[width=\linewidth]{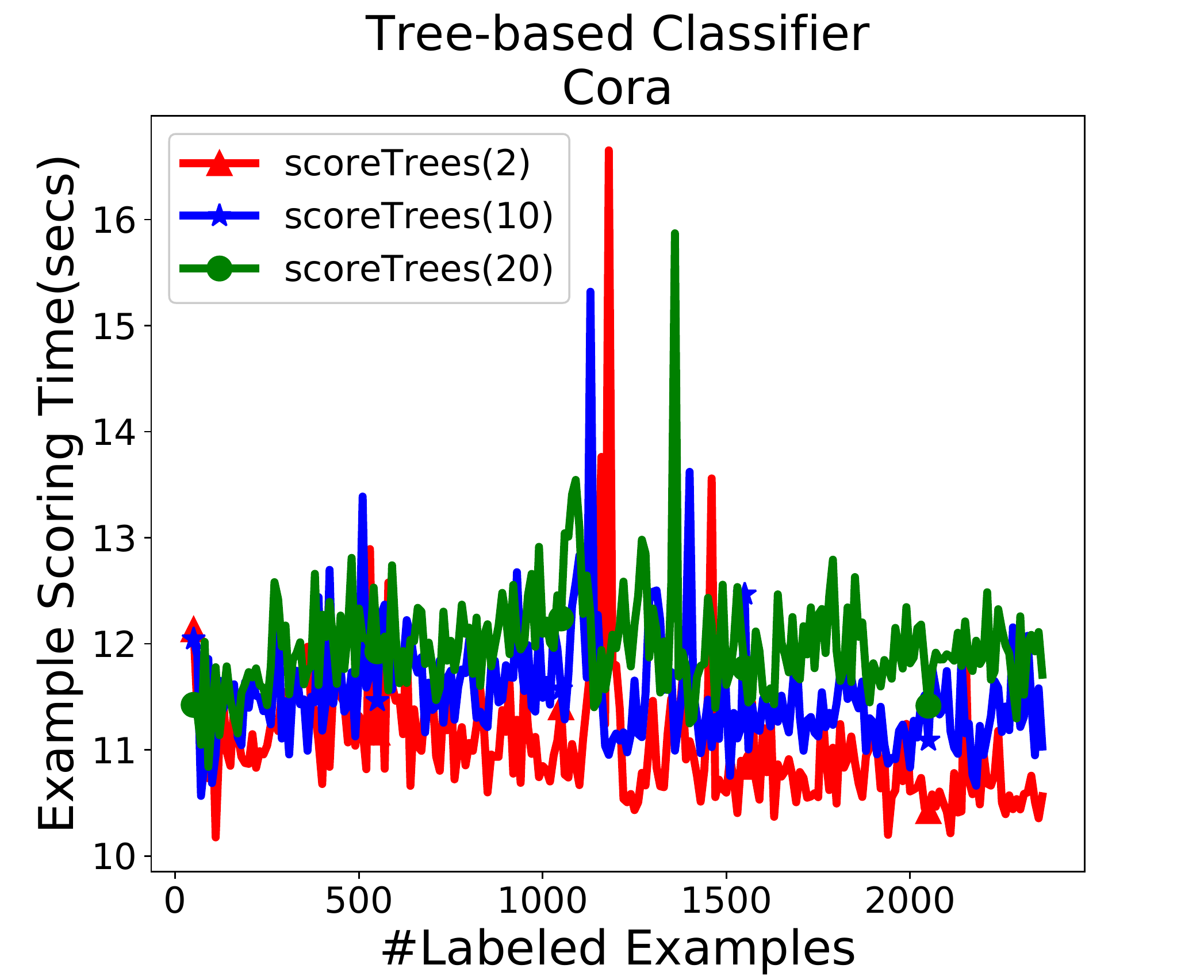}
		\caption{Tree-based}
		\label{fig:RF-Cora-exSelTime}
	\end{subfigure}
	\begin{subfigure}[t]{0.22\textwidth}
		\captionsetup{singlelinecheck = false, format= hang, justification=raggedright, font=footnotesize, labelsep=space}
		\centering
		\includegraphics[width=\linewidth]{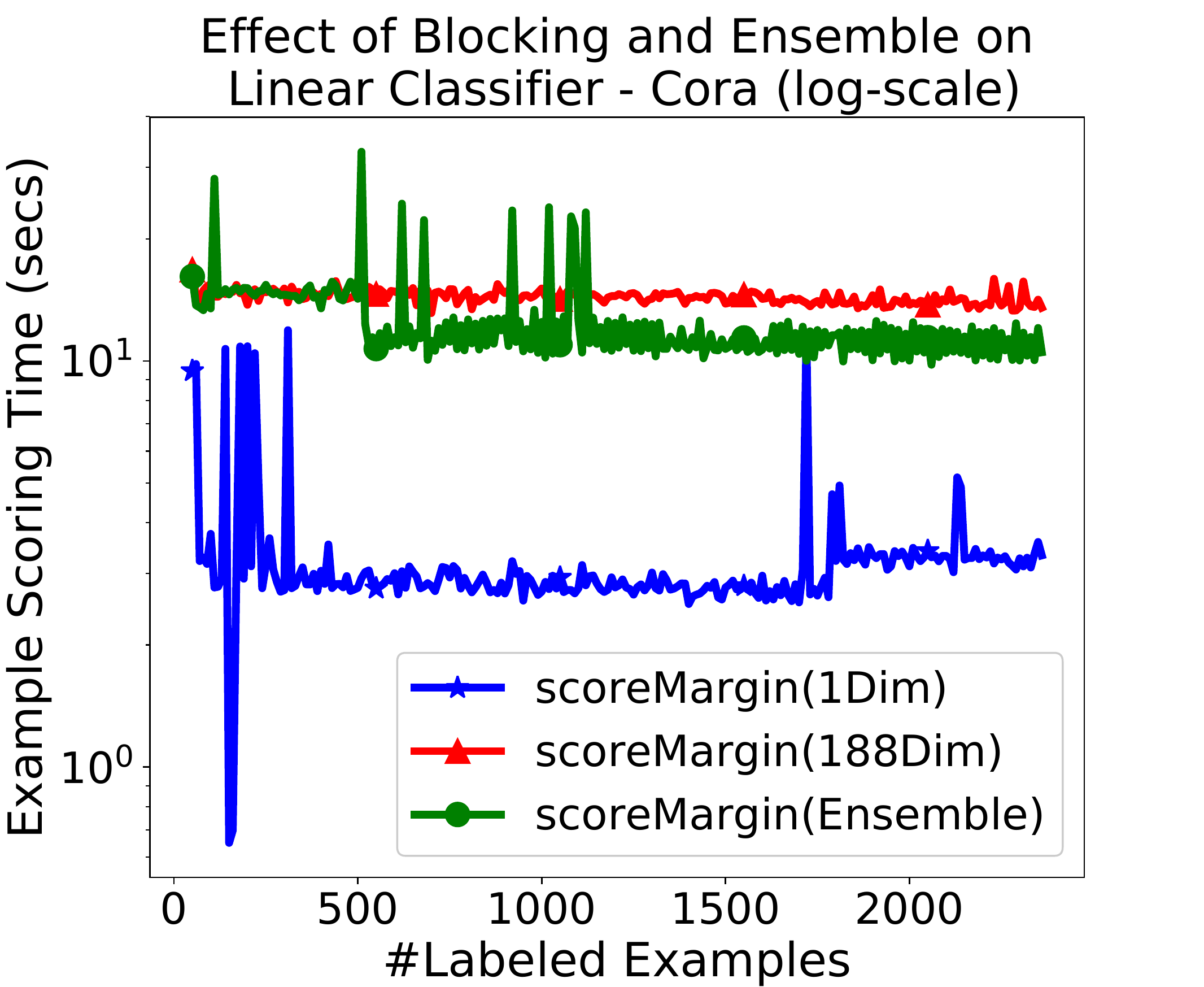}
		\caption{Linear (Enhancements)}
		\label{fig:SVM-Cora-BlockingEnsemble-exSelTime}
	\end{subfigure}
	\vspace*{-.4cm}
	\caption{Example Selection Times of various Strategies on each Classifier \textit{(Cora)}}
	\vspace*{-.4cm}
	\label{fig:exSelTime}
\end{figure*}
\begin{figure*}[htb]
	\centering
	\begin{subfigure}[t]{0.19\textwidth}
		\centering
		\includegraphics[width=\linewidth]{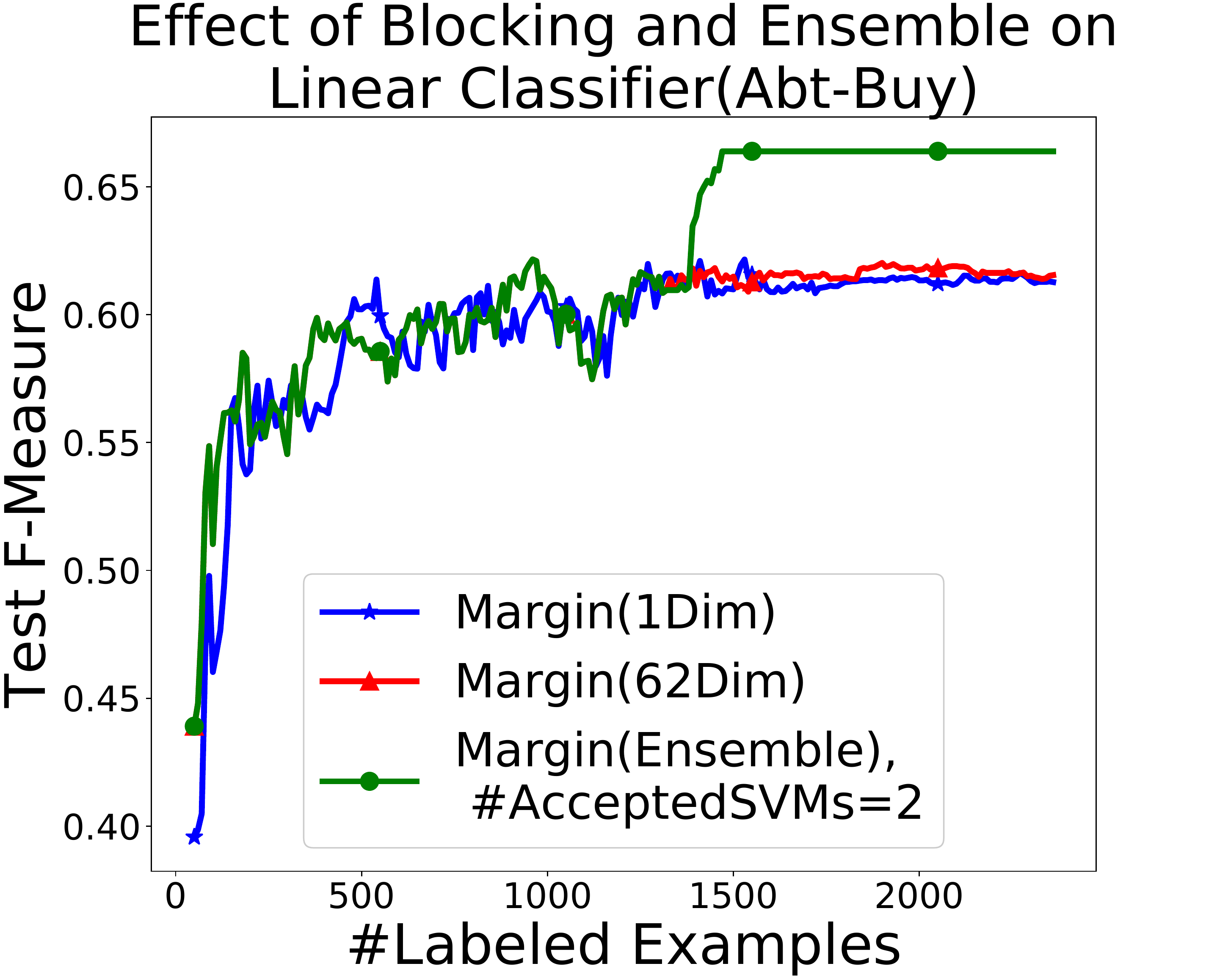}
		\caption{Abt-Buy}
		\label{fig:SVM-Abt-Buy-BlockingEnsemble-F1}
	\end{subfigure}
	\begin{subfigure}[t]{0.19\textwidth}
		\centering
		\includegraphics[width=\linewidth]{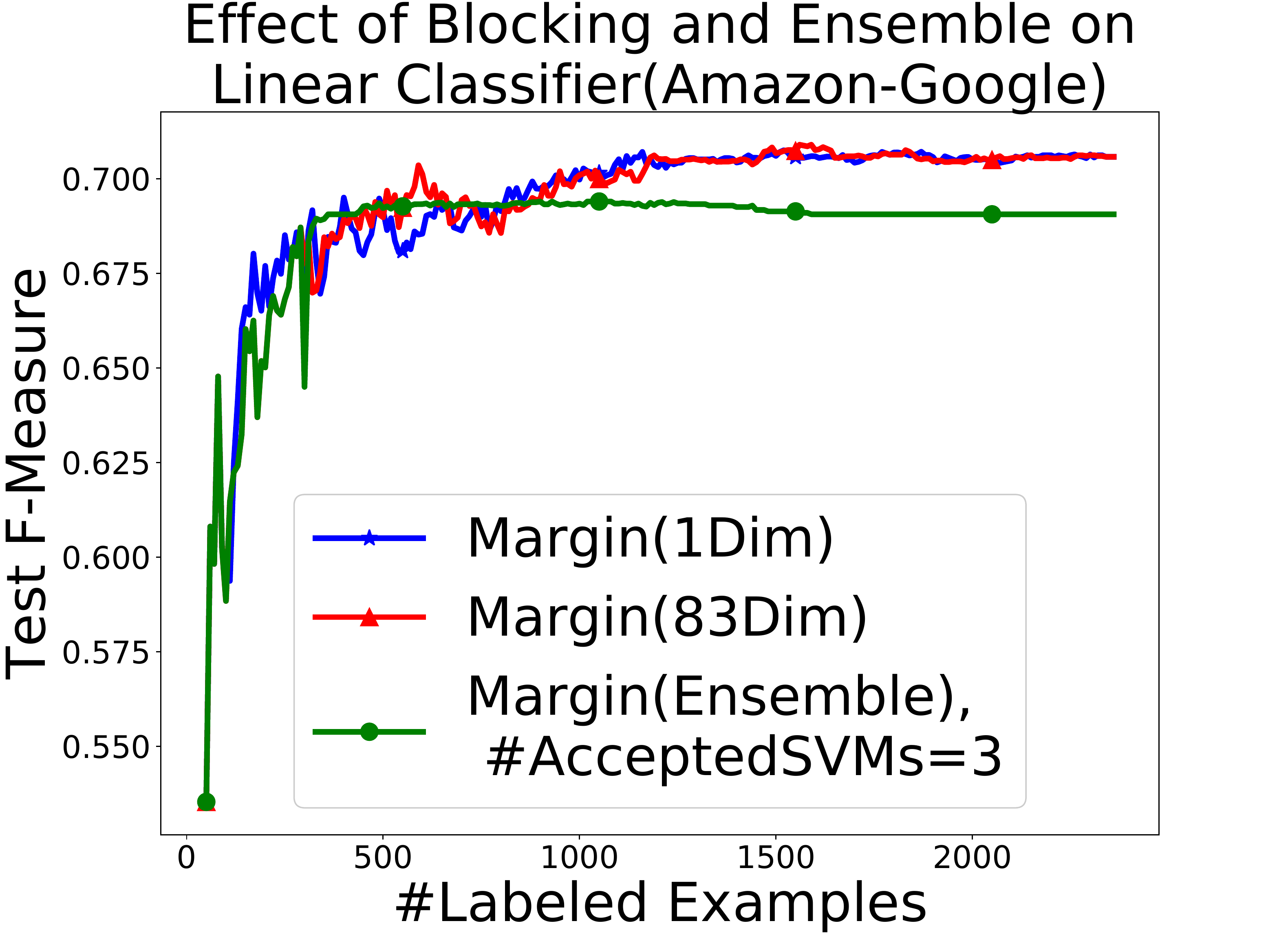}
		\caption{Amazon-Google}
		\label{fig:SVM-AmazonGoogleProducts-BlockingEnsemble-F1}
	\end{subfigure}
	\begin{subfigure}[t]{0.19\textwidth}
		\centering
		\includegraphics[width=\linewidth]{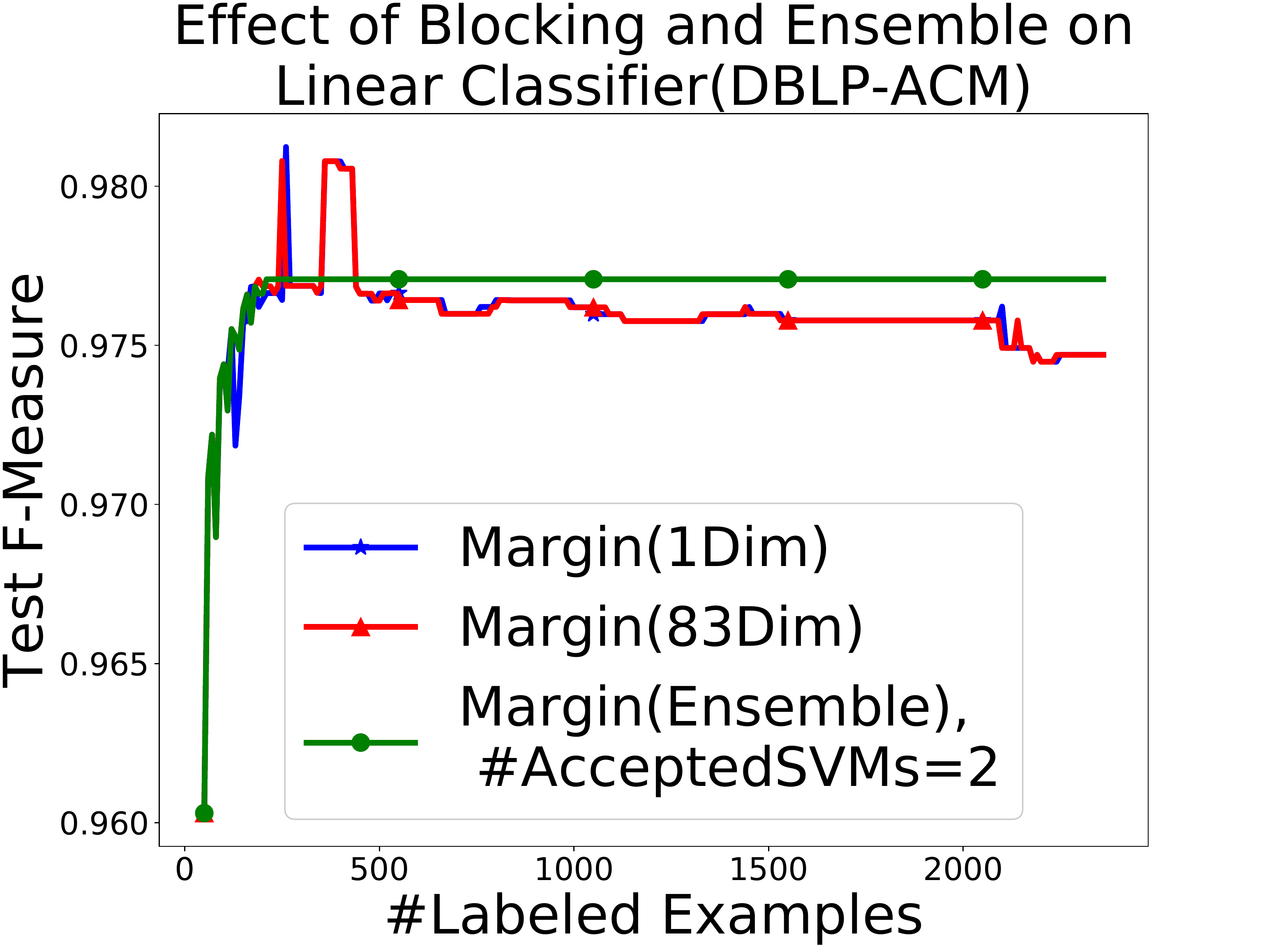}
		\caption{DBLP-ACM}
		\label{fig:SVM-DBLP-ACM-BlockingEnsemble-F1}
	\end{subfigure}
	\begin{subfigure}[t]{0.19\textwidth}
		\centering
		\includegraphics[width=\linewidth]{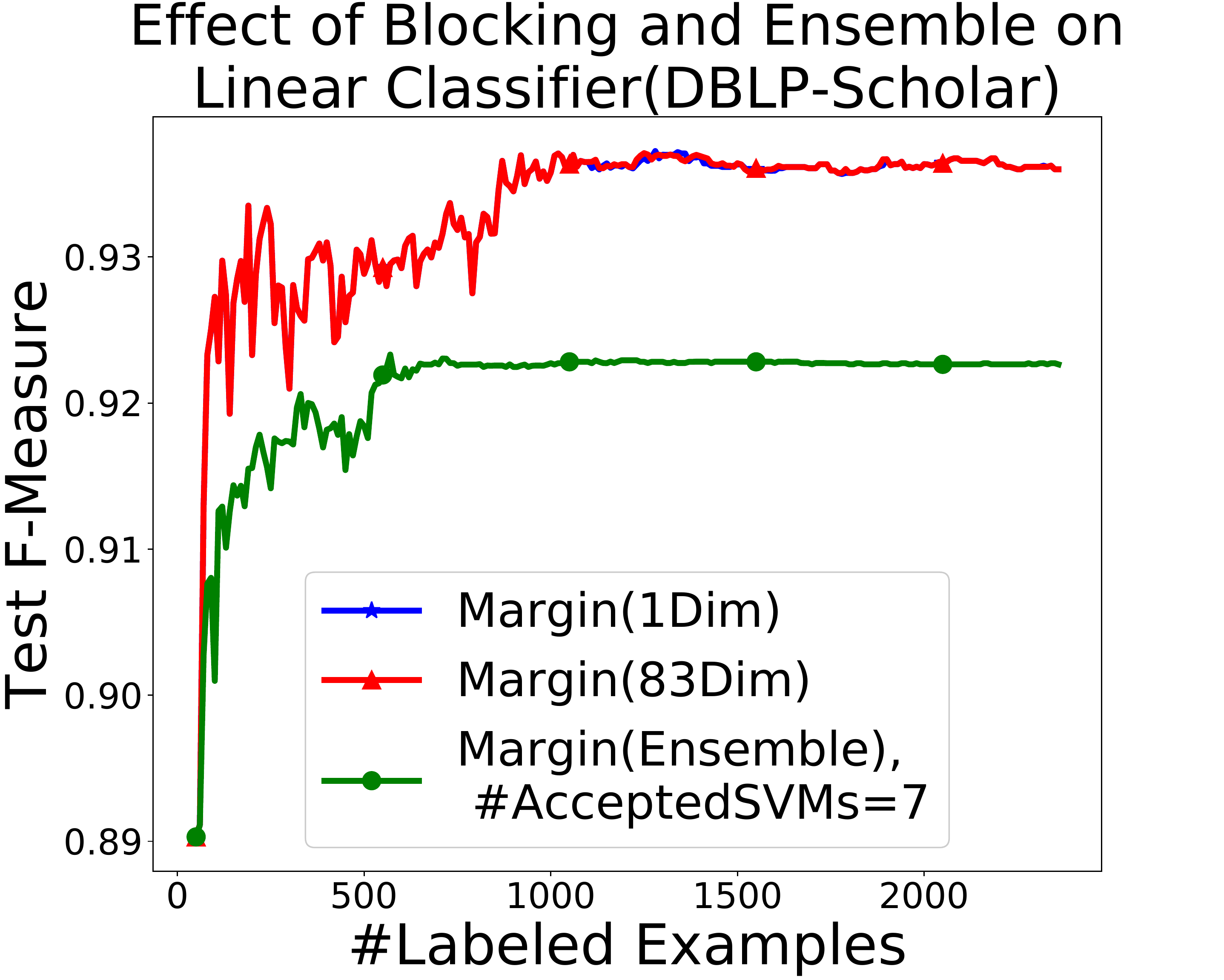}
		\caption{DBLP-Scholar}
		\label{fig:SVM-DBLP-Scholar-BlockingEnsemble-F1}
	\end{subfigure}
	\begin{subfigure}[t]{0.19\textwidth}
		\centering
		\includegraphics[width=\linewidth]{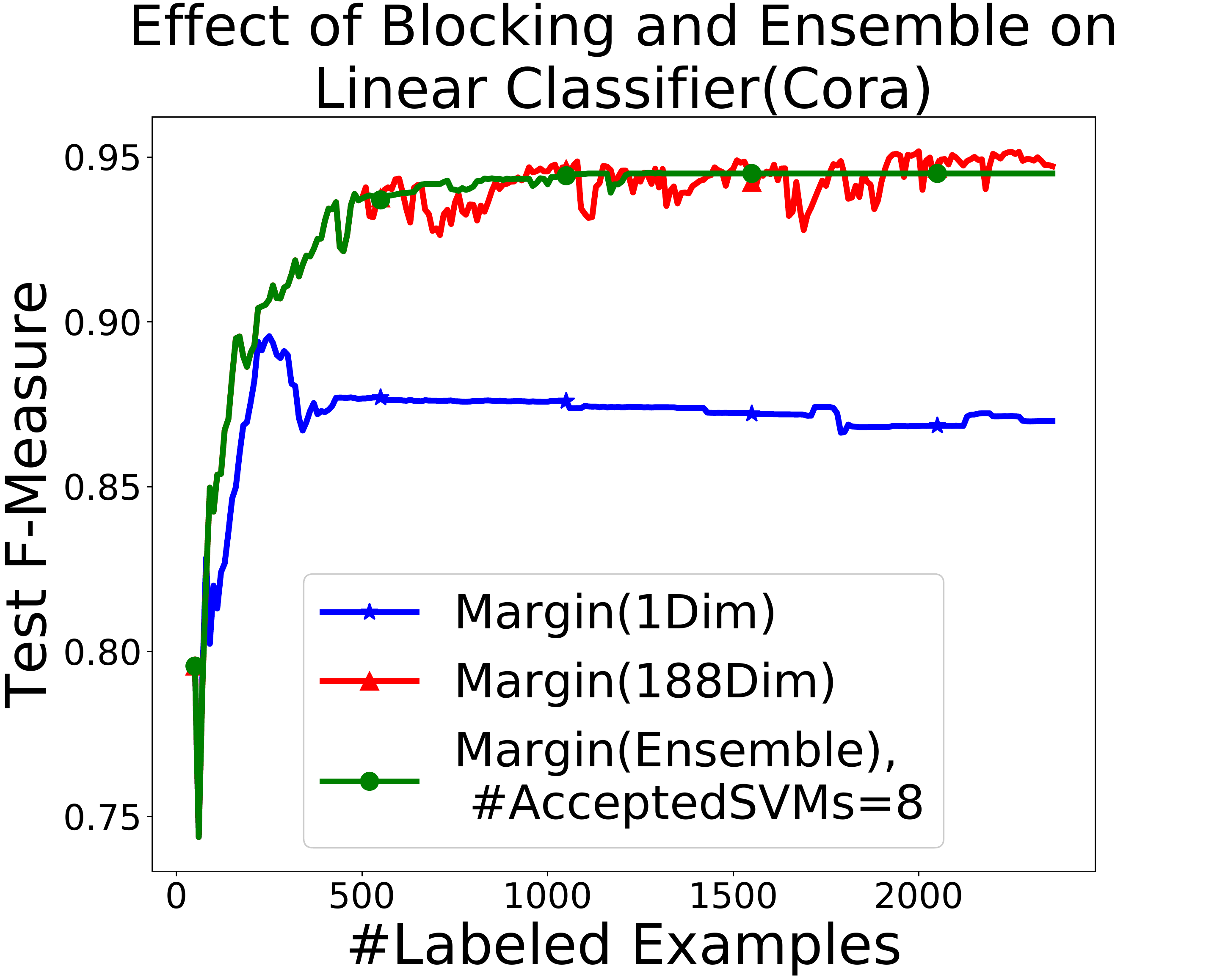}
		\caption{Cora}
		\label{fig:SVM-Cora-BlockingEnsemble-F1}
	\end{subfigure}
	\vspace*{-.4cm}
	\caption{Effect of Blocking and Active Ensemble on Linear Classifiers \textit{(Progressive F1-Scores, Perfect Oracle)}}
	\vspace*{-.4cm}
	\label{fig:SVM-BlockingEnsemble-F1}
\end{figure*}
\begin{figure*}[htb]
	\centering
	\begin{subfigure}[t]{0.19\textwidth}
		\centering
		\includegraphics[width=\linewidth]{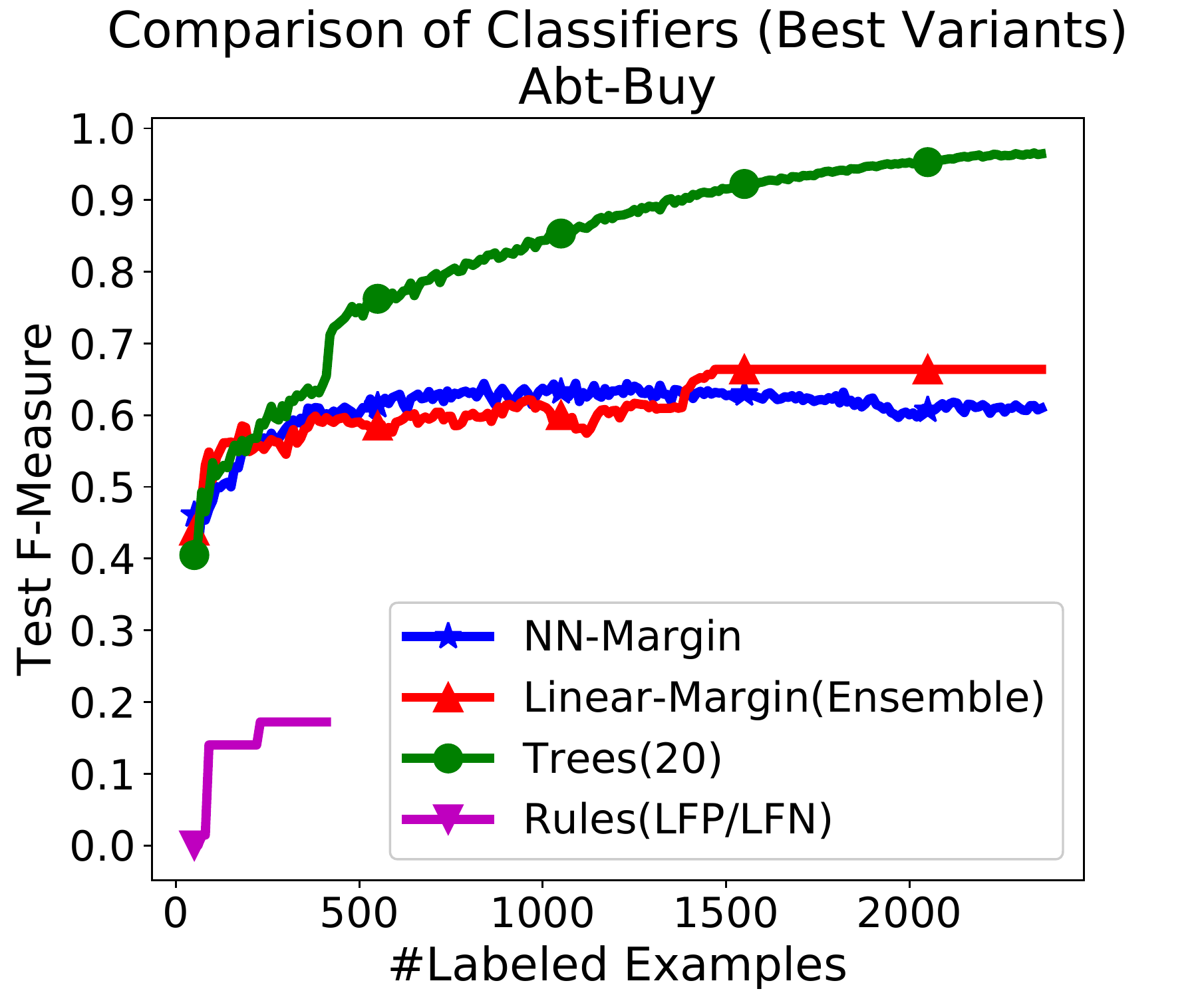}
		\caption{Abt-Buy}
		\label{fig:Abt-Buy-Classifiers-F1}
	\end{subfigure}
	\begin{subfigure}[t]{0.19\textwidth}
		\centering
		\includegraphics[width=\linewidth]{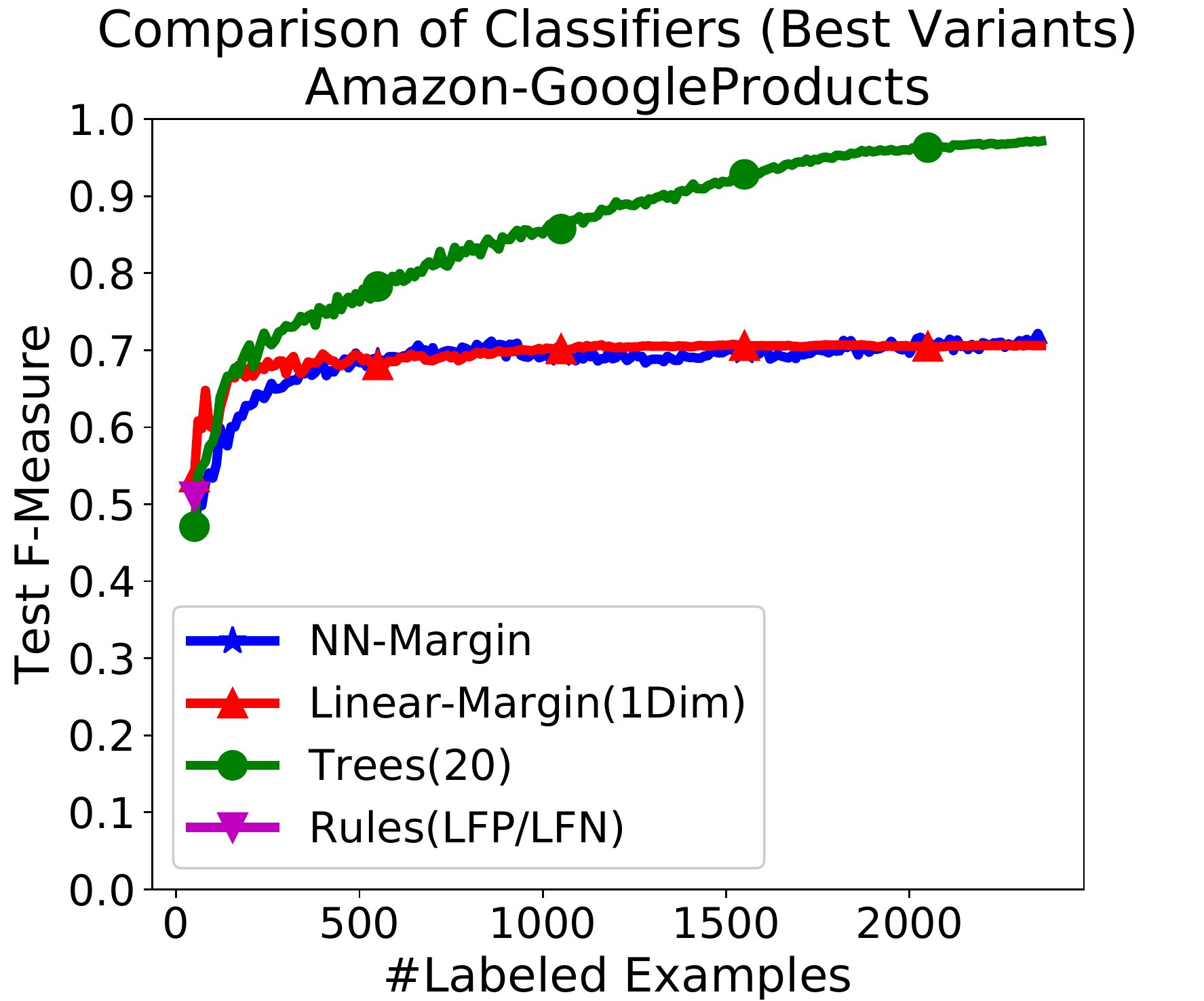}
		\caption{Amazon-Google}
		\label{fig:AmazonGoogleProducts-Classifiers-F1}
	\end{subfigure}
	\begin{subfigure}[t]{0.19\textwidth}
		\centering
		\includegraphics[width=\linewidth]{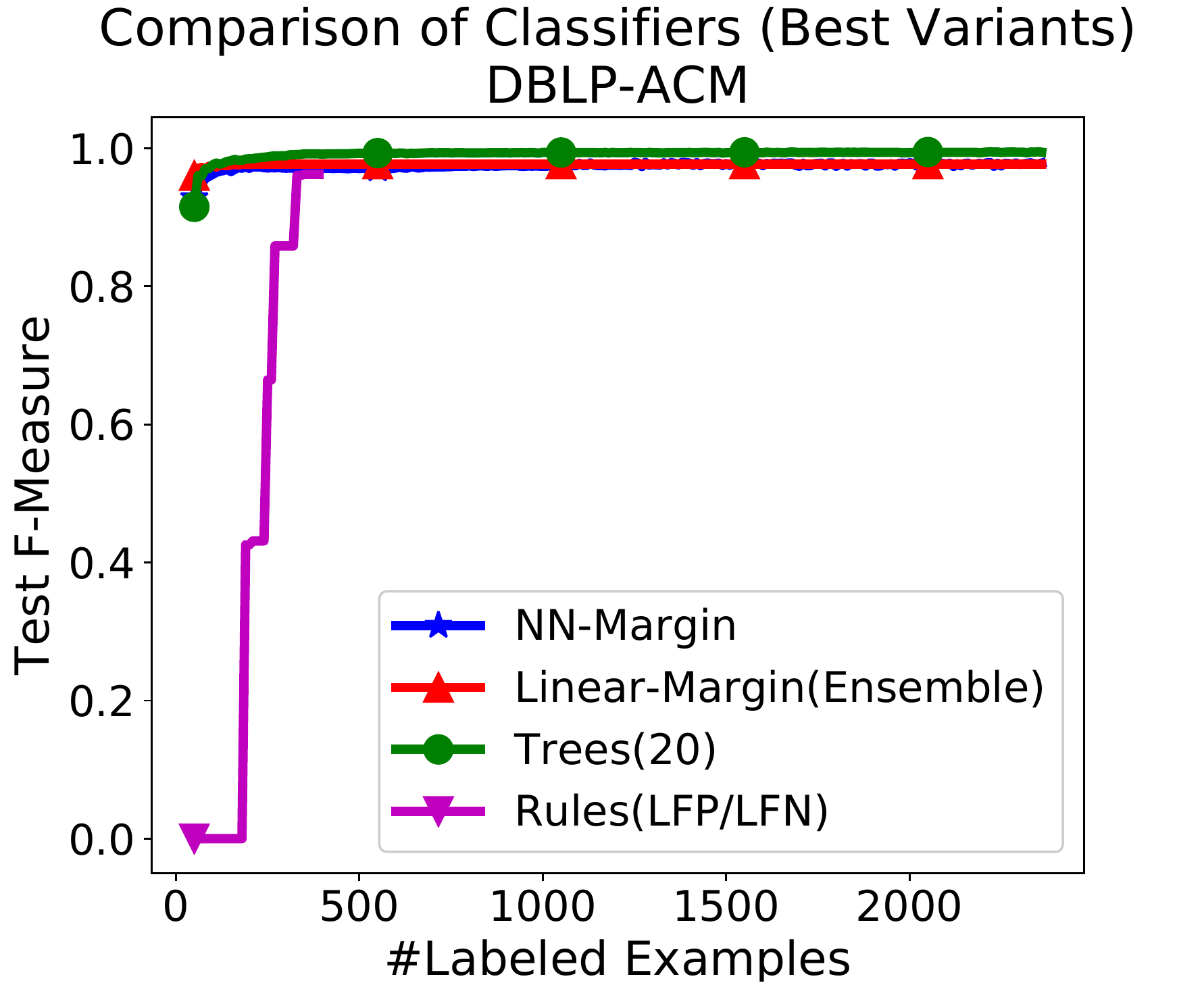}
		\caption{DBLP-ACM}
		\label{fig:DBLP-ACM-Classifiers-F1}
	\end{subfigure}
	\begin{subfigure}[t]{0.19\textwidth}
		\centering
		\includegraphics[width=\linewidth]{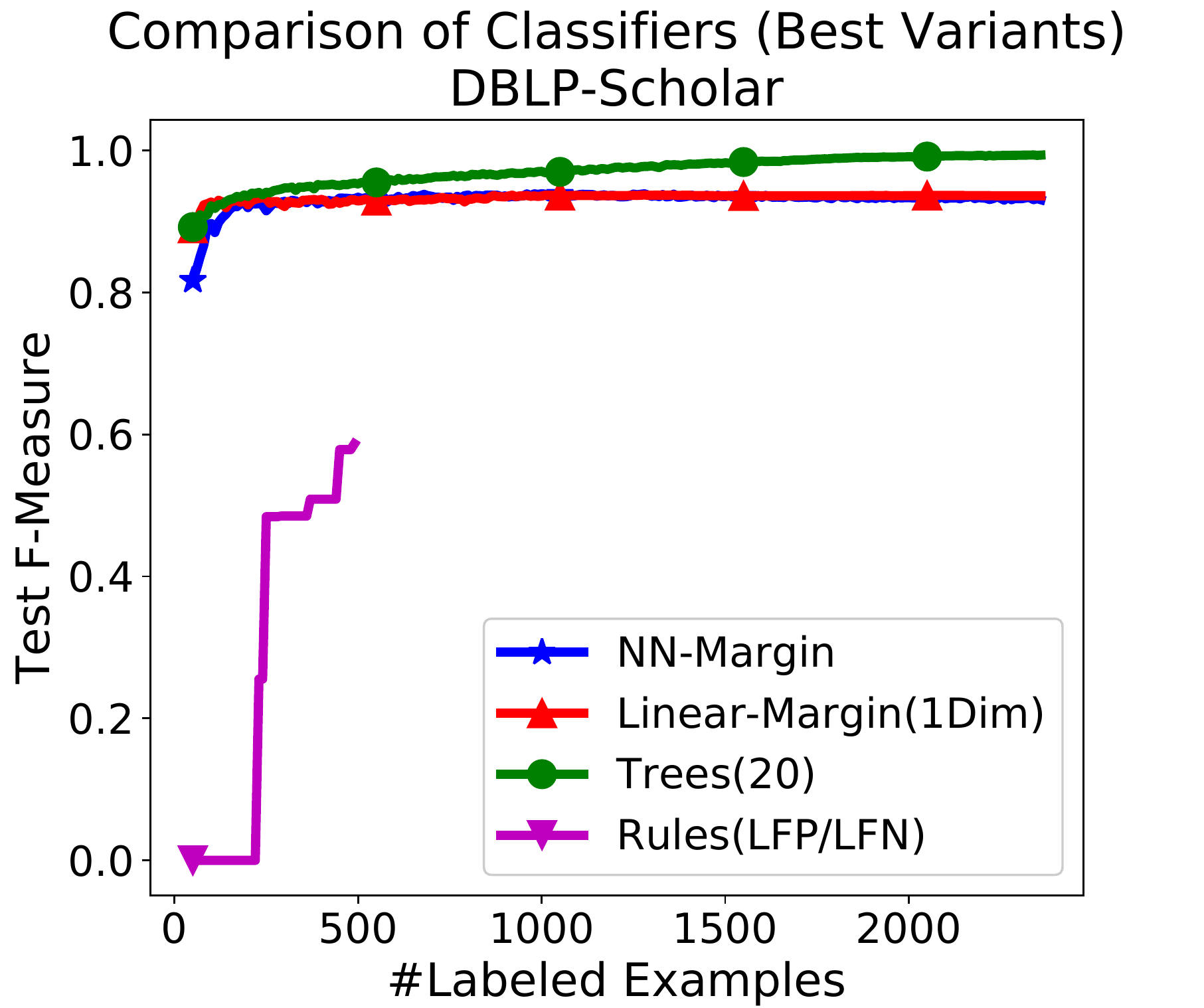}
		\caption{DBLP-Scholar}
		\label{fig:DBLP-Scholar-Classifiers-F1}
	\end{subfigure}
	\begin{subfigure}[t]{0.19\textwidth}
		\centering
		\includegraphics[width=\linewidth]{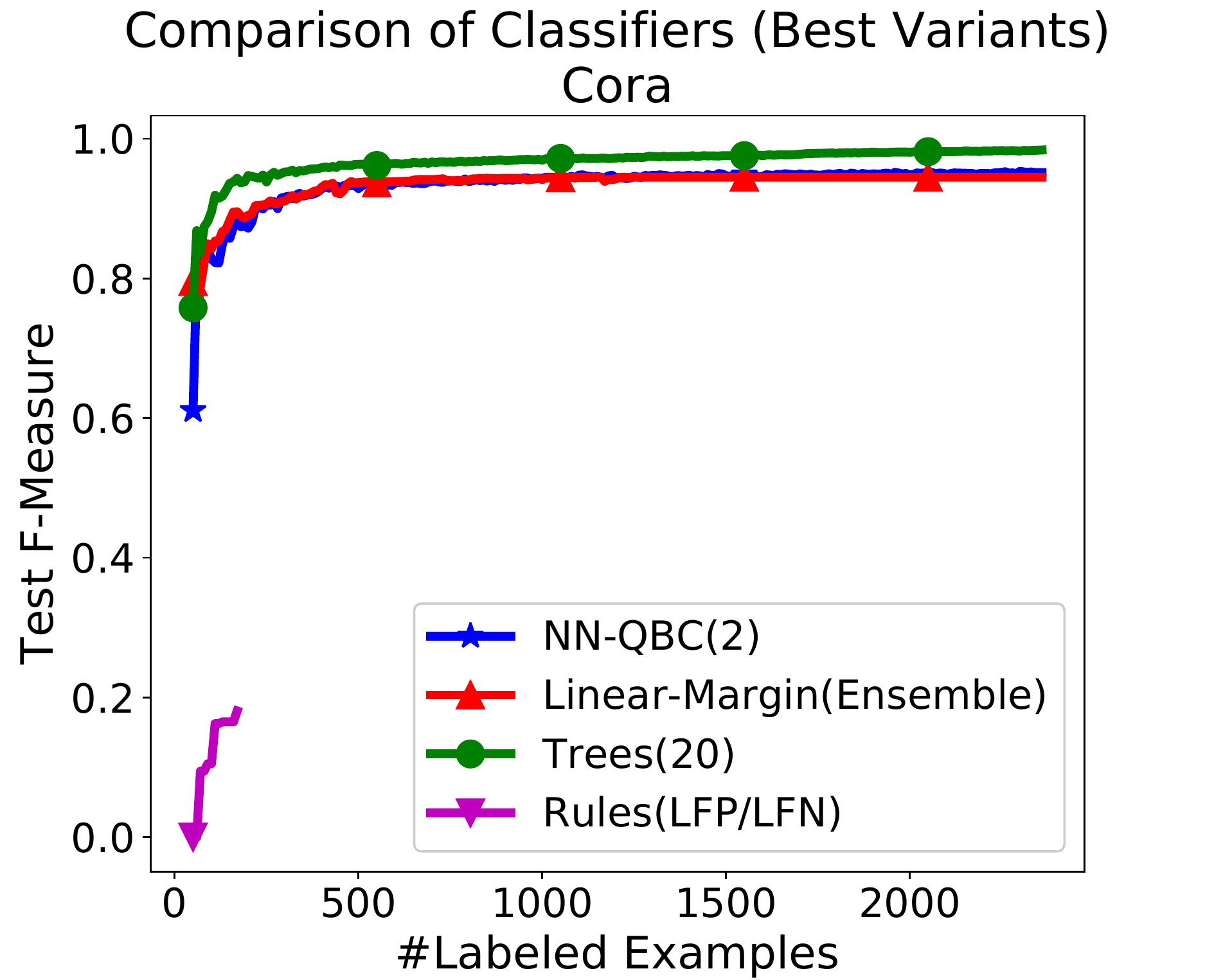}
		\caption{Cora}
		\label{fig:Cora-Classifiers-F1}
	\end{subfigure}
	\vspace*{-.4cm}
	\caption{Comparison of Classifiers with Best Selection Strategies \textit{(Progressive F1-Scores, Perfect Oracle)}}
	\vspace*{-.4cm}
	\label{fig:Classifiers-F1}
\end{figure*}
\begin{figure*}[htb]
	\centering
	\begin{subfigure}[t]{0.19\textwidth}
		\centering
		\includegraphics[width=\linewidth]{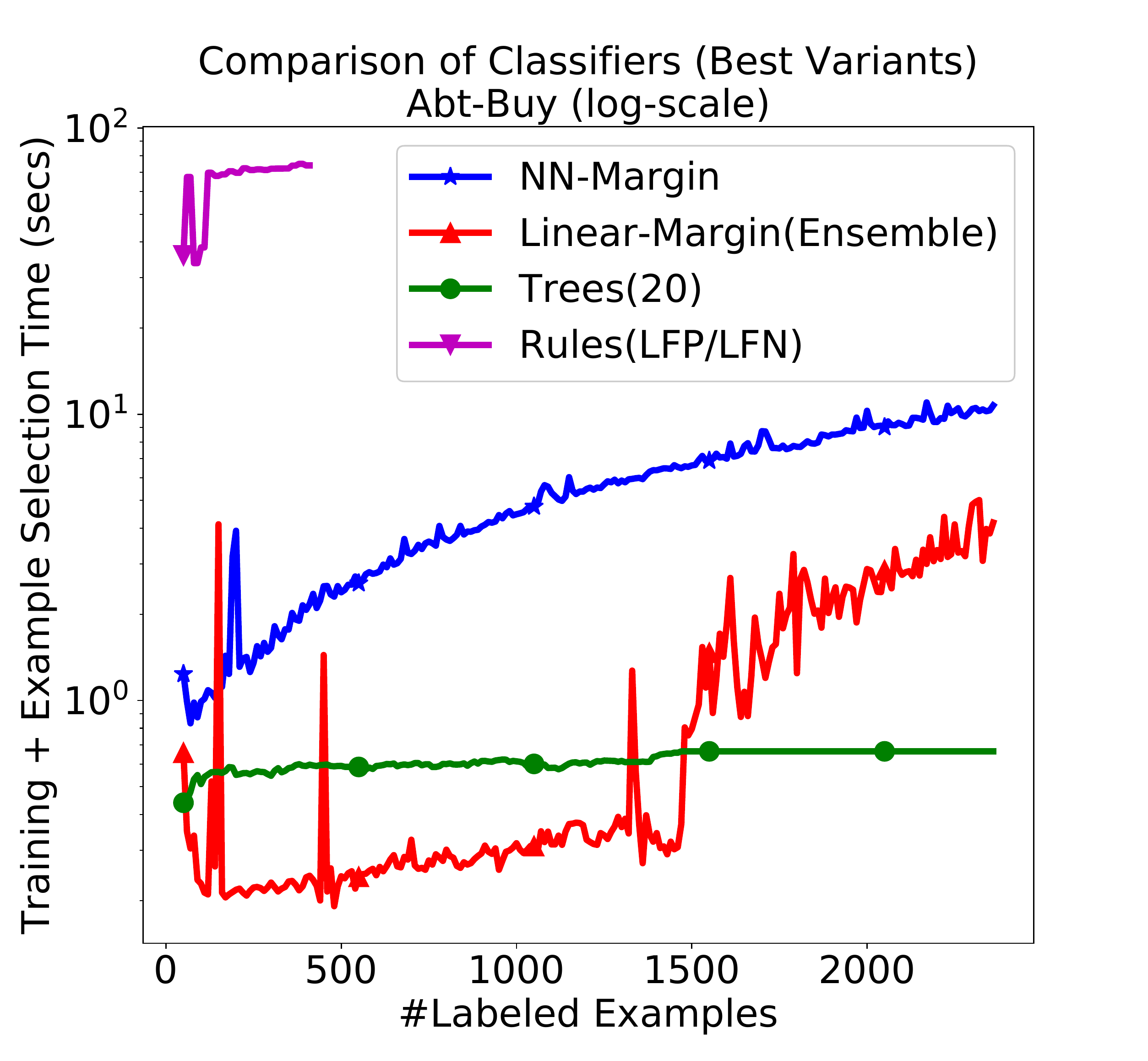}
		\caption{Abt-Buy}
		\label{fig:Abt-Buy-Classifiers-userWaitTime}
	\end{subfigure}
	\begin{subfigure}[t]{0.19\textwidth}
		\centering
		\includegraphics[width=\linewidth]{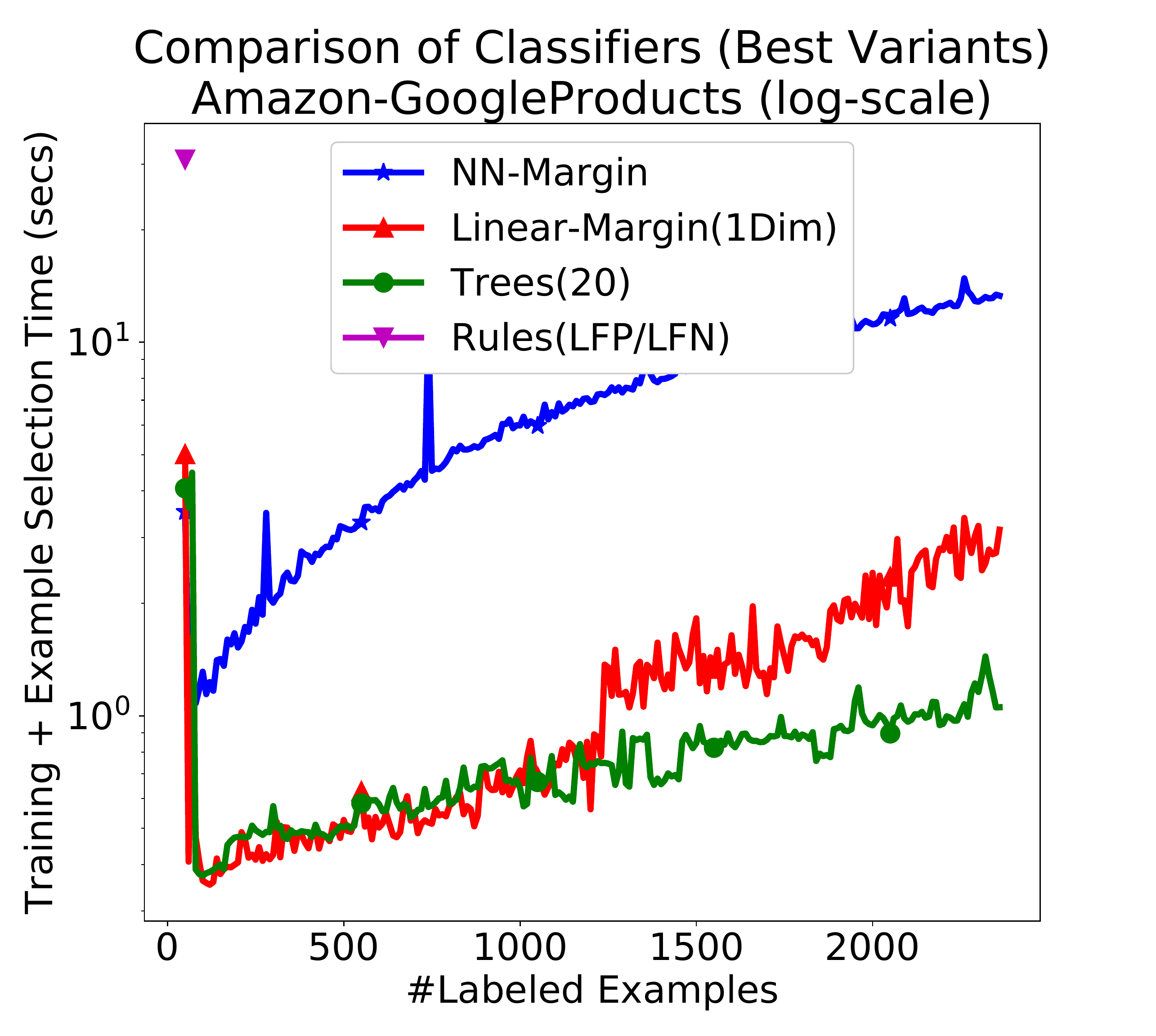}
		\caption{Amazon-Google}
		\label{fig:AmazonGoogleProducts-Classifiers-userWaitTime}
	\end{subfigure}
	\begin{subfigure}[t]{0.19\textwidth}
		\centering
		\includegraphics[width=\linewidth]{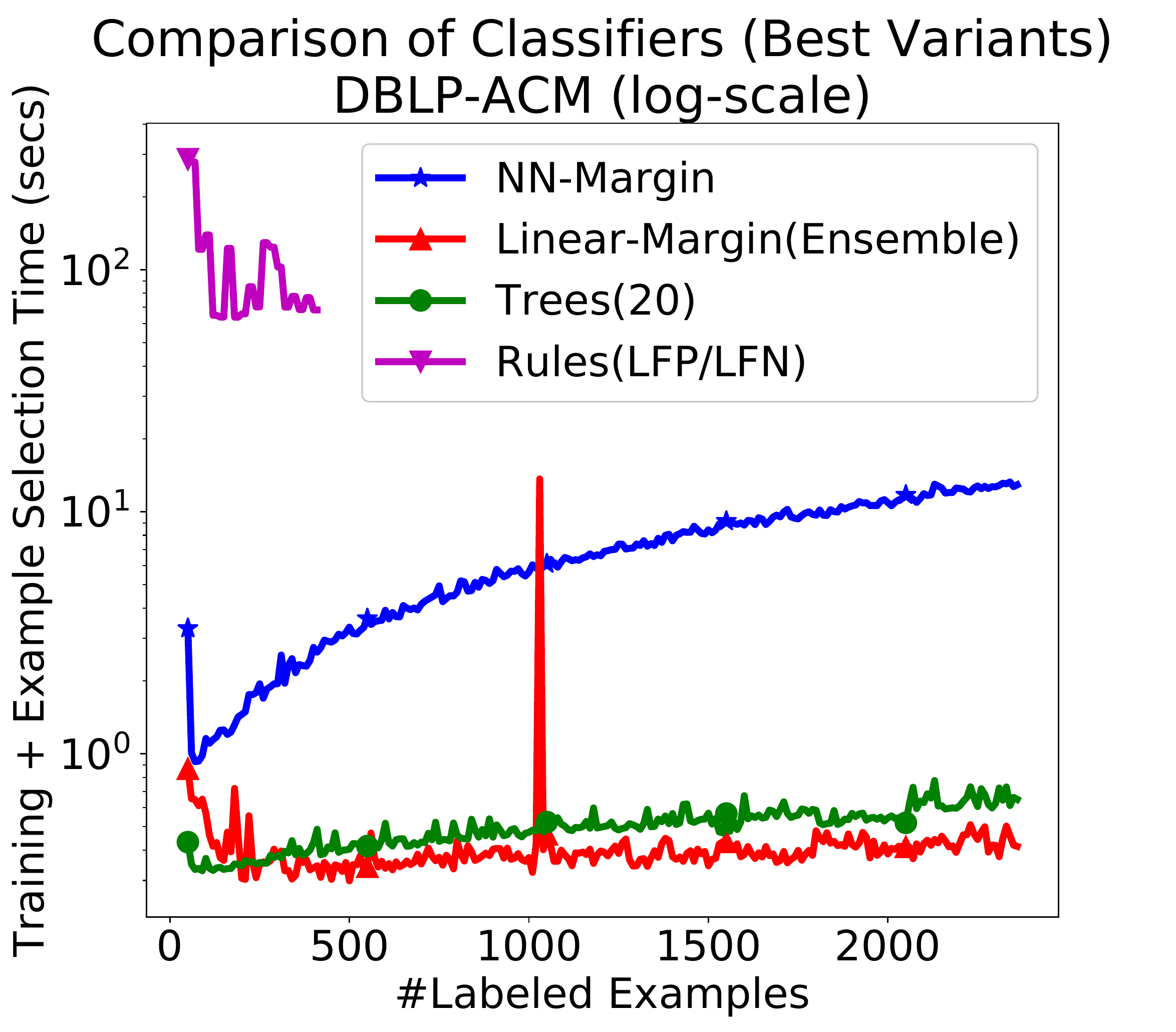}
		\caption{DBLP-ACM}
		\label{fig:DBLP-ACM-Classifiers-userWaitTime}
	\end{subfigure}
	\begin{subfigure}[t]{0.19\textwidth}
		\centering
		\includegraphics[width=\linewidth]{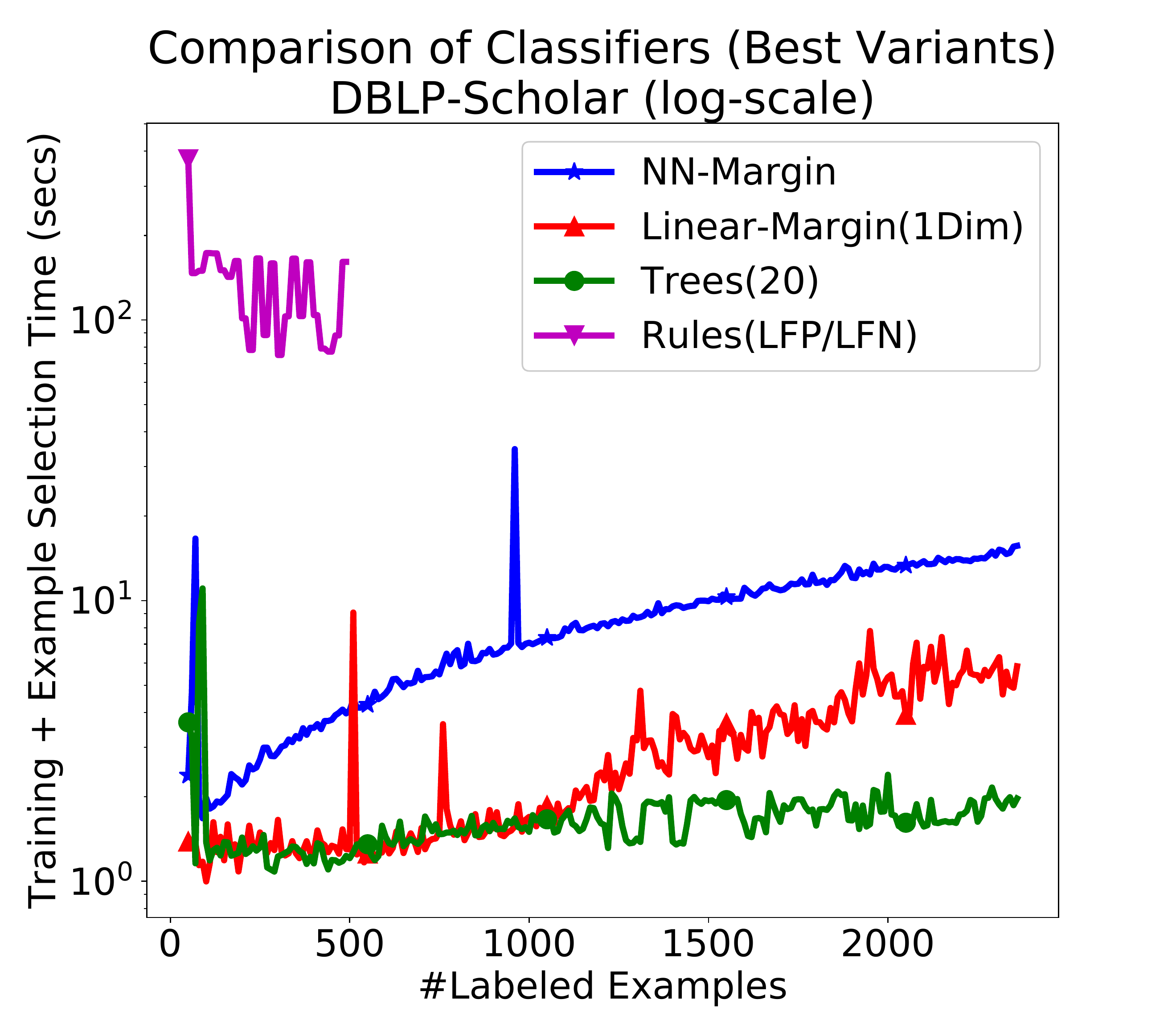}
		\caption{DBLP-Scholar}
		\label{fig:DBLP-Scholar-Classifiers-userWaitTime}
	\end{subfigure}
	\begin{subfigure}[t]{0.19\textwidth}
		\centering
		\includegraphics[width=\linewidth]{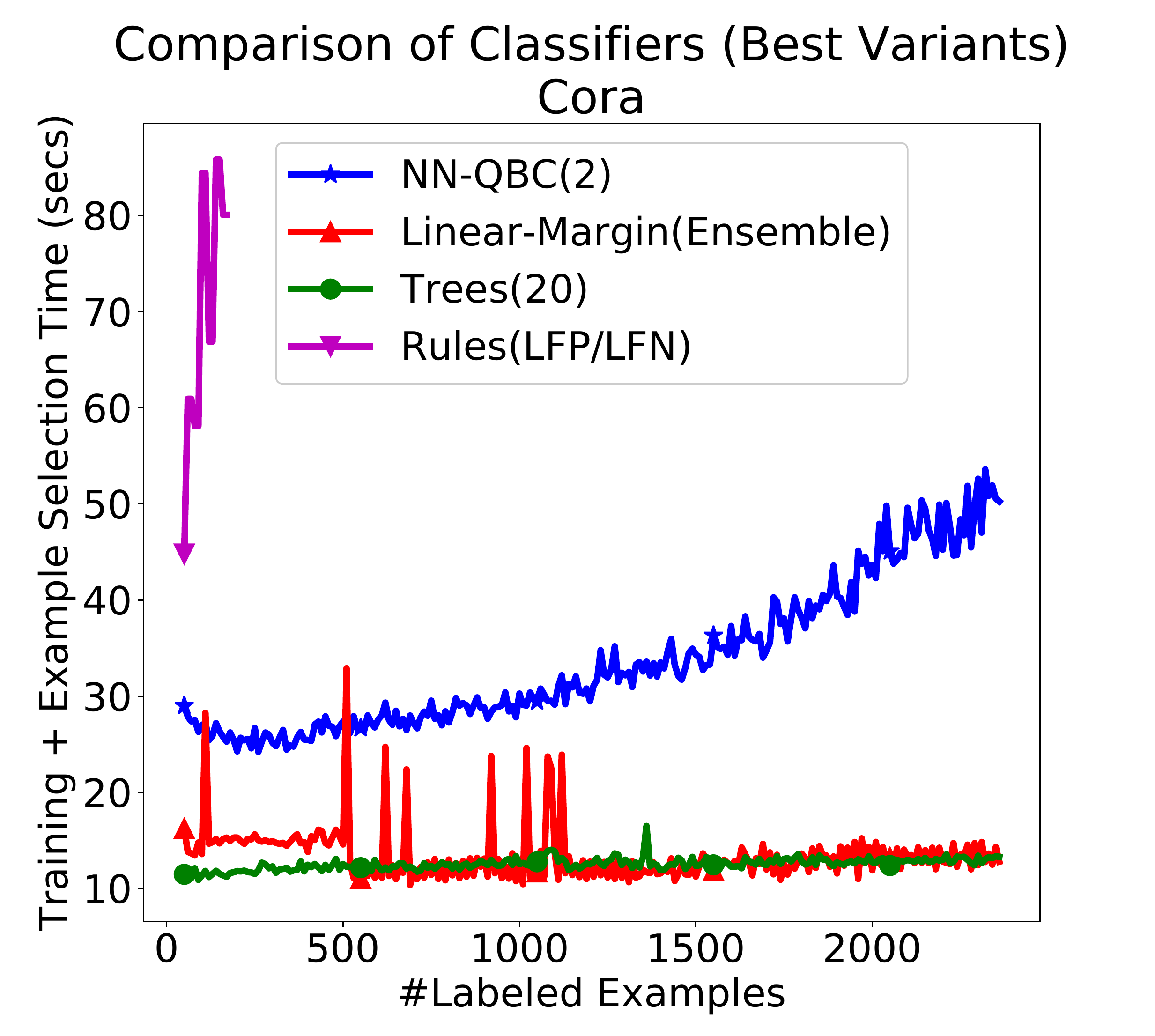}
		\caption{Cora}
		\label{fig:Cora-Classifiers-userWaitTime}
	\end{subfigure}
	\vspace*{-.4cm}
	\caption{Comparison of Classifiers with Best Selection Strategies \textit{(User Wait Time)}}
	\vspace*{-.4cm}
	\label{fig:Classifiers-userWaitTime}
\end{figure*}
\textit{Comparison of Example Selectors}: Figs.~\ref{fig:Abt-Buy-QBCvsMargin} and~\ref{fig:Cora-QBCvsMargin} show the progressive F1-scores of various example selectors applied to non-convex non-linear (neural networks), linear (SVMs) and tree-based (random forests) classifiers respectively. Upon neural networks and SVMs, we compare the learner-agnostic strategy of QBC against margin as the learner-aware strategy. Since neural networks take a long time to train, larger committees are prohibitively expensive in terms of example selection latency. Therefore, we implemented a smaller committee of size 2 for QBC on neural nets, whereas for SVMs (see Fig.~\ref{fig:SVM-Abt-Buy-F1},~\ref{fig:SVM-Cora-F1}), we implemented both QBC(2) and QBC(20) with 2 and 20 learners in the committee. In the case of random forests, margin is inapplicable and the trees in the forest are equivalent to the classifier committee in QBC but created in a learner-aware manner. The F1-scores are plotted for 233 active learning iterations until 2360 labeled examples are consumed (including the initial training set of 30 examples) because this is the maximum $\#$labels needed among all the approaches for the best convergent progressive F1 score. 

As we can see from Figs.~\ref{fig:Abt-Buy-QBCvsMargin} and~\ref{fig:Cora-QBCvsMargin}, margin-based selection achieves similar F1-scores as QBC in conjunction with all learners on both Abt-Buy and Cora. The observations are similar on Amazon-GoogleProducts, DBLP-ACM and DBLP-Scholar although we do not plot them for space reasons. The only exception is in the case of neural networks on the Cora dataset (Fig.~\ref{fig:NN-Cora-F1}) where QBC(2) outperforms margin. Likewise, we plot example selection times on Cora because it has the highest number of pairs post-blocking (114K) incurring the longest example selection time among all the datasets. The latency trends we observe in Fig.~\ref{fig:exSelTime} hold on the remaining datasets as well.  
 Figs.~\ref{fig:NN-Cora-exSelTime} and~\ref{fig:SVM-Cora-exSelTime} present the QBC selection time broken down into committee creation (dashed lines in the figures) and example scoring times (solid lines) (see latency metric in Section~\ref{sec:benchmark} for definitions). 
 
 While the committee creation times increase with more active learning iterations and labeled examples, example scoring times decline with more labels as the unlabeled set shrinks gradually. 
 Thus, margin-based strategy consumes lesser example scoring times than QBC on both neural networks and SVM. Adding the committee creation time to the scoring time of QBC leads to margin outperforming QBC by 10-100x on aggregate selection times.
In contrast to neural networks and SVMs for which we explicitly learn the committees from resampled training data, in the case of random forests, the committee is learned during the training phase and example selection only involves scoring the unlabeled examples based on committee labeling variance. Therefore the difference in example selection time (Fig.~\ref{fig:RF-Cora-exSelTime}) 
among the forests of different $\#$trees (2, 10 and 20) is not too high either. The training times are also not too different because of optimized learner-aware ensemble learning. Ensembles with 20 trees, Trees(20), achieve a near-perfect progressive F1 on all the datasets as compared to smaller tree ensembles. 

\textit{Effect of Blocking and Ensembles on SVM: }
As we have discussed in Section~\ref{sec:blocking}, blocking dimensions are used to prune the ambiguous example space during margin computation. 
Thus, if we assume that all the dimensions in the weight vector are blocking dimensions, the margin has to be computed for every example as none of the examples gets pruned away. This turns out to be equivalent to not using blocking in the first place. 
\begin{table*}[htb]
	\vspace{-2ex}
	\centering
	\vspace{-1ex}
	\begin{small}
		\begin{tabular}{|c|c|c|c|c|c|}
			\hline		
			{\textbf{\scriptsize Approach}}&{\textbf{\scriptsize Abt-Buy}}&{\textbf{\scriptsize Amazon-GoogleProducts}} &{\textbf{\scriptsize DBLP-ACM}} &{\textbf{\scriptsize DBLP-Scholar}} & {\textbf{\scriptsize Cora}}\tabularnewline
			\hline
			\color{brightgreen}{\bf \scriptsize Trees(20)} & \color{brightgreen}{\bf \scriptsize 0.963} \bf \scriptsize (2360 labels) & \color{brightgreen}{\bf \scriptsize 0.971} \bf \scriptsize (2360 labels)  & \color{brightgreen}{\bf \scriptsize 0.99} \bf \scriptsize (260 labels)  & \color{brightgreen}{\bf \scriptsize 0.99} \bf \scriptsize (1770 labels) & \color{brightgreen}{\bf \scriptsize 0.98} \bf \scriptsize (1700 labels)\tabularnewline
			\hline
			\color{brightgreen}{\bf \scriptsize Linear-Margin(Ensemble)} & \scriptsize 0.663 (1470)  & \scriptsize 0.69 (330)  & \scriptsize 0.977 (210) & \scriptsize 0.922 (560) & \scriptsize 0.945 (1220)\tabularnewline
			\hline
			\color{brightgreen}{\bf \scriptsize Linear-Margin(Blocking)} & \scriptsize 0.61 (640) & \scriptsize 0.7 (930)  & \scriptsize 0.975 (170) & \scriptsize 0.936 (920) & \scriptsize 0.89 (220) \tabularnewline
			\hline
			\color{brightgreen}{\bf \scriptsize Linear-QBC(2)} & \scriptsize 0.61 (1420)  & \scriptsize 0.7 (1550)  & \scriptsize 0.976 (170) & \scriptsize 0.935 (1090) & \scriptsize 0.941 (2190) \tabularnewline
			\hline
			\color{brightgreen}{\bf \scriptsize Linear-QBC(20)} & \scriptsize 0.61 (1620) & \scriptsize 0.7 (1260) & \scriptsize 0.976 (180) & \scriptsize 0.936 (1600)  & \scriptsize 0.95 (2130) \tabularnewline
			\hline
			\color{brightgreen}{\bf \scriptsize Non-Convex Non-Linear-Margin} & \scriptsize 0.63 (670) & \scriptsize 0.72 (2360) & \scriptsize 0.978 (1100)  & \scriptsize 0.938 (970) &  \scriptsize 0.709 (410)\tabularnewline
			\hline
			\color{brightgreen}{\bf \scriptsize Non-Convex Non-Linear-QBC(2)} & \scriptsize 0.63 (970) & \scriptsize 0.725 (1350)  & \scriptsize 0.97 (90) & \scriptsize 0.949 (740)  & \scriptsize 0.95 (1640)\tabularnewline
			\hline
			\color{brightgreen}{\bf \scriptsize Rules(LFP/LFN)} & \scriptsize 0.17 (230) & \scriptsize 0.51 (50) & \scriptsize 0.962 (350) & \scriptsize 0.586 (490) & \scriptsize 0.18 (170)  \tabularnewline
			\hline
			\color{red}{\scriptsize Mudgal et al.~\cite{MudgalDeepER}} & \scriptsize 0.628 & \scriptsize 0.693 & \scriptsize 0.984 & \color{red}{\bf \scriptsize 0.947} & \scriptsize N/A  \tabularnewline
			\hline
			\color{red}{\scriptsize Singh et al.~\cite{conciseERVLDB}} & \scriptsize N/A & \color{red}{\bf \scriptsize 0.694} & \scriptsize N/A  & \scriptsize 0.9436 & \scriptsize 0.9718 \tabularnewline
			\hline
			\color{red}{\scriptsize Kopcke et al.~\cite{ERKopcke}} & \scriptsize 0.713  & \scriptsize 0.622 &  \scriptsize 0.976 & \scriptsize 0.894 & \scriptsize N/A \tabularnewline
			\hline
			\color{red}{\scriptsize Kasai et al.~\cite{kasai:acl19}} & \scriptsize N/A & \scriptsize N/A & \color{red}{\bf \scriptsize 0.985}  & \scriptsize 0.929  & \scriptsize 0.987  \tabularnewline
			\hline
			\color{red}{\scriptsize Mozafari et al.~\cite{MozafariBootstrap}} & \scriptsize 0.56  & \scriptsize N/A & \scriptsize N/A & \scriptsize N/A  & \scriptsize N/A  \tabularnewline
			\hline
			\color{red}{\scriptsize \textsf{Corleone}~\cite{Corleone}} & \scriptsize N/A & \scriptsize N/A & \scriptsize N/A  & \scriptsize 0.921 & \scriptsize N/A  \tabularnewline
			\hline
			\color{red}{\scriptsize \textsf{Waldo}~\cite{verroios:sigmod17}} & \color{red}{\bf \scriptsize 0.8} \bf \scriptsize (2200) & \scriptsize N/A & \scriptsize N/A  & \scriptsize N/A & \color{red}{\bf \scriptsize 1.0} \bf \scriptsize (1600)  \tabularnewline
			\hline
			\color{red}{\scriptsize Whang et al.~\cite{whang:vldb13}} & \scriptsize N/A & \scriptsize N/A & \scriptsize N/A  & \scriptsize N/A & \scriptsize 0.9 (1000)  \tabularnewline
			\hline
			\color{red}{\scriptsize Chai et al.~\cite{CostEffectiveCrowdSourcing}} & \scriptsize N/A & \scriptsize N/A & \scriptsize 0.9 (150)  & \scriptsize N/A & \scriptsize 0.92 (354)  \tabularnewline
			\hline
		\end{tabular}
	\end{small}
	\caption{Best Progressive F1-Scores from our Benchmark using Perfect Oracles vs. State-of-the-art Approaches}
	\label{tbl:bestResults}
	\vspace{-4ex}
\end{table*}
We compare the margin baseline that uses all dimensions for blocking against using a single blocking dimension. 
\RTHREE{As we can notice from Fig.~\ref{fig:SVM-Cora-BlockingEnsemble-exSelTime}, using a single blocking dimension denoted by margin(1Dim), brings savings in example selection time without sacrificing quality (see Fig.~\ref{fig:SVM-BlockingEnsemble-F1}). With the exception of the Cora dataset (Fig.~\ref{fig:SVM-Cora-BlockingEnsemble-F1}) where margin(1Dim) performs worse than the baseline margin(188 Dim), blocking performs same as vanilla margin flavors - margin(62Dim) from Abt-Buy and margin(83Dim) from Amazon-GoogleProducts, DBLP-ACM and DBLP-Scholar on all the other datasets (Fig.~\ref{fig:SVM-Abt-Buy-BlockingEnsemble-F1} to \ref{fig:SVM-DBLP-Scholar-BlockingEnsemble-F1}).} 

\RTHREE{As we have described in Section~\ref{sec:ensemble}, active ensembles learned incrementally over several active learning iterations prune predicted matching pairs from both labeled and unlabeled examples. This results in the example selection time using active ensembles decreasing aggressively in the later active learning iterations as shown in Fig.~\ref{fig:SVM-Cora-BlockingEnsemble-exSelTime}. 
While active ensembles of SVM achieve slightly higher progressive F1 on Abt-Buy and DBLP-ACM (see Fig.~\ref{fig:SVM-BlockingEnsemble-F1}), they perform slightly worse than baseline margin on Amazon-GoogleProducts and DBLP-Scholar and show no effect on Cora. This is because of a uniform precision threshold of 0.85 we use across all the datasets. This threshold is conservative enough for Abt-Buy and DBLP-ACM, which can be inferred from two high precision SVMs accepted into the ensemble by the termination of active learning. However, this may not be a suitable threshold for DBLP-Scholar and Amazon-GoogleProducts on which there is no significant boost in the progressive F1-score despite accepting 7 (or 3) SVMs into the ensemble.}

\textit{Comparison of Classifiers: }
The results we have reported so far fix the classifier and vary the example selector. In Figs.~\ref{fig:Classifiers-F1} and~\ref{fig:Classifiers-userWaitTime}, we compare the best performing example selectors from each of the classifiers (margin for neural nets, margin with ensemble or blocking for linear SVMs, learner aware QBC(20) for random forests and LFP/LFN for rule learners) against each other w.r.t. progressive F1-score and user wait time (which is the sum of train time and example selection time, see Section~\ref{sec:benchmark} for definition).
Having compared these best example selectors from each classifier, we observe from Fig.~\ref{fig:Classifiers-F1} that random forest with QBC(20) labeled as \textsf{Trees(20)} outperforms all other learners upon progressive F1-scores on all the datasets.  
We find that rules lead to the largest user wait time, least progressive F1-scores and early termination. However, they perform much better on interpretability and produce an ensemble of concise, high precision DNF rules that can be easily understood and debugged by the end user. We delve into the details of rule-based results in the section on interpretability. Neural networks incur the second largest latency because of the long training they undergo, while random forests require the user to wait for the shortest time despite training an ensemble of 20 trees. This emphasizes the importance of learner-aware training rather than learner-agnostic training used in QBC. SVMs with blocking and ensembles incur the least user wait times in the beginning but with the arrival of more labels, the training time increases thus increasing the wait time between iterations.

\textit{$\#$Labels:}
\RONE{Table~\ref{tbl:bestResults} presents the best progressive F1-scores from the active learning approaches implemented in our benchmark (highlighted in green) and the best F1-scores reported by earlier related works (highlighted in red). For our benchmark results, we also present within parentheses, the minimum \# labels required by the approaches from a perfect Oracle, to converge to the corresponding F1-scores. For results on noisy Oracles, please refer to Section~\ref{sec:expNoisy}.} 
Learner-aware committees (size 20) of tree-based learners achieve the best results close to 1.0 on all the datasets but also consume the largest \# labels. However, we can also notice from Fig.~\ref{fig:Classifiers-userWaitTime} that these approaches achieve them with least user wait time. Among the active learning methods for linear classifiers, margin-based optimizations of blocking and active ensemble achieve comparable progressive F1 as QBC while requiring fewer labels and lesser user wait time on almost all the datasets. Although QBC(2) of non-convex non-linear classifiers consumes fewer labels than its margin counterpart on 3 out of 5 datasets and achieves similar F1 scores, training committees of neural nets incurs huge training times. Rule learning using LFP/LFN terminates as soon as no LFPs or LFNs are found on the learned ensemble of high precision rules. This keeps its $\#$labels low and because of the limited number of similarity functions supported by the heuristic, it achieves low progressive F1-scores.
\begin{figure*}[htb]
	\centering
	\begin{subfigure}[t]{0.2\textwidth}
		\centering
		\includegraphics[width=\linewidth]{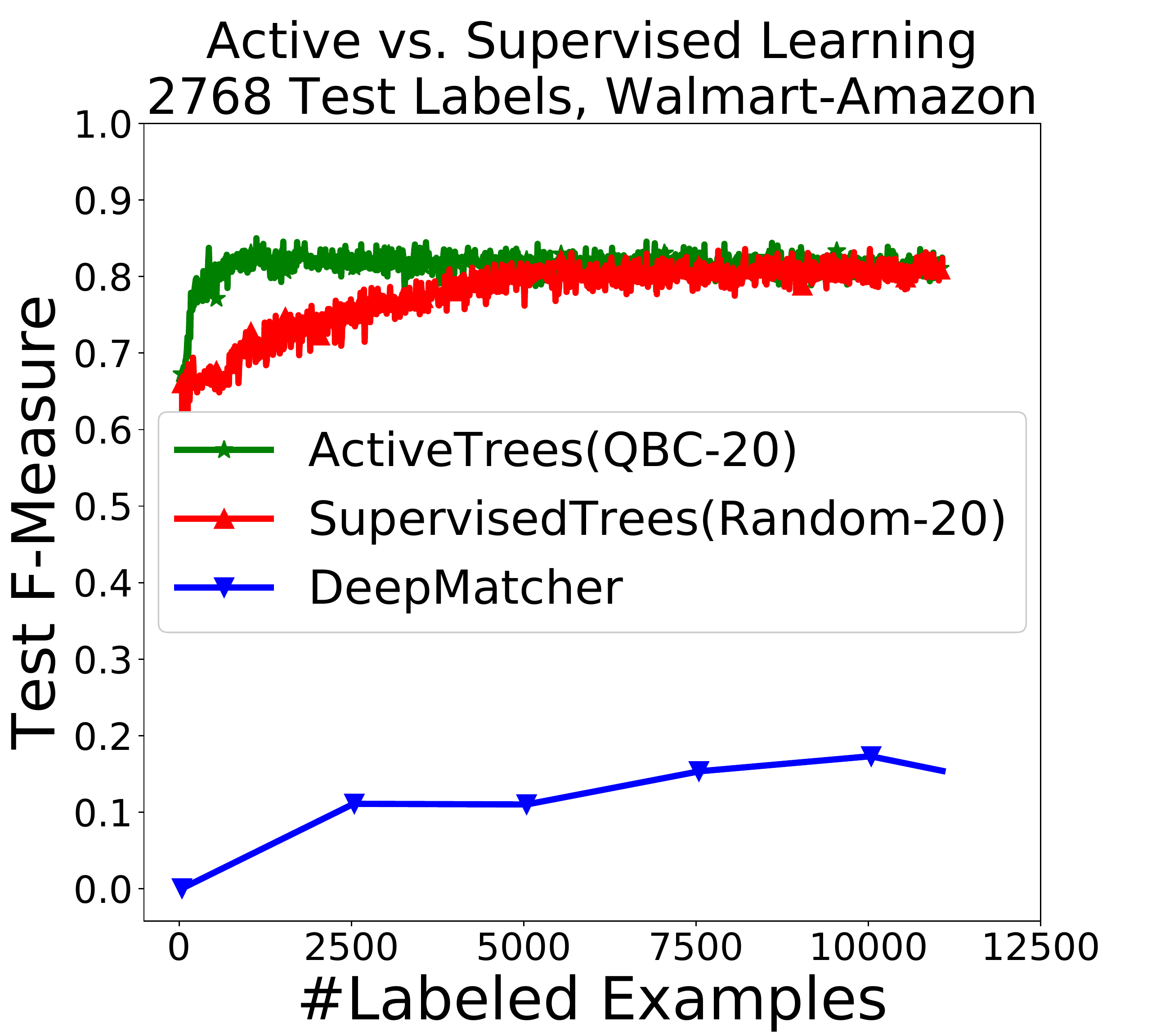}
		\caption{Walmart-Amazon}
		\label{fig:Walmart-Amazon-ActivevsSupervised-Noisy-0}
	\end{subfigure}
	\begin{subfigure}[t]{0.2\textwidth}
		\centering
		\includegraphics[width=\linewidth]{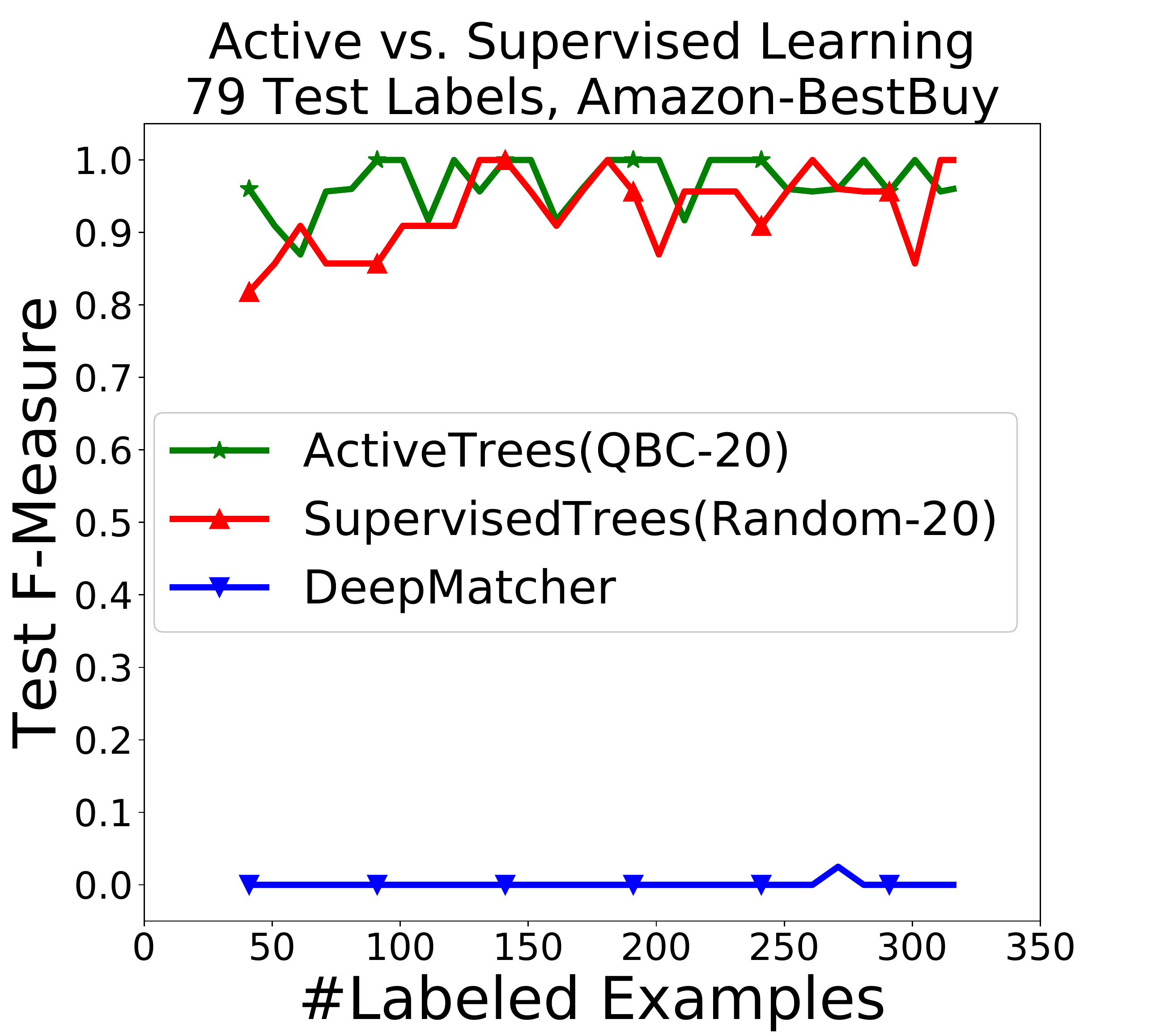}
		\caption{Amazon-BestBuy}
		\label{fig:Amazon-BestBuy-ActivevsSupervised-Noisy-0}
	\end{subfigure}
	\begin{subfigure}[t]{0.2\textwidth}
		\captionsetup{singlelinecheck = false, format= hang, justification=raggedright, font=footnotesize, labelsep=space}
		\centering
		\includegraphics[width=\linewidth]{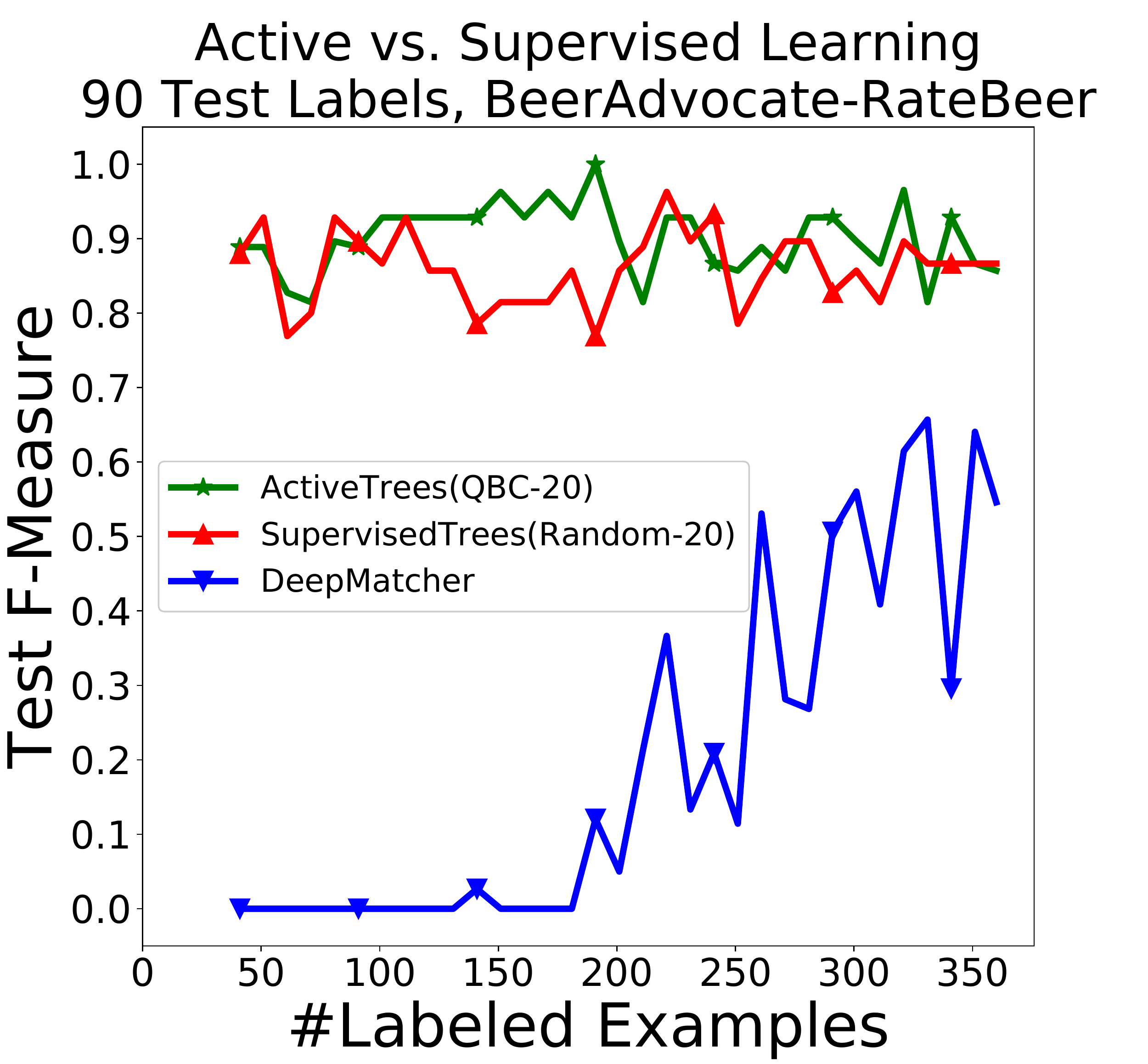}
		\caption{BeerAdvocate-RateBeer}
		\label{fig:BeerAdvocate-RateBeer-ActivevsSupervised-Noisy-0}
	\end{subfigure}
	\begin{subfigure}[t]{0.2\textwidth}
		\captionsetup{singlelinecheck = false, format= hang, justification=raggedright, font=footnotesize, labelsep=space}
		\centering
		\includegraphics[width=\linewidth]{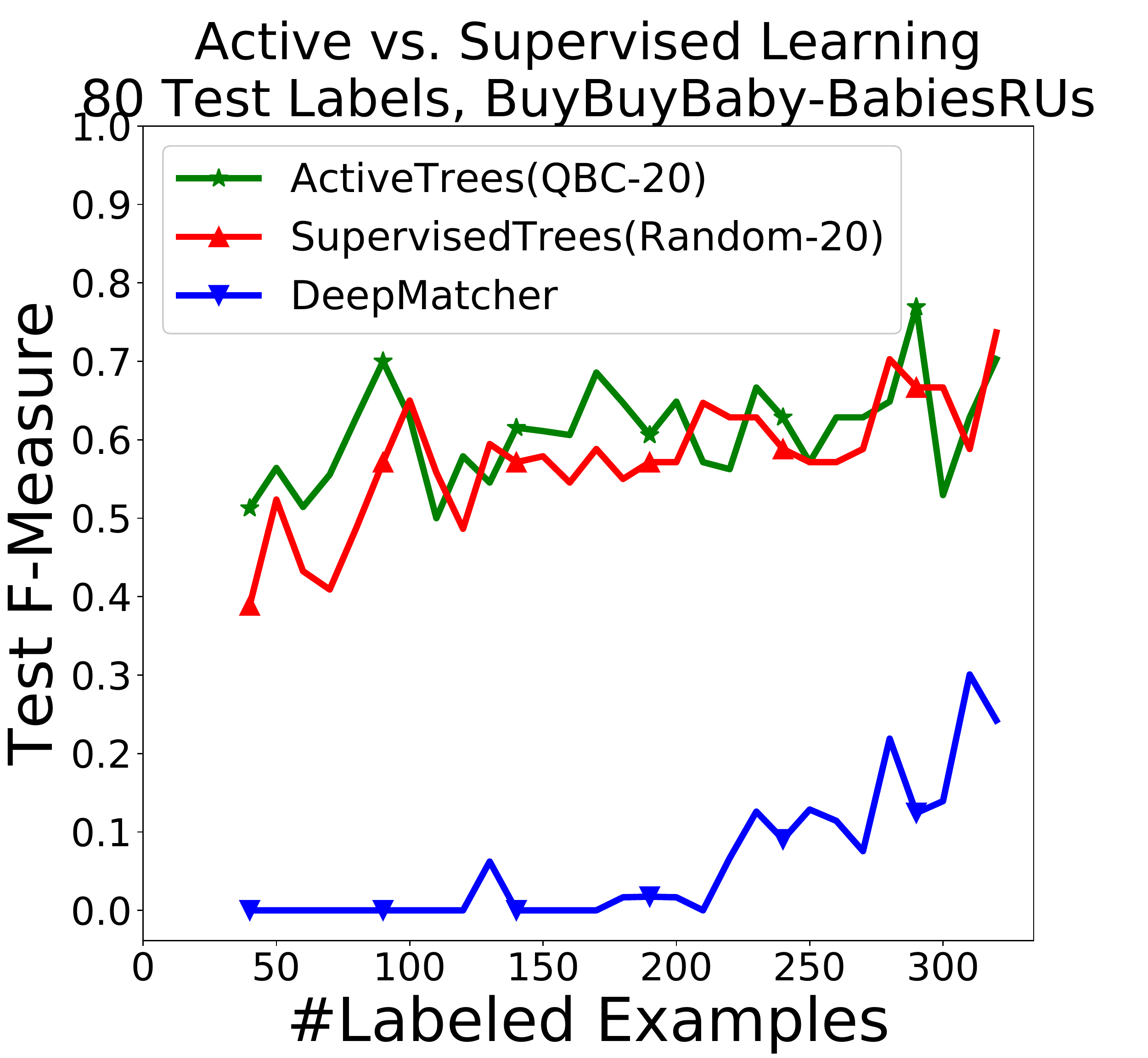}
		\caption{BuyBuyBaby-BabiesRUs}
		\label{fig:BuyBuyBaby-BabiesRUs-ActivevsSupervised-Noisy-0}
	\end{subfigure}
	\vspace*{-.4cm}
	\caption{Active vs. Supervised Learning on Magellan/DeepMatcher Datasets \textit{(Perfect Oracles, 20\% Test Labels)}}
	\vspace*{-.4cm}
	\label{fig:ActivevsSupervisedMagellan}
\end{figure*}

\textit{Results from Related Work:} \RONE{Among the related work, we found supervised, transfer and crowd-sourced learning-based approaches which have reported results on our experimental datasets. Note that each of these works use their own set of train and test splits which may vary from our test set that happens to be the entire set of post-blocking EM pairs comprising both labeled and unlabeled examples. The purpose of Table~\ref{tbl:bestResults} is thus to contrast the best F1 results from the related work with those from our benchmark. For an experimental comparison of active learning with the state-of-the-art supervised learning, see Fig.~\ref{fig:ActivevsSupervisedMagellan}.} Among the supervised learning approaches in Table~\ref{tbl:bestResults}, we mention results from \textsf{DeepMatcher}~(\citet{MudgalDeepER}),~\citet{conciseERVLDB} and~\citet{ERKopcke}. While~\citet{MudgalDeepER} use deep learning, the latter two compare rule-based models with machine learning models. However, the results we report here are not necessarily from those approaches but also from their contenders who achieved the best results on these datasets. For instance, the best result from~\citet{conciseERVLDB} is from supervised random forests. Likewise, some of the best results in~\citet{kasai:acl19} are from their implementation of supervised deep learning models that they try to surpass through transfer learning.~\citet{MozafariBootstrap}, \textsf{Corleone}~\cite{Corleone}, \textsf{Waldo}~\cite{verroios:sigmod17},~\citet{whang:vldb13} and~\citet{CostEffectiveCrowdSourcing} use crowd-sourcing approaches which are slightly different from active learning as they rely on the crowd for labels unlike active learning which assumes the presence of experts for labeling. 
\subsection{Experiments with Noisy Oracles}
\label{sec:expNoisy}
\RFIVE{In order to model crowd-sourced scenarios, we use an imperfect Oracle which perturbs the original label with a fixed probability whenever it is asked to label an example.
We vary the noise from 10$\%$ to 40$\%$ following the DeepMatcher~\cite{MudgalDeepER} settings.
It should be noted that we always perturb the original label whenever the imperfect Oracle generates a random probability that falls within the noise percentage threshold. This is a harsher criterion than real-world crowdsourced settings which regulate the noisy labels using techniques such as majority voting and label inference. Each F1-score observed with a noisy Oracle is averaged over 5 random runs using distinct random seeds to account for the randomness and ensure experimental reproducibility.}

\RFIVE{Our results on the Abt-Buy dataset in Fig.~\ref{fig:noisyOracle} show that tree ensembles produce a near-perfect F1-score using a perfect Oracle, and their performance degrades gracefully with increasing noise percentages (Fig.~\ref{fig:Abt-Buy-Noisy-RF-F1}). Tree ensembles have a relative advantage until 20$\%$ noise beyond which their F1 equalizes to that of other classifiers. In contrast, the other classifiers do not even come close to near-perfect F1-scores on perfect Oracles. While linear SVMs show a quick drop beyond 10$\%$ noise (30$\%$ and 40$\%$ overlap in Figs.~\ref{fig:Abt-Buy-Noisy-SVMEnsemble-F1} $\&$~\ref{fig:Abt-Buy-Noisy-SVMBlocking-F1}), neural networks do not suffer a steep decline in F1-scores at higher noise percentages because of regularization techniques such as drop-out and batch normalization. 
We do not present results of rule-based classifiers as they produce low progressive F1-scores even with perfect Oracles.}

\RFIVE{On similar lines, we notice from our experiments on the Magellan/DeepMatcher datasets in Fig.~\ref{fig:noisyOracleMagellan} that using the 0$\%$ noisy (perfect) Oracle, tree ensembles of size 20 produce high progressive F1-scores close to 1.0 from early on with as few as 100 labels on Amazon-BestBuy and Beer datasets (Fig.~\ref{fig:Amazon-BestBuy-Noisy-RF-F1},~\ref{fig:BeerAdvocate-RateBeer-Noisy-RF-F1}). However, the convergence on Walmart-Amazon and BabyProducts (Fig.~\ref{fig:Walmart-Amazon-Noisy-RF-F1},~\ref{fig:BuyBuyBaby-BabiesRUs-Noisy-RF-F1}) happens only after 2500 and 
300 labels respectively indicating that these are indeed challenging datasets. Upon higher noise percentages, the gradual drop in the F1-scores with an increase in labeled examples is a testimony to the fact that crowd-sourcing in practical scenarios warrant a much earlier termination and error correction techniques such as majority voting. However, the sweet spot in terms of when to terminate active learning in such scenarios may differ across datasets. For instance, while the progressive F1-scores show a monotonically declining F1-score curve on Walmart-Amazon, Amazon-BestBuy and Beer datasets, they show monotonically increasing F1-scores on the Baby Products dataset in Fig.~\ref{fig:BuyBuyBaby-BabiesRUs-Noisy-RF-F1}.}

\begin{figure}[htb]
	\captionsetup{singlelinecheck = false, format= hang, justification=raggedright, font=footnotesize, labelsep=space}
	\centering
	\begin{subfigure}[t]{0.155\textwidth}
		\centering
		\includegraphics[width=\linewidth]{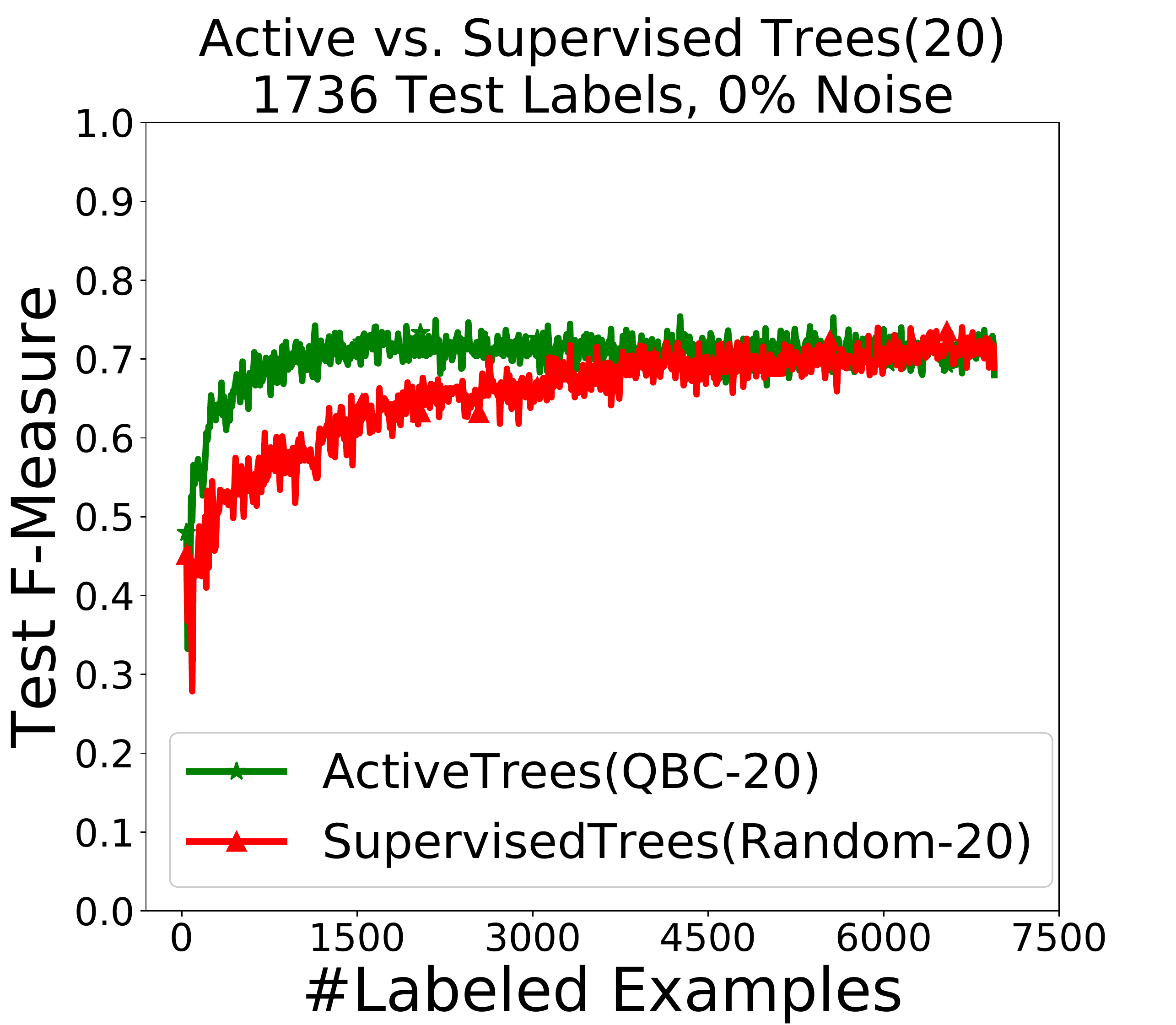}
		\caption{0\% Noisy Oracle}
		\label{fig:Abt-Buy-ActivevsSupervised-Noisy-0}
	\end{subfigure}
	\begin{subfigure}[t]{0.155\textwidth}
		\captionsetup{singlelinecheck = false, format= hang, justification=raggedright, font=footnotesize, labelsep=space}
		\centering
		\includegraphics[width=\linewidth]{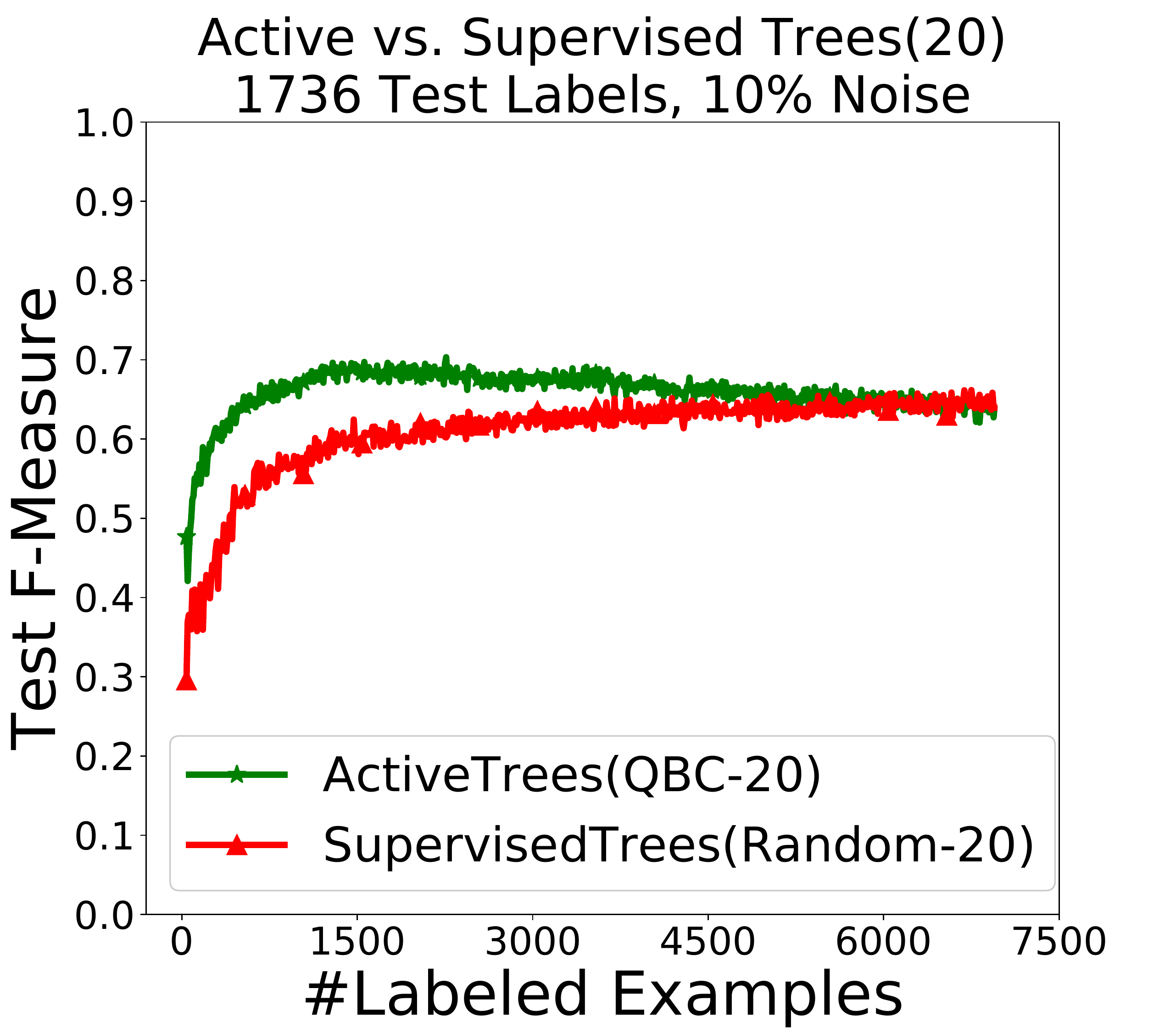}
		\caption{10\% Noisy Oracle}
		\label{fig:Abt-Buy-ActivevsSupervised-Noisy-10}
	\end{subfigure}
	\begin{subfigure}[t]{0.155\textwidth}
		\captionsetup{singlelinecheck = false, format= hang, justification=raggedright, font=footnotesize, labelsep=space}
		\centering
		\includegraphics[width=\linewidth]{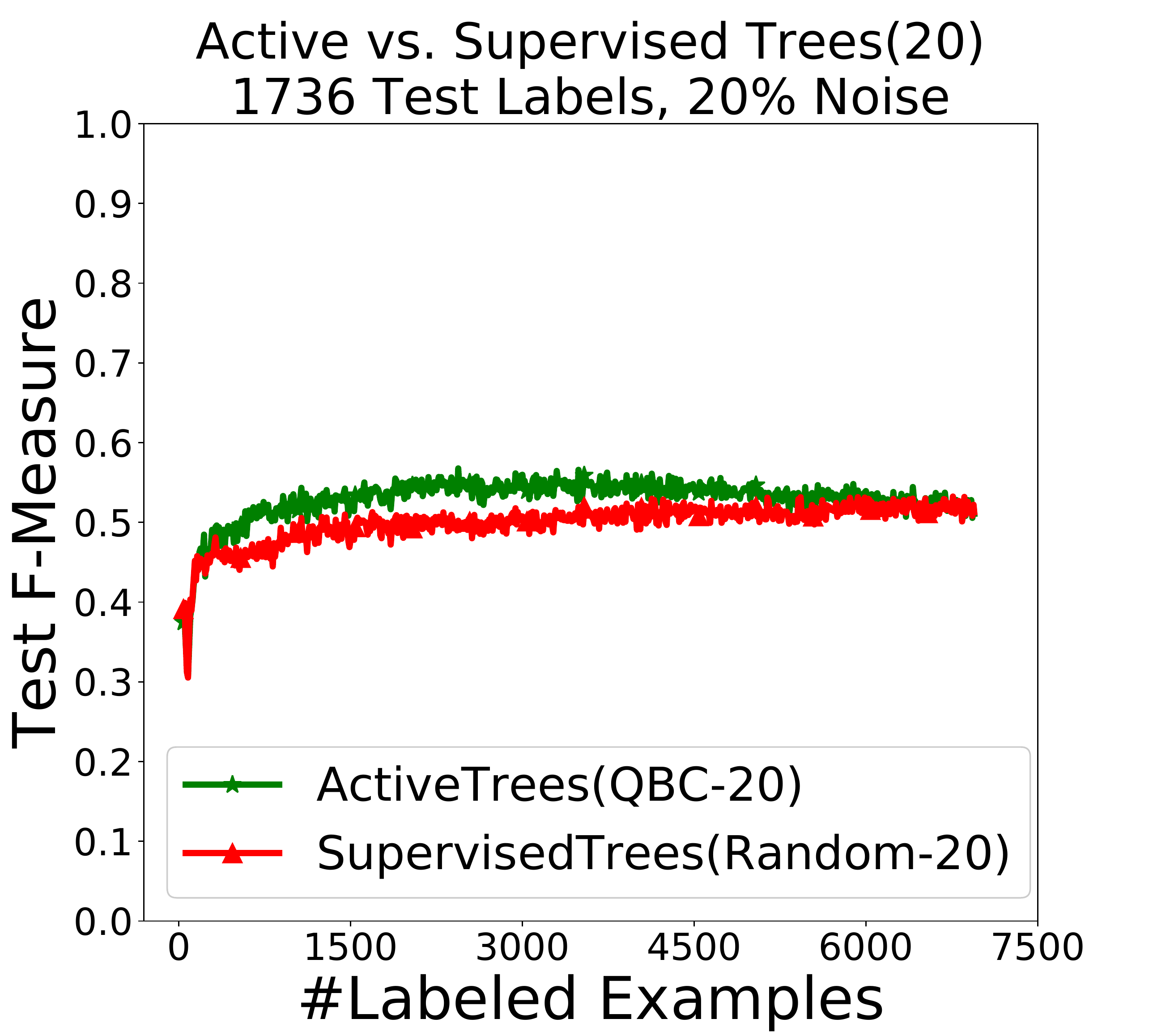}
		\caption{20\% Noisy Oracle}
		\label{fig:Abt-Buy-ActivevsSupervised-Noisy-20}
	\end{subfigure}
	\vspace*{-.4cm}
	\caption{Active vs. Supervised Trees \textit{(Abt-Buy, 20$\%$ Test Labels)}}
	\vspace*{-.4cm}
	\label{fig:ActiveVsSupervised}
\end{figure}

\textit{Comparison with Supervised Learning}: 
\RFIVE{We conduct these experiments following the conventional train-test splits of supervised learning where example selection is done out of a training set containing 80$\%$ of post-blocking examples and the evaluation is on a held-out test set of 20$\%$ of the tuple pairs which never participate in example selection. In Fig.~\ref{fig:ActiveVsSupervised}, we compare active learning against supervised learning on Abt-Buy using ensembles of 20 trees upon various imperfection levels of an Oracle. In each iteration, while active learning uses learner-aware QBC to label examples that lead to highest labeling disagreement (entropy) among the 20 decision trees, supervised learning picks random examples in each iteration. The results show that the former outperforms the latter within the first few iterations while achieving test F1-scores comparable to those that supervised learning achieves after training on the entire set of 80$\%$ training examples. This difference between supervised and active learning is insignificant at 20$\%$ noisy Oracle (see Fig~\ref{fig:Abt-Buy-ActivevsSupervised-Noisy-20}).}

\textit{Comparison with DeepMatcher~\cite{MudgalDeepER}:}
\RFIVE{Fig.~\ref{fig:ActivevsSupervisedMagellan} compares tree ensembles against a state-of-the-art supervised learning approach, DeepMatcher, upon perfect Oracles. We ran DeepMatcher with the same settings in~\cite{MudgalDeepER} by dividing the labeled examples into 3:1 (train to validation set size ratio) and evaluating it on the same 20$\%$ test set as tree ensembles. Similar to supervised tree ensembles, DeepMatcher picks a random set of unlabeled examples to label in each iteration. While we did not notice a significant variation among the test F1-scores from different executions of tree ensembles, we noticed a standard deviation of 0.002 (Amazon-BestBuy), 0.016 (Walmart-Amazon), 0.06 (Baby Products) and 0.125 (Beer) across 5 runs of DeepMatcher over all the learning iterations. Hence we report an average F1-score across these 5 runs in Fig.~\ref{fig:ActivevsSupervisedMagellan}. The results confirm that active learning using random forests requires fewer labels to achieve significantly higher test F1-scores than supervised learning. While supervised and active tree ensembles perform similarly on the smaller datasets, DeepMatcher, requires the entire 80$\%$ training labels to achieve its best test F1-scores.}
\subsection{Interpretability: Rules vs. Trees}
\label{sec:expsec3}
\begin{figure}[htb]
	\centering
	\begin{subfigure}[t]{0.21\textwidth}
		\captionsetup{singlelinecheck = false, format= hang, justification=raggedright, font=footnotesize, labelsep=space}
		\centering
		\includegraphics[width=\linewidth]{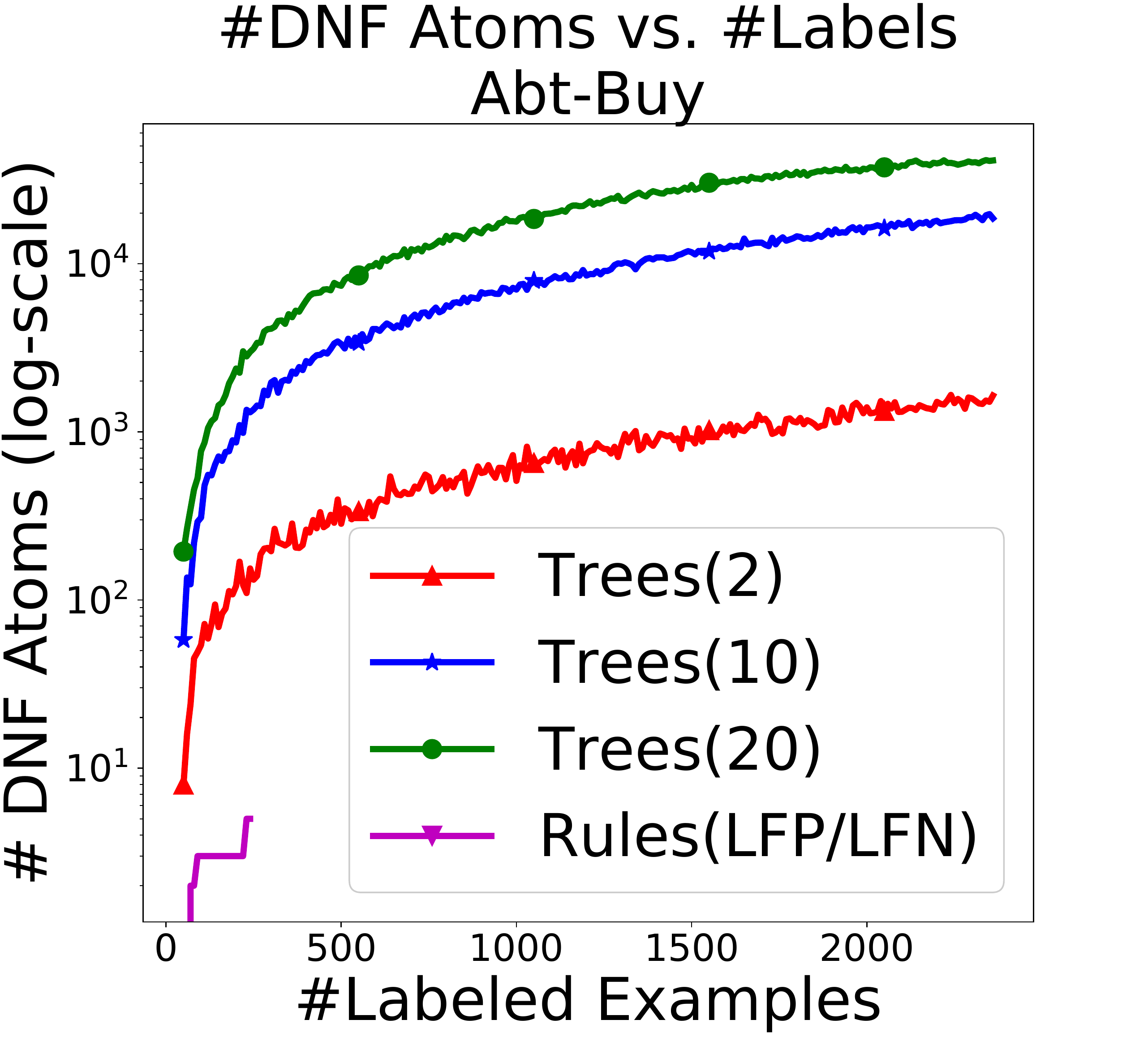}
		\caption{$\#$Atoms (Trees vs. Rules)}
		\label{fig:Abt-Buy-numAtoms}
	\end{subfigure}
	\begin{subfigure}[t]{0.19\textwidth}
		\captionsetup{singlelinecheck = false, format= hang, justification=raggedright, font=footnotesize, labelsep=space}
		\centering
		\includegraphics[width=\linewidth]{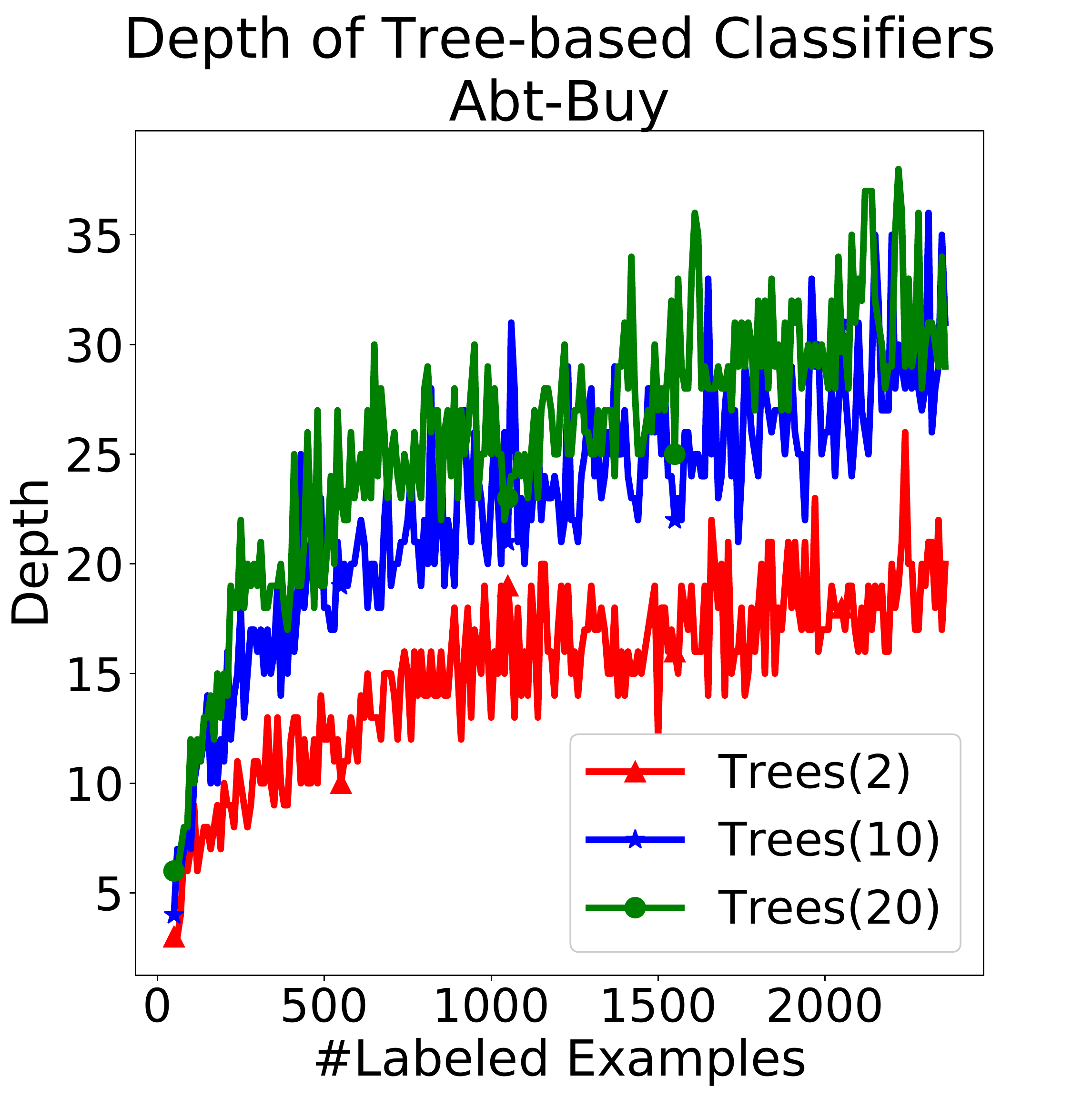}
		\caption{Tree Ensemble Depth}
		\label{fig:Abt-Buy-forestDepth}
	\end{subfigure}
	\vspace*{-.4cm}
	\caption{Interpretability Experiments}
	\vspace*{-.3cm}
	\label{fig:interpretability}
\end{figure}
While model interpretability has been established to be crucial for supervised learning-based EM by earlier works such as~\citet{conciseERVLDB}, it is also important for active learning.
In the case of supervised learning, interpretable models are used for explainability purposes in order to understand why a particular model produces higher quality of matches than a different model and also for debugging purposes to reduce false positives and false negatives, and thereby enhance precision and recall. While all these benefits also exist for active learning, a direct usage of interpretable models is to decide whether or not to accept a model into the active ensemble in a learning iteration and when to terminate active learning under the absence of ground truth.
In this section, we contrast the $\#$atoms in rule DNFs learned by LFP/LFN with those of the DNF formulae obtained using random forests. We convert the trees learned by random forests into DNF formulae by traversing the path from the root of the tree until all the leaf nodes whose predicted label is 1 or $matching$. The path turns into a conjunction of rule-based predicates, and the disjunction or union of all such formulae leads to a DNF. We do not further optimize the DNFs into concise Boolean formulae unlike~\citet{conciseERVLDB} as the latter may seem concise but need to be mentally unrolled into DNFs by a human. This is because, DNFs are more intuitive to a human. It is therefore possible that there are overlapping atoms across different conjunctive predicates in a DNF and they are counted with repetition to compute $\#$atoms for both rules and random forests. As mentioned in Section~\ref{sec:benchmark}, an atom can be defined~\cite{conciseERVLDB} as a DNF predicate or a similarity function evaluated on a pair of attributes from two records and compared against a numerical threshold. We can observe from Figs.~\ref{fig:Abt-Buy-numAtoms} $\&$~\ref{fig:Abt-Buy-forestDepth} that \# DNF atoms in the learned trees as well as their depths increase with more active learning iterations, since larger tree ensembles contain more atoms than the smaller ones. The depth of a tree ensemble is the maximum among the depths of all the trees in the random forest. We can notice from Fig.~\ref{fig:Abt-Buy-numAtoms} that rules have significantly fewer atoms than random forests on all the datasets and are thus easily interpretable by a human. 

Following is the ensemble of rules learned by LFP/LFN active learning heuristic for the Abt-Buy dataset. Each of these rules has a test precision $\ge$ 0.88 and is accepted into the ensemble at a distinct iteration. Similar concise DNF rule ensembles were obtained on other datasets as well. We do not present the DNF rules for trees as they are prohibitively large.
\newline\newline
\textbf{\scriptsize \underline{Abt-Buy} (\# DNF Atoms = 5):\newline
	\emph{\underline{Rule 1}}: Abt.price = Buy.price\newline
	$\wedge$ JaccardSim(Abt.name, Buy.name) $\ge$ 0.4\newline
	$\vee$\newline
	\emph{\underline{Rule 2}}: JaccardSim(Abt.name, Buy.name) $\ge$ 0.7\newline
	$\vee$\newline
	\emph{\underline{Rule 3}}: JaccardSim(Abt.name, Buy.name) $\ge$ 0.6\newline 
	$\wedge$ JaccardSim(Abt.description, Buy.description) $\ge$ 0.1}

\subsubsection{LFP/LFN vs. QBC (Rules) on Social Media Dataset}
\label{sec:socialMedia}
We use an EM dataset from~\citet{LFPLFN}, where the goal is to match 467,761 employee records from a large enterprise 
to a set of 50M user profiles from a social media platform. The attributes comprise name, location, email address, occupation, gender and a URL to the personal homepage for each user profile. We use this dataset to compare the learner-agnostic QBC with committee sizes ranging from 2 to 20, against the learner-aware heuristic of LFP/LFN on rule-based classifiers. 
\begin{figure}[htb]
	\centering
	\vspace*{-.4cm}
	\hspace*{-0.9cm}
	\includegraphics[width=1.2\linewidth]{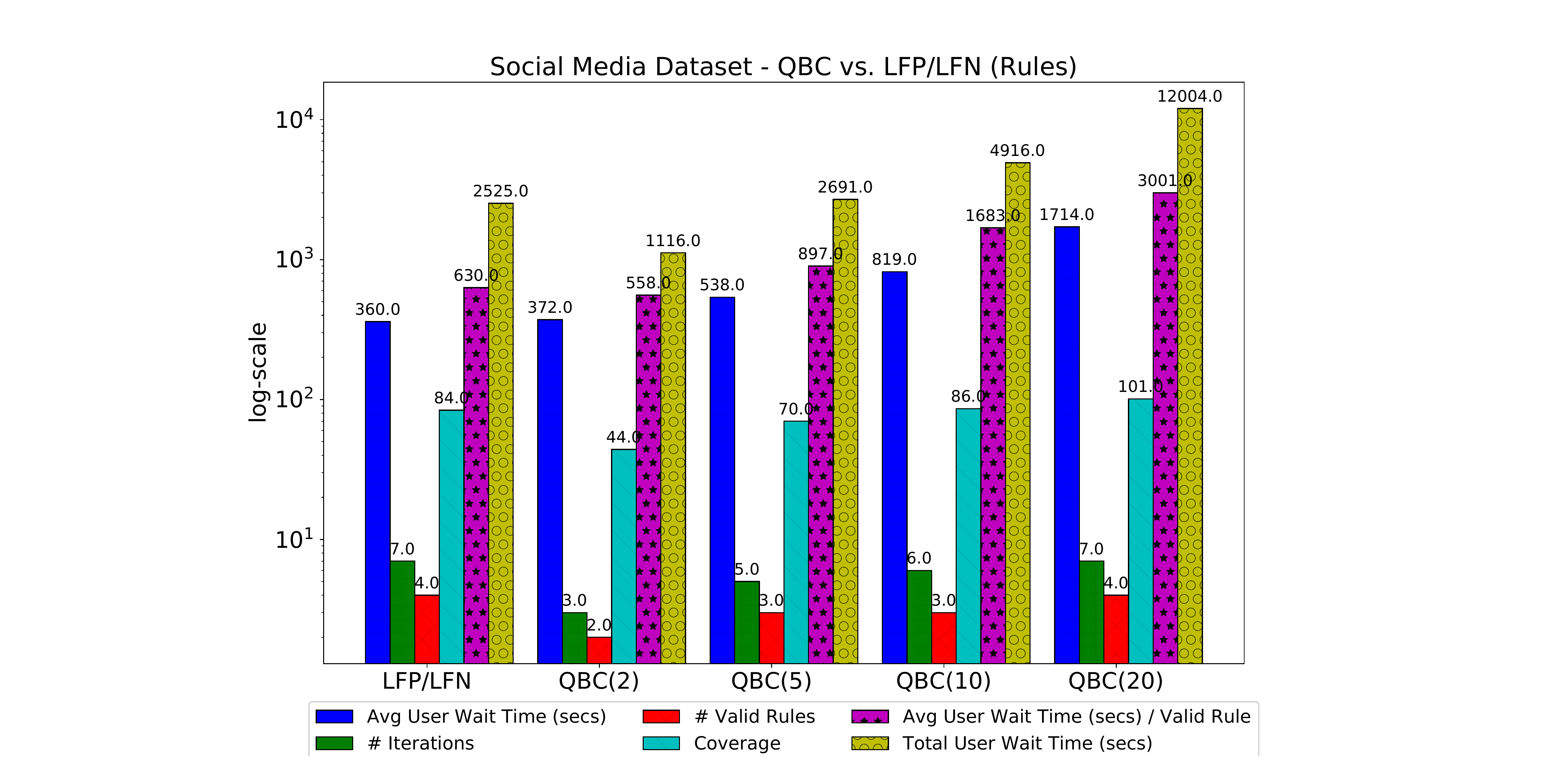}
	\vspace*{-.6cm}
	\caption{QBC vs. Heuristic Strategies (Rules)}
	\vspace*{-.6cm}
	\label{fig:LFPLFNvsQBC}
\end{figure}
\RONE{In the absence of ground truth for this real-world dataset, 
we evaluate example selection strategies indirectly   
based on the actual rules they produce. Each learned rule is interpreted by a human expert and manually validated by labeling adversarial examples (i.e., LFPs). Once accepted by the human expert, it can be reasonably assumed that they reached high precision. Thus, the quality of a selection strategy is determined by the number of manually accepted (validated) rules along with their aggregate number of predicted matches (called coverage) on the dataset. Since active learning on this dataset requires human validation of the model at each step, it is essential that the model used is interpretable. Therefore, we do not conduct this experiment with the remaining learners.}
From Fig.~\ref{fig:LFPLFNvsQBC}, we note that LFP/LFN performs comparably to larger bootstrap committees of sizes 10 and 20 on coverage and 
\# valid rules, respectively, while being 1.9x and 4.7x faster w.r.t. the total user wait time (across all iterations) which includes rule learning, rule execution and example selection times. QBC of committee size 2 is faster than LFP/LFN but produces fewer valid rules with less coverage. Fig.~\ref{fig:LFPLFNvsQBC} also plots average user wait time taken 
to learn a valid rule, average user wait time per iteration and \# iterations.
\section{Conclusion}
\label{sec:conclusion}
In this paper, we proposed a unified active learning benchmark framework that can mix-and-match several learners with multiple example selectors for entity matching (EM). Using the framework, 
we found that active learning upon learner-aware ensembles of tree-based models achieves close to perfect progressive F1-scores on all the public EM datasets we experimented with. Our best active learning methods require fewer $\#$labels for a convergent F1-score than their supervised learning counterparts up until 10$\%$ labeling noise and also surpass a state-of-the-art supervised learning algorithm on perfect Oracles.
We also found that tree-based learners achieve high quality at the expense of interpretability and applications where concise, highly precise EM rules are required may still resort to rule-based learning. Our experiments on a real-world social media dataset lacking ground truth emphasize the need for interpretable models which are manually validated in each active learning iteration. 
%
\bibliographystyle{ACM-Reference-Format}
\balance
\bibliography{ms}

\end{document}